\documentclass[a4paper,11pt]{article}
\usepackage{jheppub} 
\usepackage{comment}
\usepackage{tikz}
\usepackage{hyperref}
\usepackage{mycommands}
\usepackage{amssymb}
\newcommand{\sumint}{\mathop{\mathpalette\dosumint\relax}}
\newcommand{\dosumint}[2]{%
  \vphantom{\sum}%
  \ooalign{%
    \hidewidth$\displaystyle#1\sum$\hidewidth\cr
    $\displaystyle#1\int$\cr}%
}
\usepackage{color}

\title{Toward a first-principles description of transverse momentum dependent Drell-Yan production in proton-nucleus collisions}

\affiliation[a]{Key Laboratory of Quark and Lepton Physics (MOE) \& Institute of Particle Physics, Central China Normal University, Wuhan 430079, China}
\affiliation[b]{Theoretical Division, Los Alamos National Laboratory, Los Alamos NM 87545, United States
}

\author[a,b]{Weiyao Ke}
\emailAdd{weiyaoke@ccnu.edu.cn}
\author[b]{John Terry}
\emailAdd{jdterry@lanl.gov}
\author[b]{and Ivan Vitev}
\emailAdd{ivitev@lanl.gov}

\abstract{
In this paper, we study the parton dynamics in Drell-Yan collisions involving proton-nuclei interactions in the limit of small transverse momentum, emphasizing the role of the cold nuclear matter effects. The distribution of transverse momentum that enter into these collisions differs from that in Drell-Yan collisions with free nucleons in two distinct ways: the intrinsic parton structure of the TMDs are altered, the perturbative dynamics undergo additional modification due to interactions with the nuclear medium. In this paper, we focus on the perturbative dynamics, which we demonstrate enter from forward scattering between the parton constituents of the proton and the nuclear medium. We then derive these partonic contributions to the TMD Drell-Yan cross section up to next-to-leading order in the strong coupling constant and to the first order in the medium opacity. We demonstrate that the collinear and rapidity divergences related to parton showers in matter lead to i) an in-medium renormalization group equation that encodes the transverse momentum dependence of parton energy loss, and ii) a Balitsky-Fadin-Kuraev-Lipatov evolution equation for the forward scattering cross section.  We discuss the relation of our results to the phenomenological extraction of nuclear TMDs and apply the new formalism to Drell-Yan production at small transverse momenta in $p$+$A$ reactions.
}

\begin{document}
\preprint{
LA-UR-24-21409\\
\makebox[\textwidth][r]{INT-PUB-24-034}
}
\maketitle
\flushbottom

\section{Introduction}\label{sec:intro}

The Drell-Yan (DY) process~\cite{Drell:1970wh}, in which hadronic collisions produce a final-state lepton pair
\begin{align}
    a(P_a)+b(P_b) \rightarrow \left[\ell^++\ell^-\right]\left(Q\right)+X\,,
\end{align}
has served as one of the primary windows into the structure of nucleons. Fundamental to achieving the goal of extracting meaningful information for the nucleon structure are the  Quantum Chromodynamics (QCD) factorization and resummation formalisms. In the case of the Drell-Yan process with free nucleons, this factorization formalism has been well understood for some time~\cite{Collins:1984kg} and all modern analyses of collinear Parton Distribution Functions (PDFs) have relied on this formalism along with Drell-Yan data~\cite{NNPDF:2021njg,Hou:2019efy,Bailey:2020ooq,Alekhin:2017kpj}. However, a major goal of the nuclear physics community in the past decade has been to understand the three-dimensional structure of nucleons, which is encoded in Transverse Momentum Dependent PDFs (TMD PDFs). Advances in our understanding of perturbative calculations~\cite{Duhr:2022yyp,Moult:2022xzt} of these distributions has resulted in very high precision extractions of TMD PDFs~\cite{Anselmino:2013lza,Bacchetta:2017gcc,Scimemi:2017etj,Bertone:2019nxa,Scimemi:2019cmh,Bacchetta:2019sam,Moos:2023yfa,Bacchetta:2024qre}, where Drell-Yan collision data has served as a primary input. These high order perturbative computations are enabled by the development of Soft-Collinear Effective Theory (SCET)~\cite{Bauer:2000yr,Bauer:2001yt,Beneke:2002ph,Bauer:2001ct} and the application of this Effective Field Theory (EFT) for TMD observables~\cite{Echevarria:2011epo,Echevarria:2012js}. Despite the progress in our understanding of TMD observables with free nucleons, our understanding of the three-dimensional structure of nuclei has only begun to emerge.

The early experiments by the EMC collaboration with nuclear targets revealed that bound nucleons exhibit non-trivial modifications to their structure~\cite{EuropeanMuon:1983wih}. As nuclear matter accounts for the vast majority of the mass of the visible universe, understanding its properties from first principles has remained one of the utmost goals of the nuclear physics community~\cite{Accardi:2012qut,AbdulKhalek:2021gbh,Anderle:2021wcy,Albacete:2013ei,Citron:2018lsq,Achenbach:2023pba}. 
Despite the clear need for a transparent framework which allows one to factorize perturbative and non-perturbative (NP) contributions to the Drell-Yan process involving nuclei, this has not yet been accomplished. Various approximations involving parameterization  of cold nuclear matter contributions have been used to circumvent this problem for PDFs~\cite{Eskola:1998df,AbdulKhalek:2022fyi,Eskola:2021nhw}, fragmentation functions (FFs)~\cite{Sassot:2009sh,Zurita:2021kli}, and more recently for TMD processes~\cite{Alrashed:2021csd,Barry:2023qqh,Alrashed:2023xsv,Fang:2023thw,Gao:2023ulg}. In these approximate schemes, the perturbative evolution has been assumed to be the same as in the vacuum while the cold nuclear medium contributions have been absorbed into the non-perturbative initial condition. As energetic partons traverse a QCD medium however, the energy scale of the interaction with the medium can be either perturbative or non-perturbative and, thus, will simultaneously alter the factorization and resummation formalism for the process and introduce additional non-perturbative contributions. However, the community currently lacks a proper TMD formalism for this process. In this paper, we establish this new factorization and resummation framework.

The physics of elastic, coherent, and inelastic scattering  in large nuclei has been studied extensively in the past decades, focusing on specific nuclear effects. Already at tree level, multiple parton interactions in cold QCD matter lead to cross section enhancement at small to intermediate transverse momenta and may underlie the Cronin effect~\cite{Wang:1998ww,Zhang:2001ce,Accardi:2001ih,Vitev:2002pf,Kopeliovich:2002yh,Ke:2022gkq}. In addition, they induce acoplanarity and broadening of particle and jet correlations in the final-state~\cite{Bordes:1980ec,Qiu:2003pm,Kang:2011bp,Gyulassy:2018qhr,Jia:2019qbl,Barata:2020rdn}.        
For small values of Bjorken-$x$, scattering on nuclei becomes coherent and can lead to cross section suppression, proving a physical picture for shadowing~\cite{Frankfurt:2002kd,Qiu:2003vd,Qiu:2004da,Frankfurt:2011cs}. At one loop and beyond,  bremsstrahlung processes in matter contribute to cross section suppression in $e$+$A$ and $p$+$A$ reactions~\cite{Gavin:1991qk,FNALE772:2000fmo,Arleo:2002ph,Neufeld:2010dz,Xing:2011fb,Guo:2000nz,Arleo:2003jz,Accardi:2009qv,Arleo:2013zua,Kang:2015mta}. More recently, the final-state radiative corrections in deep inelastic scattering have been investigated using an EFT approach~\cite{Li:2020rqj,Li:2023dhb}, including a first-principles renormalization group (RG) analysis~\cite{Ke:2023ixa}, albeit limited to collinear factorization. In this paper we demonstrate how this approach can be extended to TMD distributions and the Drell-Yan process. We note that Quantum Electrodynamics interactions of the charged leptons mediated by Glauber photons can also lead to broadening and radiative corrections~\cite{Tomalak:2023kwl,Tomalak:2024lme}. Since they scale as $(\alpha/\alpha_s)^n$, they have been estimated to be on the order of per-mille to percent level and we do not consider them here.    

\begin{figure}[hbt!]
\centering
\includegraphics[scale=1]{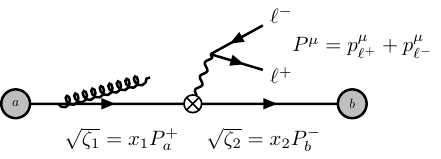}~~~\raisebox{-.75cm}{\includegraphics[scale=1]{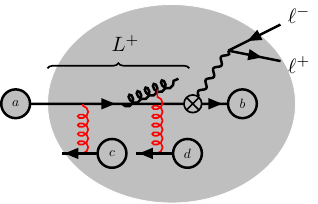}}
\caption{Illustration of the Drell-Yan process in $p$+$p$ (left) and $p$+$A$ collisions (right). For the $p$+$A$ diagram, $a$ and $b$ are nucleons that participate in the hard process, while $c$ and $d$ are spectator nucleons that participate in the forward scattering in the initial state of the hard process.} 

\label{fig:illustration}
\end{figure}

The physical picture that underlies the factorization and resummation formalism we will derive is illustrated on the left and right sides of figure~\ref{fig:illustration} for a  NLO contribution to DY in $p$+$p$ and $p$+$A$ collisions, respectively. In nuclear collisions, the TMD observable is sensitive to two types of nuclear matter effects, those that can be obtained through a perturbative calculation and those that are non-perturbative. The perturbative cold nuclear matter (CNM) effects are a consequence of partonic interactions, illustrated by the vertical gluon lines, between the proton-collinear parton and the anti-collinear scattering centers in the nuclear medium.  The intrinsic non-perturbative nuclear effects, however, account for all information below the scale $\Lambda_{\rm QCD}$. These effects can be the Fermi motion of nucleons, inter-nucleon correlations, and the intrinsic non-perturbative motion of partons in bound nucleons different from the one in free nucleons. 
In the past, the authors have done extensive work towards a perturbative treatment of the dynamical cold nuclear matter effects in $p$+$A$ collisions. These include the dynamical shadowing effect from coherent multiple interactions~\cite{Qiu:2004da}, the Cronin transverse momentum broadening ~\cite{Cronin:1974zm,Vitev:2003xu}, the cold nuclear matter energy loss~\cite{Vitev:2007ve}, and the complete set of in-medium splitting functions in the initial state~\cite{Ovanesyan:2015dop}. These results were then added independently in phenomenological studies of inclusive jet production~\cite{Kang:2015mta}, light and heavy meson productions~\cite{Vitev:2003xu,Ke:2022gkq}, and di-hadron productions in proton-nucleus collisions~\cite{Kang:2011bp}. Nevertheless,  a coherent framework that naturally incorporates in-medium parton showers, momentum broadening, and their correlations is still missing.
In this paper, we will use EFT techniques to arrive at a three-dimensional description of the perturbative CNM effects in $p$+$A$ collisions. Furthermore, this new formalism allows for a systematic treatment of the radiative corrections to the transverse momentum broadening, which is not included in previous studies. It naturally builds upon the transverse-momentum-dependent factorization framework and is applied to phenomenology.

The rest of this paper is organized as follows. In section~\ref{sec:TMD-vac}, we define the kinematic variables, review the Drell-Yan TMD factorization for $p$+$p$, and present the cross section at NLO+NNLL accuracy. Section~\ref{sec:CNMDY-overview} gives a high-level discussion of CNM effects from an EFT point of view. 
In section~\ref{sec:TMD-med-scales}, we discuss the power counting in $p$+$A$ collisions and consider two possible separations of scales.
In section~\ref{sec:TMD-NLO-matching}, based on the in-medium power counting, we compute the NLO TMD matching coefficients at first order in opacity for the collinear sector, from which we identify the collinear and rapidity divergences.
We will also use a previously developed approach to treat the collinear divergence, which leads to an in-medium RG equation that encodes TMD parton energy loss.
Section~\ref{sec:TMD-NLO-soft} demonstrates the cancellation of the rapidity divergence once we include the calculation for the soft and the anti-collinear sectors, which leads to the Balitsky-Fadin-Kuraev-Lipatov (BFKL) evolution~\cite{Balitsky:1978ic,Fadin:1975cb} with respect to the rapidity scale. We complete the reorganized NLO calculation to first order in opacity in section~\ref{sec:final-formula} and discuss the inclusion of multiple Glauber exchanges at higher-order opacity. We also comment on the connection of this paper to the study of jet broadening in heavy ion collisions from series of studies~\cite{Wu:2011kc,Liou:2013qya,Blaizot:2014bha,Vaidya:2020lih,Vaidya:2021vxu,Caucal:2021lgf,Caucal:2022fhc,Caucal:2022mpp}.
Section~\ref{sec:result} discusses the choice of the CNM parameters and presents phenomenological applications to the study of TMD Drell-Yan process in $p$+$A$.
Finally, section~\ref{sec:summary} summarizes the paper.

\section{TMD factorization and resummation in elementary collisions}\label{sec:TMD-vac}

\subsection{Kinematics}
In this paper, we make a frame choice such that the incoming hadron moves in the collinear direction while the incoming nucleon moves in the anti-collinear direction. In this frame, we can parameterize the four momenta of these particles as
\begin{align}
P_a^\mu = \bar{n}\cdot P_a \frac{n^\mu}{2}+\frac{m^2}{\bar{n}\cdot P_a} \frac{\bar{n}^\mu}{2}\,,
\qquad
P_b^\mu = n\cdot P_b \frac{\bar{n}^\mu}{2}+\frac{m_N^2}{n\cdot P_b} \frac{n^\mu}{2}\,,
\end{align}
where we use the convention that $a$ and $b$ denote a free projectile hadron (proton or pion) and a bound nucleon, respectively.  We take the SCET convention that $n^\mu = \hat{t}^\mu+\hat{z}^\mu$ and $\bar{n}^\mu = \hat{t}^\mu-\hat{z}^\mu$. Here we use $m$ and $m_N$ to denote the mass of the collinear hadron and anti-collinear nucleon. For simplicity, these momenta can be written as
\begin{align}
P_a^\mu = \left(P_a^+, \frac{m^2}{P_a^+}, \bm{0}_T\right)\,, \qquad
P_b^\mu= \left(\frac{m_N^2}{P_b^-}, P_b^-, \bm{0}_T\right),
\end{align}
where $p^+ = \bar{n}\cdot p$ and $p^- = n\cdot p$ for an arbitrary momentum $p$. Unless stated otherwise, we will now drop mass corrections. The Drell-Yan observable is differential in the invariant mass $Q$, the transverse momentum $\mathbf{P}_T$\footnote{In this paper, a subscript $T$ denotes a vector that is transverse to the beam axis, while a subscript $\perp$ means a vector transverse to a reference parton}, and the rapidity $y$ of the lepton pair. In the lab frame, the pair has the four-momentum
\begin{align}
    Q^\mu = \left(M_T\, e^y, M_T\, e^{-y}, \bm{P}_T\right)\,,
\end{align}
where $M_T = \sqrt{P_T^2+Q^2}$ is the transverse mass of the pair. The light-cone momenta of the partons are controlled by the momentum fraction variables
\begin{align}
    x_1 = \frac{M_T^2}{2 P_a\cdot Q}\,,
    \qquad
    x_2 = \frac{M_T^2}{2 P_b\cdot Q}\,.
\end{align}

Thus the partonic momenta can be written as
\begin{align}
    p_a^\mu = \left(x_1 P_a^+, p_a^-, \bm{p}_{1 T}\right)\,,
    \qquad
    p_b^\mu = \left(p_b^+, x_2 P_b^-, \bm{p}_{2 T}\right)\,,
\end{align}
where the small components of the momentum are fixed by the on-shell conditions at tree level but are off-shell at higher order.

\subsection{Factorization}
The TMD factorization applies in the kinematic region where $\Lambda_{\rm QCD} \lesssim P_T \ll Q$. The large difference between the transverse momentum and invariant mass is handled in SCET by introducing a power counting parameter $\lambda \sim P_T/Q$. In this region, the infrared (IR) dynamics of QCD are asymptotically captured by the three QCD modes with the scalings
\begin{align}
    p_c \sim Q\left(1, \lambda^2, \lambda\right)\,,
    \qquad
    p_{\bar{c}} \sim Q\left(\lambda^2, 1, \lambda\right)\,,
    \qquad
    p_s \sim Q\left(\lambda, \lambda, \lambda\right)\,,
\end{align}
where we use the typical convention that the third component denotes the magnitude of the transverse momentum. Under an operator product expansion in this region, the factorization takes on the simple form
\begin{align}\label{eq:TMDDY-dsigma}
\frac{d\sigma}{d\mathcal{PS}} &= \frac{4\pi \alpha_{\rm em}^2}{3N_c Q^2 s} H(Q,\mu) \sum_q  c_q(Q) \int \frac{d^2\bfb}{(2\pi)^2} e^{i\bfb\cdot \bm{P}_T} \nonumber \\
&\times \mathcal{B}_{q/a}\left(x_1, b, \mu, \frac{\zeta_1}{\nu^2}\right) \mathcal{B}_{\bar{q}/b}\left(x_2, b, \mu, \frac{\zeta_2}{\nu^2}\right)\, S(b, \mu, \nu)  \,,
\end{align}
where $d\mathcal{PS} = dy\, dQ^2\, d^2 \mathbf{P}_T$ and $\bm{b}$ is the Fourier conjugate to $\bm{P}_T$, see for instance \cite{Bacchetta:2022awv, Bacchetta:2024qre} for a discussion on this cross section. $c_q(Q)$ is the effective squared charge taking into account the interference between $\gamma$ and $Z$ bosons exchange. The expression for $c_q(Q)$ can be found in Ref.~\cite{Bacchetta:2024qre, Bacchetta:2024qre}. For $Q$ much smaller than the Z-boson mass, $c_q(Q)\approx e_q^2$.
We have introduced the usual functions $H$, $\mathcal{B}$, and $S$, which are the hard matching function, the beam function, and the global soft function. The vacuum beam function and soft functions can be defined in terms of the matrix elements
\begin{align}
    \mathcal{B}_{q/a}\left(x, b, \mu, \frac{\zeta_1}{\nu^2}\right) =&  \int \frac{d^4 z}{2\pi}\, e^{i z \cdot p}\delta\left(n\cdot z \right) \delta^{2}\left(\bm{z}_T- \bm{b} \right)\nonumber\\
    &\times\frac{1}{2 N_c}\operatorname{Tr}\left[\left \langle P_a \left |  \bar{\chi}_n(z) \frac{\slashed{\bar{n}}}{2} \chi_n\left(0\right) \right | P_a \right \rangle \right] \,,\\
    S(\bfb,\mu,\nu) =& \frac{1}{N_c}\langle0|\operatorname{Tr}\left[W_{\softstaple}(\bfb)\right]|0\rangle\,.
\end{align}
$\chi_n$ is the gauge-invariant collinear quark field. $W_{\softstaple}$ is a staple-like Wilson line of soft gluon fields defined in Ref.~\cite{Boussarie:2023izj}.
The trace is taken in the color space for both $\mathcal{B}_{q/a}$ and $S$, and it also sums the spin of quark in $\mathcal{B}_{q/a}$.
The separation of the matching function to the IR modes is controlled by the renormalization group scale $\mu$, while the separation of the IR modes into either collinear, soft, or anti-collinear is controlled by the rapidity renormalization group scale $\nu$. Lastly, the collinear modes depend on the Collins-Soper (CS) scales $\sqrt{\zeta_1} = p_a^+$ and $\sqrt{\zeta_2} = p_b^-$ with $\sqrt{\zeta_1\zeta_2}=Q^2$.

In the kinematic region where $P_T \gg \Lambda_{\rm QCD}$, the beam functions can be further factorized in an operator production expansion. In this case, the perturbatively generated transverse momentum is captured by the matching function $C_{q/i}$, while the IR dynamics of the collinear emissions are captured by the collinear parton distribution function (PDF) $f_{i/h}$. The beam function then takes on the form
\begin{align}
\mathcal{B}_{q/a}\left(x_1, b, \mu, \frac{\zeta_1}{\nu^2}\right) = \sum_i \int_{x_1}^1 \frac{dx}{x}C_{q/i}\left(x, b, \mu, \mu_i, \frac{\zeta_1}{\nu^2}\right)f_{i/a}\left(\frac{x_1}{x}, \mu_i\right) + \mathcal{O}(b^2\Lambda_{\rm QCD}^2),
\end{align}
where the summation $i$ runs over quarks, anti-quarks, and gluons. To simplify the discussion, we move the expressions for the matching coefficient function to the appendix~\ref{app:vac-matching}. In this expression, $\mu_i$ denotes the scale at which the beam function is matched onto the PDF, which is normally taken to be $\mu_i = \mu \sim 1/b$ and cancels between the matching coefficient and the PDF. Lastly, the terms in the $\mathcal{O}$ denote contributions associated with the non-perturbative intrinsic transverse momentum of the bound partons, which can be parametrized and determined from a global QCD analysis~\cite{Anselmino:2013lza,Bacchetta:2017gcc,Scimemi:2017etj,Bertone:2019nxa,Scimemi:2019cmh,Bacchetta:2019sam,Moos:2023yfa}.

\subsection{Resummation of large logarithms}
\begin{figure}[hbt!]
\centering
\includegraphics[height=.475\textwidth]{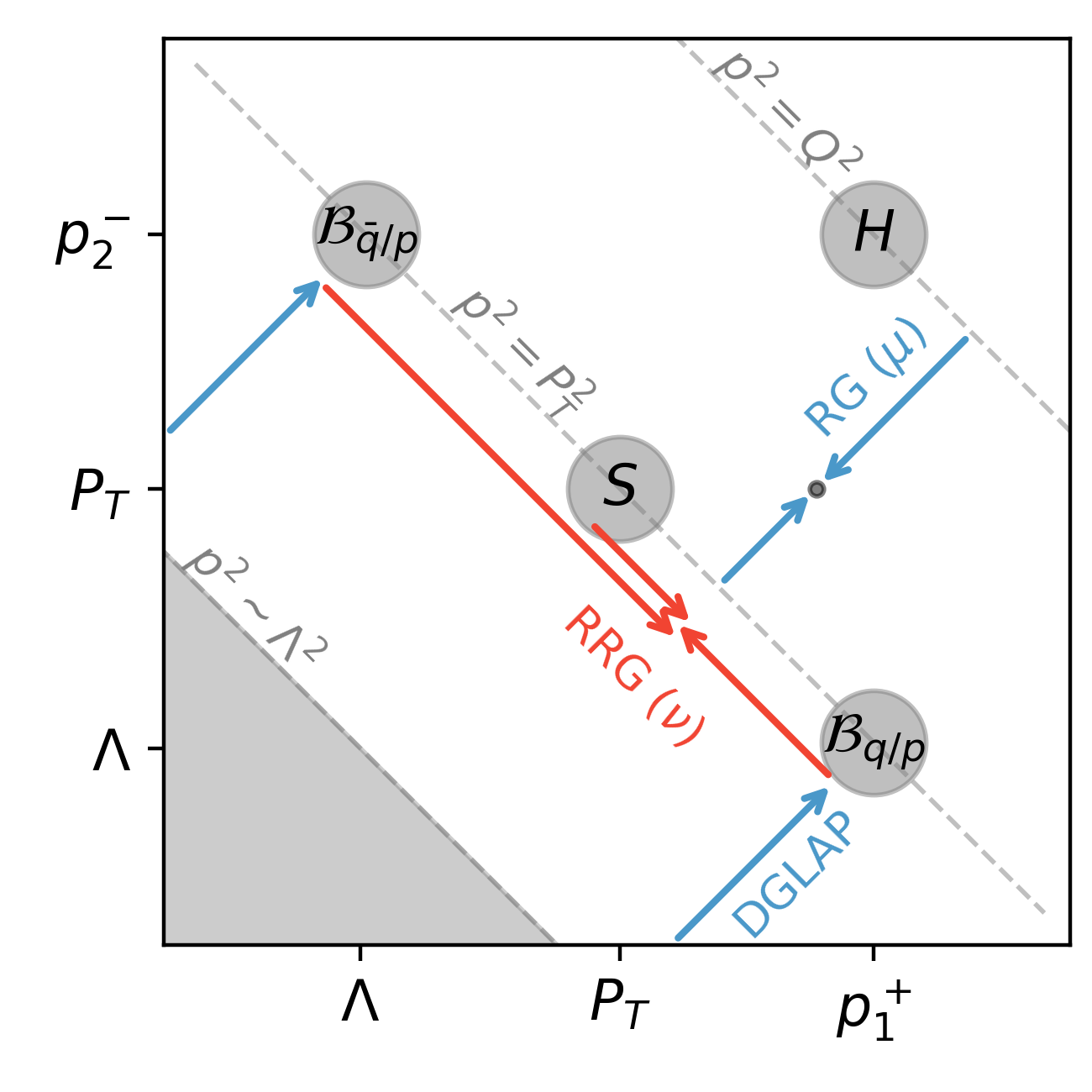}
\caption{Modes in TMD factorization in elementary collisions organized by their plus and minus components (put in log scale). The arrow represents the resummation of large logarithms in each mode. 
}
\label{fig:TMD-DY-pp}
\end{figure}

Each SCET mode captures the asymptotic behavior of QCD in a particular region of invariant mass and rapidity. In the standard SCET language, the SCET modes can be organized in terms of their light-cone components as shown in figure~\ref{fig:TMD-DY-pp}. In this language, we define ``natural" scales for each mode, namely the SCET modes and the matching coefficient will resolve contributions at the RG scale
\begin{align}
    \mu_B \sim \mu_S \sim 1/b\,,
    \qquad
    \mu_H \sim Q\,,
\end{align}
where the subscript denotes the relevant contribution to the factorized cross section. Based on this, each contribution is responsible for performing the resummation of logs of the form $\ln\left(\mu/\mu_i\right)$ where $i$ is either $H$, $B$, or $S$. As the SCET modes also depend on a rapidity scale, these functions have natural rapidity scales
\begin{align}
    \nu_B \sim \sqrt{\zeta} \sim Q\,,
    \qquad
    \nu_S \sim \mu\,,
\end{align}
and rapidity RG (RRG) is responsible for resumming logs of the form $\nu/\nu_i$.

The resummation of these large logs is accomplished by solving the renormalization group (RG) equations and rapidity renormalization group (RRG) equations of each sector. The RG equations are~\cite{Collins:2011zzd,Aybat:2011zv,Boussarie:2023izj}
\begin{align}
\frac{d}{d \ln\mu} \ln H(Q,\mu) &= \gamma_\mu^H(Q,\mu)\,, \\
\frac{d}{d\ln\mu} \ln \mathcal{B}\left(x, b, \mu, \frac{\zeta}{\nu^2}\right) &= \gamma_\mu^B\left(\mu,\frac{\zeta}{\nu^2}\right)\,, \\
\frac{d}{d \ln\mu} \ln S\left(b, \mu, \nu\right) &= \gamma_\mu^S\left(\mu, \frac{\mu}{\nu}\right)\,.
\end{align}
The RRG equations are~\cite{Collins:2011zzd,Aybat:2011zv,Chiu:2012ir,Boussarie:2023izj}
\begin{align}
\frac{d}{d \ln\nu} \ln \mathcal{B}\left(x, b, \mu, \frac{\zeta}{\nu^2}\right) &= \gamma_\nu^B(b,\mu)\,, \nonumber\\
\frac{d}{d \ln\nu} \ln S\left(b, \mu, \nu\right) &= \gamma_\nu^S(b,\mu)\,.
\end{align}
All  anomalous dimensions satisfy the following consistency conditions
\begin{align}
\gamma_\mu^H(Q,\mu) + \gamma_\mu^S\left(\mu,\frac{\mu}{\nu}\right) + \gamma_\mu^B\left(\mu, \frac{\zeta_1}{\nu^2}\right) + \gamma_\mu^B\left(\mu, \frac{\zeta_2}{\nu^2}\right) &= 0, \\
\gamma_\nu^S(b,\mu) + 2\gamma_\nu^B(b,\mu) &= 0,
\end{align} 
that guarantee $\mu$ and $\nu$ invariance of the cross section.
Each sector evolved from its natural scales to the same set of $\mu$ and $\nu$ to minimize the uncertainty of perturbation theory. In this study, the vacuum TMD factorization calculation is accurate to NLO+NNLL accuracy and the required anomalous dimensions are also listed in appendix~\ref{app:vac-matching}.

\section{Overview of dynamical CNM effects in Drell-Yan}\label{sec:CNMDY-overview}
\begin{figure}
\centering
\includegraphics[width=.8\textwidth]{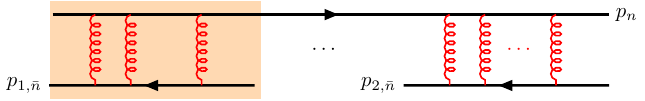}
\caption{This figure illustrates multiple Glauber exchanges between a collinear jet parton and anti-collinear medium partons. Multiple collisions can happen between a jet parton and a medium parton, or between a jet parton and multiple medium partons. The number of the latter is referred to as the opacity.}
\label{fig:opacity-definition}
\end{figure}

In nuclear collisions, the CNM effects are derived by considering the interactions of the nuclear medium with the active partons. This is illustrated in figure~\ref{fig:opacity-definition}, where a collinear active parton $p_n$ interacts with two anti-collinear partons $p_{1,\nbar}$ and $p_{2,\nbar}$ from the spectator nucleons of the target. 
Such forward scatterings are mediated by the so-called Glauber gluons shown in red in figure~\ref{fig:opacity-definition}.
Because SCET is constructed by considering on-shell (anti)collinear and soft modes, the Glauber interactions cannot knock these modes far off their mass shell. The momentum scaling of the Glauber gluon in this physical picture is
\begin{align}
q^\mu \sim Q(\lambda^\alpha, \lambda^\beta, \lambda).
\end{align}
In this expression $(\alpha,\beta)=(2,2),(1,2),(2,1)$ for Glauber interactions between the collinear-anti-collinear, collinear-soft, and anti-collinear-soft regions, respectively. Based on this power counting, the LO exchange of Glauber gluons cannot alter the large-momentum component of the collinear and anti-collinear partons, but contribute to $P_T$ at leading power. However, at NLO, the Glauber interaction induces radiative corrections that lead to energy loss of the collinear parton. This, in turn, can alter the rapidity spectrum of the cross section.

\paragraph{Multiple-Glauber exchange and the opacity expansion.}
As discussed in Ref.~\cite{Rothstein:2016bsq}, for example, multiple Glauber gluon exchanges are not power-suppressed by the momentum scaling parameter $\lambda$, and one has to consider arbitrary multiples of Glauber gluons between the collinear and anti-collinear parton. This is illustrated in the shaded box of figure~\ref{fig:opacity-definition}. It is was found that the effect of summing over all exchanges is equivalent to the single Glauber exchange diagram times a unitary matrix in the color space. 

In the medium, the opacity is an additional expansion parameter, which counts the number of independent medium partons that participate in the forward scattering. For example, the illustration in figure~\ref{fig:opacity-definition} corresponds to a contribution for opacity $\chi=2$. In addition to independent interactions, there can be correlations between multiple interactions that 
are accounted for by an opacity series. Each additional order produces a multiplicative factor $\chi\sim\rho^- \sigma L^+$ associated to its contribution to the cross section, where $\rho^-$ measures the density of the medium, $\sigma$ is the parton level forward-scattering cross section and $L^+$ is the path-length (see the right panel of figure~\ref{fig:illustration}). Equivalently, we can define the mean-free-path $\lambda_i^+ = (\rho^- \sigma)^{-1}$ with $i=q,g$. Then, the opacity parameter $\chi \sim L^+/\lambda_i^+$ can be interpreted as the average number of medium partons interacting with the jet parton. For a thin medium, the opacity is assumed to be a small number and the organization of correlated interactions in powers of the opacity $\chi^n$ is particularly well-suited.

\paragraph{SCET with Glauber gluon (SCET$_{\rm G}$) and beyond.}
Many techniques have been developed to compute radiative correction associated with the jet-medium interaction. They are mainly driven by the need to understand the jet-quenching phenomena in heavy-ion collisions. 
One systematic way is given by SCET with Glauber gluons (SCET$_{\rm G}$)~\cite{Ovanesyan:2011kn,Ovanesyan:2011xy}. In SCET$_{\rm G}$, the medium partons are treated as color source terms without any dynamics, which generate a background Glauber gluon field. Radiative corrections are then calculated at each order of opacity. 
This approach provides a good description of the medium modifications to observables dominated by the collinear degrees of freedom, for instance, quenching of hadron/jet spectra~\cite{Kang:2014xsa,Kang:2017frl,Li:2018xuv,Ke:2023ixa}. For TMD observables, as we will show in this paper, it is critical to consider soft and anti-collinear modes for renormalization group (RG) consistency. This means that one has to go beyond a background treatment of the medium partons to include their dynamics, see for example Refs.~\cite{Vaidya:2021vxu,Singh:2024vwb}.

In this paper, we calculate the first order in opacity ($\chi^1$) correction to the three-dimensional collinear beam function using SCET$_{\rm G}$ to NLO. Additionally, we calculate the soft and anti-collinear sectors, which are contributions beyond the scope of SCET$_{\rm G}$. In these computations, we discover the appearance of novel rapidity and collinear divergences in each sector. We then show that the rapidity and collinear divergences are either canceled among different perturbative sectors or absorbed into the redefinition of non-perturbative quantities. The NLO cross section expression is then improved by the (rapidity) renormalization group equations and partial summation of contributions from higher orders in opacity.
At the leading logarithmic level, the final formula can be cast into a simple modification to the standard TMD factorization for Drell-Yan process in $p$+$p$.
We will demonstrate how the concept of parton energy loss and a non-trivial transverse momentum broadening, as well as their interplay, emerge from this final formula.

\section{The structure of the first order in opacity correction and scale separation}\label{sec:TMD-med-scales}
\subsection{Factorization structure at first order in opacity }
In the opacity expansion, the Gyulassy-Levai-Vitev (GLV) formalism~\cite{Gyulassy:2000fs,Gyulassy:2000er}, the $p$+$A$ cross-sections are expanded in powers of $\chi$,
\begin{align}
\frac{d\sigma}{d\mathcal{PS}} = \sum_{n = 0}^\infty \frac{1}{n!}\chi^n \, \frac{d\sigma_n}{d\mathcal{PS}}\,.
\end{align}
The first term in this expansion is the cross-section in the absence of nuclear matter, which is given in Eq. (\ref{eq:TMDDY-dsigma}). Terms at order $\chi^n$ come from the $n^{\rm th}$ order correlation between the collinear parton with medium partons along its path of propagation. 
Therefore, for DY this is effectively an opacity expansion of the proton beam function passing the medium
\begin{align}
\mathcal{B}_{q/a} = \mathcal{B}_{q/a,0} +  \chi \, \mathcal{B}_{q/a,1}+ \cdots  \, ,
\end{align}
where the beam function of the proton in the absence of nuclear matter is given by the first term while the second term contains the first non-trivial contribution from matter and is the focus of this paper. For simplicity in this expression, we have dropped explicit dependence on scales and kinematics. 

The opacity-one correction to the cross section can then be written as
\begin{align}
\frac{d\sigma_1}{d\mathcal{PS}} =& \frac{4\pi \alpha_{\rm em}^2}{3N_c Q^2 s} H(Q,\mu) \sum_q  c_q(Q) \int \frac{d^2\bfb}{(2\pi)^2} e^{i\mathbf{P}_T\cdot\bfb} \nonumber\\
& \times \sum_{N\in A} \mathcal{B}_{q/p,1}\left(x_1, b, \mu, \frac{\zeta_1}{\nu^2}; \mu_{E}, \mathcal{L}_1\right)\,   \mathcal{B}_{\bar{q}/N}\left(x_2, b, \mu, \frac{\zeta_2}{\nu^2}\right)\, S(b, \mu, \nu)\,. 
\end{align}
$N$ labels the target nucleon that participates in the hard process. An additional set of medium energy scale $\mu_{E}$ and rapidity logarithm $\mathcal{L}_1$ will show up at first order in opacity. They are introduced in Eqs.~(\ref{eq:energyscale-separation}) and~(\ref{eq:N=1-log}) and their meaning will also become clear in the respective sections.

\begin{figure}
    \centering
    \includegraphics[scale=1.2]{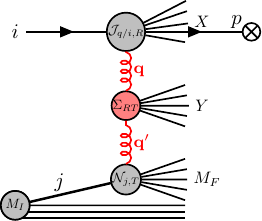}\quad\quad\quad
    \includegraphics[scale=1.2]{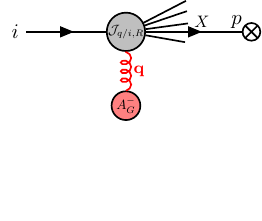}
    \caption{Left: the factorization structure of the forward scattering cross-section between parton $i$ and $j$, producing a final-state $n$-collinear quark that participates in the hard process. Right: in calculating the collinear function, one can apply the background field method,  that underlies  the SCET$_{\rm G}$ approach.}
    \label{fig:Opacity-one-factorization}
\end{figure}

In studies of jet functions in a finite-temperature medium, it has been shown that the first order in opacity correction to the jet function can be factorized into the convolution of a collinear function, a medium function, and a forward scattering cross section~\cite{Vaidya:2020lih,Vaidya:2021vxu}.
From the similarity of the forward scatterings in the final and initial states, we infer that the opacity-one correction to the beam function can be formulated analogously.
At a perturbative transverse momentum, the beam function at opacity one factorizes into
\begin{align}
\mathcal{B}_{q/p,1} &=  \sum_{i=q,g}\sum_{j=q,\bar{q},g}\sigma_{ij\rightarrow q}\otimes f_{i/p} \otimes f_{j/N}\cdot \rho_0^-L^+,
\end{align}
where $f_{i/p}$ is the collinear distribution function of the proton, $\rho_0=0.15$ fm$^{-3}$ is the saturation density of the nuclear matter~\cite{Horowitz:2020evx}, and $f_{j/N}\cdot\rho_0^-L^+$ is the area density to find parton $j$ in the medium. Finally, $\sigma_{ij\rightarrow q}$ is a parton level forward differential cross-section between parton $i$ and $j$ with an $n$-collinear quark $q$ identified in the final state,
\begin{align}
&\sigma_{ij\rightarrow q}(x_1,\bfb) = \int d^2\bfp \, e^{-i\bfb\cdot\bfp} \int \frac{d^2\bfq}{(2\pi)^2} \int \frac{d^2\bfq'}{(2\pi)^2} \sum_{R,T} \nonumber\\
&  \left[\mathcal{J}_{q/i,R}(x_1,\bfp, \bfq)\frac{g_s^2}{\bfq^2}\right]\times \left[\left(\frac{g_s^2}{\bfq^2}\right)^{-1} \Sigma_{RT}(\bfq, \bfq')\left(\frac{g_s^2 }{{\bfq'}^2}\right)^{-1}\right] \times \left[\frac{g_s^2 }{{\bfq'}^2} \mathcal{N}_{j,T}(\bfq')\right] \, .
\label{eq:Opacity-one-factorization-1}
\end{align}
Figure~\ref{fig:Opacity-one-factorization} is a schematic illustration showing the structure of the calculation of $\sigma_{ij\rightarrow q}$. The term $\Sigma_{RT}(\bfq, \bfq')$ is the forward cross-section between currents of color representations $R$ and $T$, with an arbitrary number of soft particles in the final state (denoted as $Y$).
The term $\mathcal{J}_{q/i,R}$ is associated with the collinear parton $i$ interacting with the Glauber gluon under color representation $R$ and producing a final state containing an identified quark carrying momentum fraction $x_1$ and transverse momentum $\bfp$ with the rest of the collinear final state collectively denoted as $X$. 
Lastly, $\mathcal{N}_{j,T}(\bfq')$ is associated to the target parton $j$ interacting with the Glauber gluon under color representation $T$ and transitioning to all possible final states.
The additional powers of $\bfq^2$ and ${\bfq'}^2$ inserted into different sectors in Eq. (\ref{eq:Opacity-one-factorization-1}) are for later conveniences in discussing the renormalization of each sector.
At leading order, $\mathcal{J}, \mathcal{N}$ and $\Sigma$ are given by
\begin{align}
\mathcal{J}_{q/i,R}^{(0)}(x_1, \bfp, \bfq) &= \delta_{iq}\delta_{RF}\delta(1-x_1)\delta^{(2)}(\bfp-\bfq) \, , \\
\mathcal{N}_{j,T}^{(0)}(\bfq') &=\delta_{Tj} \, , \\
\Sigma_{RT}^{(0)}(\bfq, \bfq') &=  \frac{g_s^2 C_R}{\bfq^2+\xi^2} \frac{1}{d_A} (2\pi)^2 \delta^{2}(\bfq-\bfq') \frac{g_s^2 C_T}{{\bfq'}^2+\xi^2}\, . \label{eq:LO-Glauber-cross-section}
\end{align}
The subscript $F,A$ indicate whether the collinear or anti-collinear function is coupled to the Glauber gluon through fundamental or adjoint representations.
The inverse range of the interaction $\xi$ regulates the infrared behavior of the forward cross-section.

\subsection{The background field approach for the collinear function $\mathcal{J}_{q/i, R}$}

To obtain $\mathcal{J}_{q/i, R}$, it is sufficient to replace the interaction with the medium by a background field as shown on the right panel of figure~\ref{fig:Opacity-one-factorization}. This background field approach is referred to as SCET$_{\rm G}$ in the literature~\cite{Ovanesyan:2011kn,Ovanesyan:2011xy}. It provides an efficient path to the computation of $\mathcal{J}_{q/i, R}$, and at the same time helps to elucidate the relevant energy scales in the problem.
Of course, to obtain NLO corrections to $\Sigma_{RT}$ and $\mathcal{N}_{j,T}$, we will go beyond the scope of SCET$_{\rm G}$.

The SCET$_{\rm G}$ Lagrangian is related to the usual SCET Lagrangian through the relation
\begin{align}
    \mathcal{L}_{\rm SCET_G}\left(\chi_n, \mathcal{B}_n, A_G \right)= \mathcal{L}_{\rm SCET}\left(\chi_n, \mathcal{B}_n \right)+\mathcal{L}_{\rm G}\left(\chi_n, \mathcal{B}_n, A_G \right)\,,
\end{align}
where $\mathcal{L}_G$ is given in terms of the quark and gluon SCET building blocks interacting with the background field
\begin{align}
    \mathcal{L}_G \left(\chi_n, \mathcal{B}_n, A_G \right) = \sum_i\, \sum_{p\,, p'} e^{i\left(p-p'\right)\cdot x} \, \Gamma^{\mu, a}_i \left(\chi_n, \mathcal{B}_n, A_G \right)\, A_{G\, \mu, a}(x)\,.
\end{align}
In this expression,  $\Gamma$ denotes operators associated with the interaction of the Glauber vector potential $A_G$ with the collinear sector. The Feynman rules for these interactions are summarized in appendix~\ref{sec:app:SCETG_feynman_rule}.
In SCET${}_{\rm G}$, the Glauber fields are generated as matrix elements of source currents in the nuclear medium
\begin{align}
    A_{G}^{\mu, a}\left(x\right) = \int d^4y\, D^{\mu \nu}\left(x-y\right)\,  \left \langle M_F \left | J_{\nu}^a(y) \right| M_I \right \rangle\,,
\end{align}
where $D^{\mu\nu}(x)$ is the position-space Green's function of the Glauber gluon while $J$ is the QCD current to generate the gluon. 
In this expression, $M_I$ and $M_F$ denote the initial and final-state states of the medium. For the purposes of this paper, it is useful to define the momentum space expression for the vector potential, which can be obtained by Fourier transforming the Green's function of the gluon to obtain
\begin{align}
A_G^{\mu, a}(q^+, q^-, \bm{q}) &= \frac{i}{\bfq^2+\xi^2}  \int d^4 x\, e^{i q \cdot x}\, \langle M_F|J^{\mu, a}(x)|M_I\rangle\,, \label{eq:medium-background-field} 
\end{align}
where power counting has been applied such that $q^2 = -\bm{q}^2 +\mathcal{O}\left(\lambda^3\right)$. Confinement requires that, at the distances separated by the size of a nucleon, correlations of the gluon field vanish implying an effective mass $\xi$ in the Glauber gluon propagator. In this expression, we write the components of the Glauber momentum explicitly for later convenience. Due to the power counting that the plus momentum component of the collinear field is $\mathcal{O}(\lambda^0)$, while the Glauber gluon scales at most like $\mathcal{O}(\lambda)$, the observable does not depend on the plus component ($q^+$) of the Glauber gluon. As a result, the $q^+$ integral can always be performed to $A_G$ and one only needs the information of the medium color sources along the $x^-=0$ light cone. Furthermore, at leading power only the minus component of the background field couples to the collinear field. Combining the two arguments, the relevant component of the vector potential is given by 
\begin{align}
n\cdot A_G^{a}(q^-, \bm{q}) &= \frac{i}{\bfq^2+\xi^2}  \int d^4 x\delta(x^-/2) e^{i q \cdot x}\, \langle M_F|n\cdot J^{ a}(x)|M_I\rangle\,. \label{eq:medium-background-field-2}
\end{align}

In the opacity expansion, $\mathcal{L}_G$ is treated as a perturbation to the SCET Lagrangian. 
In the interaction picture, one expands to the first non-trivial order, which contains two insertions of $\mathcal{L}_G$, and arrives at the first order term in the opacity expansion of the projectile beam function,
\begin{align}
\label{eq:beam-unaveraged}
    &\chi\,\mathcal{B}_{q/a,1}\left(x_1, b, \mu, \frac{\zeta_1}{\nu^2}; \mu_E, \mathcal{L}_1\, |\, M_I, M_F\right) \\ \nonumber
    &=  \sumint_X \int \frac{d^4 z}{2\pi}\, e^{i (n\cdot z) \cdot (x_1 \nbar\cdot P_a)/2}\, \delta\left(\nbar\cdot z \right) \delta^{(2)}\left(\bm{z}_T- \bm{b} \right) \\
    &\hspace*{.4cm}\times \frac{1}{2 N_c}\Bigg\{ \operatorname{Tr}\left[\left \langle P_a\left |  \bar{\chi}_n\left(z\right)\, i\int d^4  x \mathcal{L}_G(x) \right| X\right \rangle \frac{\slashed{\bar{n}}}{2} \left \langle X\left | \chi_n\left(0\right) \,  (-i)\int d^4y \mathcal{L}_G(y)\right | P_a\right \rangle \right] \nonumber
    \\
    &\hspace*{1cm}+\operatorname{Tr}\left[\left \langle P_a\left |  \bar{\chi}_n\left(z\right)\, \frac{i^2}{2}\int d^4  x \int d^4y \mathcal{T}\left\{\mathcal{L}_G(x)\mathcal{L}_G(y)\right\} \right| X\right \rangle \frac{\slashed{\bar{n}}}{2} \left \langle X\left | \chi_n\left(0\right) \, \right | P_a\right \rangle \right] +h.c. \Bigg\}  \,, \nonumber
\end{align}
with the summation of the final states given by 
\begin{align}
\sumint_X =  \prod_{i \in X}  \int \frac{d^3p_i}{(2\pi)^3 2E_i} (2\pi)^4\delta^{(4)}\left(P_a - p - \sum_{j\in X}p_j\right).
\end{align}
Within the curly brackets is a squared amplitude for annihilating the initial state quark while interacting twice with the background field.
The first trace corresponds to the squared amplitude of the direct Glauber exchange, and the second trace and the Hermitian conjugate correspond to the interference between vacuum diagrams and virtual Glauber exchange.
Higher-order terms in the opacity expansion will contain higher powers of the background field. 

The expression in Eq. (\ref{eq:beam-unaveraged}) is for a specific given set of initial and final states of the medium that define the background field $A_G$.
To describe measured cross-sections, we can perform an ensemble average over $M_I$ and summation over all possible $M_F$.
\begin{align}
\mathcal{B}_{q/a,1}\left(x_1, b, \mu,\frac{\zeta}{\nu^2}; \mu_E, \mathcal{L}_1\right) &= \left\langle \mathcal{B}_{q/a,1}\left(x_1, b, \mu, \frac{\zeta}{\nu^2}; \mu_E, \mathcal{L}_1\, |\, M_I, M_F\right)\right\rangle_{\rm M_I, M_F}. \label{eq:in-medium-beam-function}
\end{align}
Because the Drell-Yan measurement is inclusive over the nuclear final states, we sum over $M_F$ and perform an ensemble average over $M_I$ with some density matrix\footnote{In Ref.~\cite{Vaidya:2020cyi}, this ensemble average and dynamics is treated within the framework of the open quantum system.}. This operation is denoted by the bracket $\langle \cdots \rangle_{\rm M_I,M_F}$. 
We assume the ensemble-averaged properties of the medium can be factorized from the calculation of the projectile matrix-element
\begin{align}
\chi\mathcal{B}_{q/a,1}\left(x_1, b, \mu,\frac{\zeta}{\nu^2}; \mu_E, \mathcal{L}_1\right) &= \sumint_X \int \frac{d^4 z}{2\pi}\, e^{i (n\cdot z) \cdot (x_1 \nbar\cdot P_a)/2}\, \delta\left(\nbar\cdot z \right) \delta^{(2)}\left(\bm{z}_T- \bm{b} \right) \nonumber \\
&\hspace*{.4cm}\times \int d^4 x\, d^4 y\, D^{bc}(x,y,z)\, W^{bc}(x,y) + \cdots 
\label{eq:factorize-medium-correlation}
\end{align}
where the $D^{bc}$ is a collinear coefficient from the squared amplitude of the direct Glauber exchange and $W^{bc}$ is the ensemble-averaged correlator of the background field. The ellipses represent contributions from the virtual Glauber exchange. The expressions for these functions are given by
\begin{align}
D^{bc}(x,y,z) &= \frac{1}{2 N_c} \operatorname{Tr}\left[\left \langle P_a\left |  \bar{\chi}_n\left(z\right)\, i\Gamma_G^b(x) \right| X\right \rangle \frac{\slashed{\bar{n}}}{2} \left \langle X\left | \chi_n\left(0\right) \,  (-i)\Gamma_G^c(y)\right | P_a\right \rangle \right],\\
W^{bc}(x,y) &= \left\langle \left[n\cdot A^{a}(x)\right]^\dagger n\cdot A^{b}(y) \right\rangle_{M_I, M_F}.
\end{align}
One can evaluate the ensemble-averaged correlator $W^{ab}$ in the so-called dilute limit, where the average separation between the color sources is much larger than the color correlation length $\xi^{-1}$. The correlator $W^{ab}(x,y)$ is non-vanishing only if the two fields have the same plus light-cone coordinate. This is known as taking the ``contact limit'' of two interactions with the background field. Its derivation is provided in appendix~\ref{sec:app:ensemble-avg} and here we quote the final result expressed in the momentum space
\begin{align}
W^{bc}(q^-, \bfq, \bar{q}^-, \bar{\bfq})
&= \delta^{bc}    \frac{d_A}{g_s^2 C_R}  \sum_T\Sigma^{(0)}_{RT}(\bfq, \bar{\bfq})\mathcal{N}_{j,T}^{(0)}(\bfq)  \nonumber\\
&\hspace*{.4cm}\times\int dz^+ \int dx_t f_{j/N}(x_t) \rho_N^-(z^+,\bfz_\perp) e^{i(q-\bar{q})^-z^+/2}\, .
\label{eq:correlator}
\end{align}
$q$ and $\bar{q}$ denote the momentum variable in the amplitude and its conjugate, respectively. $\rho_N^-(z^+,\bfz_\perp)$ is the density of nucleon along the path length. For a large nucleus, we can approximate it by the saturation density $\rho_0$.
This correlator is diagonal in the color space, which helps to simplify the color trace
\begin{align}
D^{bc}W^{bc} = \frac{1}{d_A}D^{bb}W^{cc}.
\end{align}
Now, $D^{bb}$ can be further decomposed according to the color Casimir factor
\begin{align}
D^{bb} \equiv \sum_R g_s^2 C_R D_R.
\end{align}
Comparing to Eq. (\ref{eq:LO-Glauber-cross-section}), one can extract the desired expression of the collinear function at first order in opacity as 
\begin{align}
\mathcal{J}_{q/q,R}(x_1,b)&= \int\frac{d^2\bfq}{(2\pi)^2} \int \frac{dq^-}{2\pi}\int \frac{d\bar{q}^-}{2\pi} \int \frac{dz^+}{L^+}\frac{\rho_N^-(z^+, \bfz)}{\rho_0^-} e^{i(q-\bar{q})^-z^+/2}\nonumber\\
& \times  \sumint_X \int \frac{d^4 z}{2\pi}\, e^{i (n\cdot z) \cdot (x_1 \nbar\cdot P_a)/2}\, \delta\left(\nbar\cdot z \right) \delta^{(2)}\left(\bm{z}_T- \bm{b} \right)D_R(-q^-, -\bfq, -\bar{q}^-, -\bfq; z)  \nonumber\\
& + [\textrm{virtual Glauber exchange contributions}]  \, .
\label{eq:factorize-medium-correlation-contact-limit}
\end{align}
Eq. (\ref{eq:factorize-medium-correlation-contact-limit}) provides a practical way to compute the NLO collinear function. Moreover, the phase factor in the lase line reveals an important feature, known as the Landau-Pomeranchuk-Migdal (LPM) effect~\cite{Landau:1953um,Migdal:1956tc}, of the medium correction. It is closely related to the in-medium scale separation and power counting.

\subsection{The Landau-Pomeranchuk-Migdal effect and the Gunion-Bertsch regions}
We will now discuss the implications of the phase
factor in Eq. (\ref{eq:factorize-medium-correlation-contact-limit}). After the integration over $z^+$, the argument of the phase is of order $(q^--\bar{q})^-L^+$. 
As a result, the medium size enters as an additional scale in the collinear function $\mathcal{J}_R$.
It introduces a critical line as defined by the condition
\begin{align}
1 = q^-L^+ \sim \lambda^2 Q L^+, \label{eq:critical-line}
\end{align}
which is shown as the dotted horizontal line on both panels of figure~\ref{fig:TMD-DY-pA-modes}. Eq. (\ref{eq:critical-line}) also introduces a new semi-hard scale to the problem $p^+p^- \sim p^+/L^+$. Its interplay with the $P_T$ scale will be discussed in the next subsection.

To understand the significance of the critical line, let us consider real emissions, where the inverse of the small lightcone momentum $p^- \sim q^-$ of the mother parton is often referred to as the formation time of the emission\footnote{For virtual correction, $1/p^-$ is interpreted as the lifetime of the quantum fluctuation.}
\begin{align}
\tau_f = \frac{1}{p^-} = \frac{x(1-x)p^+}{\bfp^2}\, .
\end{align}
It is the timescale during which the interference between different emission amplitudes remains important. Therefore, the critical line can also be written as $L^+/\tau_f = 1$.
The emission pattern changes drastically for radiation kinematics below and above this critical line:
\begin{itemize}
\item For splittings below the line, we have $\tau_f \gg L^+$, i.e., 
the emission process is coherent over the entire path length.
Furthermore, diagrams with multiple collisions contribute coherently to the total splitting amplitude, each coming with a phase factor $e^{i p^-L^+}$ of order unity.
Their interferences are destructive and lead to the so-called Landau-Pomeranchuk-Migdal effect of QCD. We will refer to the region below the critical line as the LPM region.
Due to the LPM effect, the emission probability acquires additional $[x(1-x)]^{-1}$ divergences relative to the vacuum splitting function~\cite{Ke:2023ixa}.
\item For splittings above the line, $p^-L^+ \gg 1$ suggest the formulation is short $\tau_f \ll L^+$. The phase factor $e^{i p^- L^+}$ becomes rapidly oscillating. We can take the limit $L^+\rightarrow \infty$ to get the leading-power contribution.
Thus, qualitatively, gluons are emitted from an almost on-shell quark line that extends from the infinite past to the infinite future.
Due to the short formation time and dilute nature of the medium, such radiations are induced by collision with a single medium parton. Such bremsstrahlung is incoherent with respect to emissions from other Glauber exchanges and the hard vertex. We refer to the incoherent region as the Gunion-Bertsch (GB) regime~\cite{Gunion:1981qs}.
\end{itemize}
From the qualitative differences, one can conclude that only emissions in the Gunion-Bertsch region cause the rapidity logarithm. While in the LPM region, the emission spectrum is qualitatively modified to $1/[x(1-x)^2]$, which cannot contribute to a rapidity logarithm. Furthermore, the collinear logarithm is given by radiations in the LPM region. This has been studied in detail in our earlier paper~\cite{Ke:2023ixa}. Such emissions are responsible for the radiative energy loss for collinear partons in cold nuclear matter.

\subsection{Medium-generated energy scales and scale separation}
We now establish the separation of the scales and identify the sectors that can be studied perturbatively with partonic degrees of freedom at first order in opacity.
The spatial extent of the hard Drell-Yan process in the ``+'' direction is on the order of $\sqrt{\zeta_1}/Q^2$. The same length scale of non-perturbative fluctuation in the proton is $\sqrt{\zeta_1}/\Lambda_{\rm QCD}^2$.
We will study the problem under the following separation of time scales.
\begin{align}
\sqrt{\zeta_1}/\Lambda_{\rm QCD}^2 \gg L^+ \gg \sqrt{\zeta_1}/Q^2. 
\label{eq:timescale-separation}
\end{align}
The first inequality in  Eq. (\ref{eq:timescale-separation}) requires that the quark is highly boosted relative to the nuclear size so that we can neglect the hadron-level $p$+$A$ interaction. 
The second inequality ensures that the timescale of the hard process is small so that it can be viewed as point-like compared to the extensive nuclear medium.
Eq. (\ref{eq:timescale-separation}) translates into the following separation of energy scales
\begin{align}
\Lambda_{\rm QCD}^2 \ll \mu_{E}^2 \ll Q^2\,,
\label{eq:energyscale-separation}
\end{align}
where $\mu_{E}^2 = p_1^+/L^+$ is a medium-generated semi-hard scale related to the location of the critical line.
Another medium-generated scale is the averaged transverse momentum broadening $\langle \Delta P_T^2 \rangle \propto \rho^-L^+$. For a large or a dense medium, it is possible that $\langle \Delta P_T^2 \rangle$ also becomes a semi-hard scale. 
From an order of magnitude estimate use can use for example the E772 experiment~\cite{PhysRevLett.66.2285}, where an 800 GeV proton collides with an $A\approx 200$ nucleus. At $x_a=0.1$ and using average $\langle L^+\rangle$, we find $\mu_{E}^2\approx 3.0$ GeV$^{2}$, while the measured $\langle \Delta P_T^2 \rangle\approx 0.1$ GeV$^{2}$. Therefore, the scenario $\mu_{E}^2 \gg \langle \Delta P_T^2 \rangle \gtrsim \Lambda_{\rm QCD}^2$ might be more realistic at large $x$ and is adopted in this paper.
The appearance of the semi-hard scale is the foundation that cold nuclear matter effects can be, at least partly, understood perturbatively.

\begin{figure}[ht!]
\centering
\includegraphics[height=.475\textwidth]{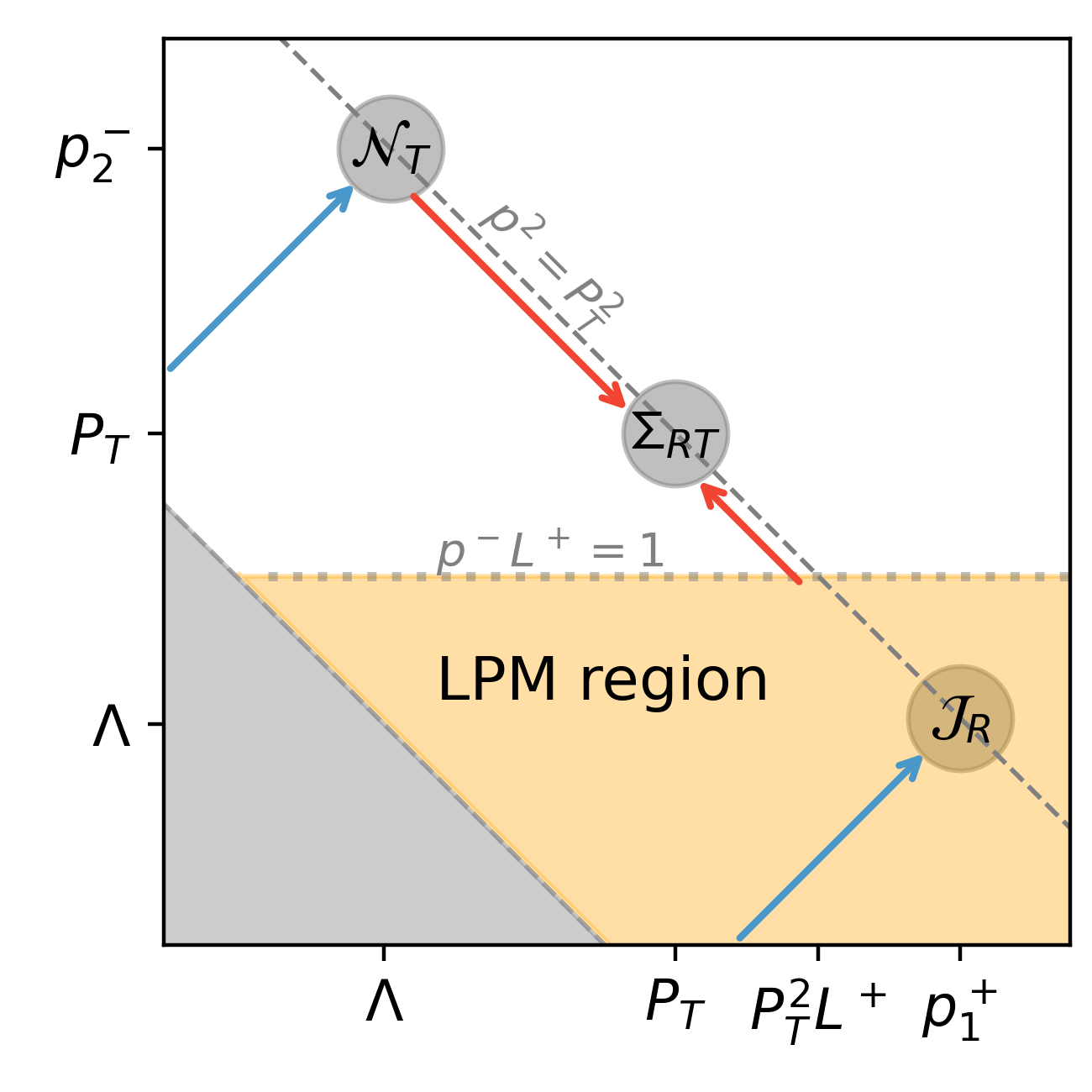}
\hskip1em
\includegraphics[height=.475\textwidth]{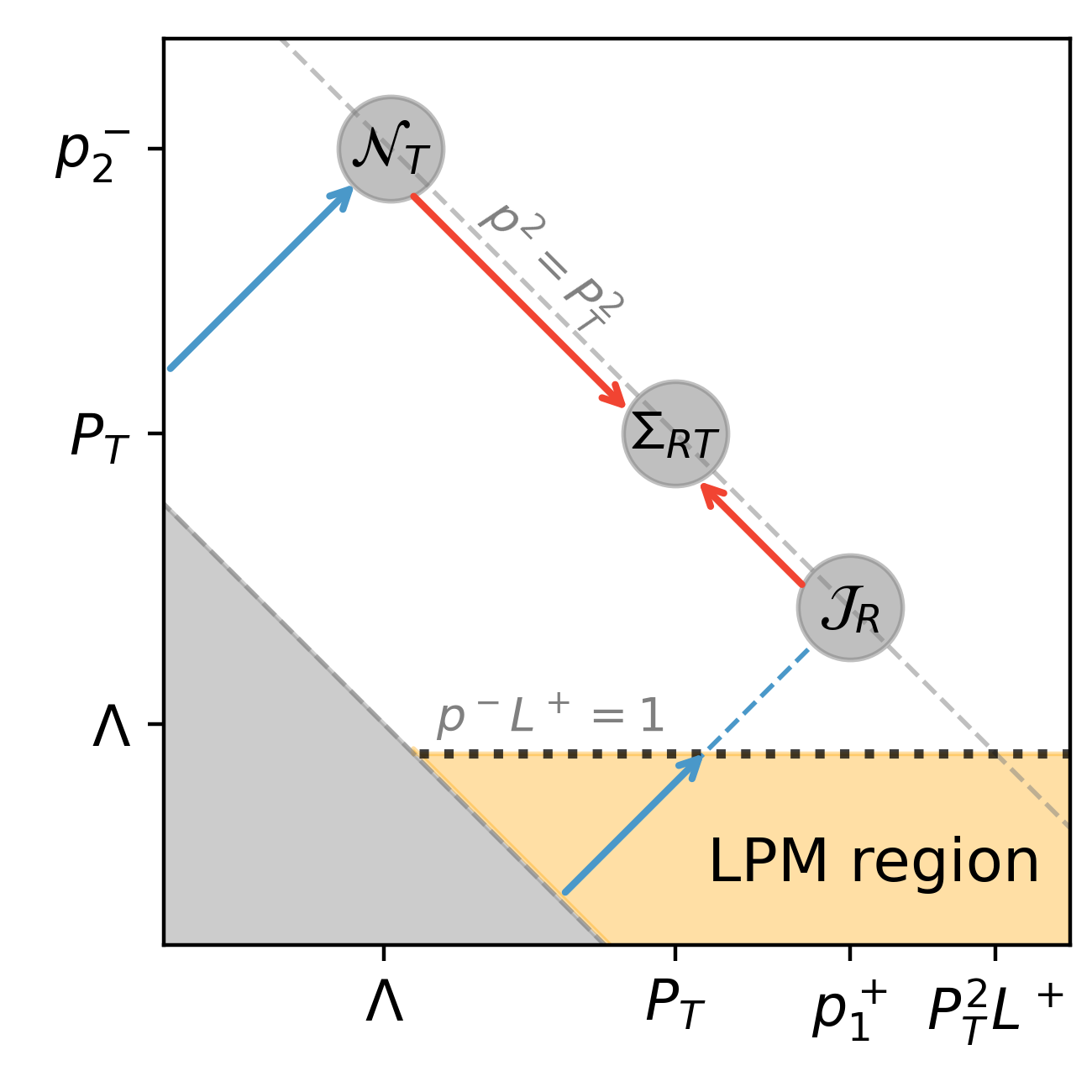}
\caption{Illustration of two different hierarchies of scales in the first order in opacity analysis. Left: $\mu_{E}\gg \mu_b$. Right: $\mu_{E}\ll \mu_b$. The plus and minus momenta have been put in the log scale. The shaded region in yellow denotes the LPM region.}
\label{fig:TMD-DY-pA-modes}
\end{figure}

To fully specify the hierarchy of scales in the calculation of $\mathcal{J}_R$, we now discuss the relation between the medium-generated scale $\mu_{E}$ and the scale of the TMD physics $\mu_b \sim P_T$.
\begin{itemize}
\item $\mu_{E}\gg \mu_b$: this scenario is illustrated on the left panel of figure~\ref{fig:TMD-DY-pA-modes}. The collinear sector $\mathcal{J}_R$ is located in the LPM region. 
The collinear logarithm is $\ln(\mu_b^2/\Lambda_{\rm QCD}^2)$. However, the phase space for the rapidity logarithm is restricted to the Gunion-Bertsch region.
\item $\mu_{E}\ll \mu_b$: this is illustrated on the right panel of figure~\ref{fig:TMD-DY-pA-modes}. The collinear sector is located in the Gunion-Bertsch region. 
The rapidity logarithm is unaffected, but the medium-induced collinear logarithm is restricted below $\mu_{E}$ and gives $\ln(\mu_{E}^2 / \Lambda_{\rm QCD}^2)$.
\item An interpolation formula will connect calculations in the two limiting cases $\mu_E\gg \mu_b$ and $\mu_E\ll \mu_b$ and be applied to the entire phase space.
\end{itemize}

For collisions with moderate center of mass energy (for example, those achieved in nuclear fixed-target experiments), the soft radiations from the Glauber exchange ($\Sigma_{RT}$) and the anti-collinear sector ($\mathcal{N}_T$) both reside in the Gunion-Bertsch regime, as shown in both panels of figure~\ref{fig:TMD-DY-pA-modes}.
There is no complication arising from the LPM effect in these two sectors. The calculations are straightforward and presented in appendix~\ref{app:anticollinear} and appendix~\ref{app:soft} with only the main results summarized in section~\ref{sec:TMD-NLO-soft}.

\subsection{Decomposition of the first order in opacity corrections at NLO}
We write down the LO+NLO parts of the partonic forward cross-section, including the calculation of $\mathcal{J}_R$ using the background field approach, and radiative correction to $\Sigma_{RT}$ and $\mathcal{N}_T$ 
\begin{align}
\sigma_{q/q,T}^{(0)}+\sigma_{q/q,T}^{(1)} &= \left(\mathcal{J}_{q/q,F}^{(0)}+\mathcal{J}_{q/q, F}^{(1),\rm rap}\right)\otimes\Sigma^{(0)}_{FT} \otimes \mathcal{N}_T^{(0)} +  \nonumber\\
&+ \mathcal{J}_{q/q,F}^{(1),\rm coll}\otimes\Sigma_{FT}^{(0)} \otimes \mathcal{N}_T^{(0)}  + \mathcal{J}_{q/q,A}^{(1),\rm coll}\otimes\Sigma_{AT}^{(0)} \otimes \mathcal{N}_T^{(0)} \nonumber\\
&+ \mathcal{J}_{q/q,A}^{(1),\rm rap}\otimes\Sigma_{AT}^{(0)} \otimes \mathcal{N}_T^{(0)}  + \mathcal{J}_{q/q,F}^{(0)}\otimes\Sigma_{FT}^{(1)} \otimes \mathcal{N}_T^{(0)} +\mathcal{J}_{q/q,F}^{(0)}\otimes\Sigma_{FT}^{(0)} \otimes \mathcal{N}_T^{(1)} \nonumber\\
&+\Delta \sigma_{q/q,T}^{\rm NLO} \, , 
\label{eq:NLO-decomposition-q2q}
\end{align}
and the gluon-to-quark conversion is
\begin{align}
\sigma_{q/g,T} &= \mathcal{J}_{q/g,F}^{(1)}\otimes\Sigma^{(0)}_{FT}\otimes \mathcal{N}_T^{(0)} + \mathcal{J}_{q/g,A}^{(1)}\otimes\Sigma^{(0)}_{AT}\otimes \mathcal{N}_T^{(0)}  + \Delta \sigma_{q/g,T}^{\rm NLO} \, .
\label{eq:NLO-decomposition-g2q}
\end{align}
For the case of the quark-to-quark channel, the NLO collinear function $\mathcal{J}_{q/q,R}^{(1)}(x,b)$ has been further decomposed into pieces that 1) contain rapidity divergence as labeled by superscript ``rap'', 2) only contain collinear divergence as labeled by the superscript ``coll'', and 3) finite NLO leftovers $\Delta \sigma_{q/q,T}^{(1)}$.   

\paragraph{The factorized scattering term.} The first term of Eq. (\ref{eq:NLO-decomposition-q2q}) contains the LO cross-section and an NLO piece $\mathcal{J}_{q/q,F}^{(1), \rm rap}$ that corresponds to quark scattering with the target and exhibits a rapidity divergence. 
In section~\ref{sec:app:collinear_sector}, we will show that $\mathcal{J}_{q/q,F}^{(1), \rm rap}$ is just the NLO collinear matching coefficient in the vacuum.
Therefore, this term has the simple physical interpretation that the forward scattering is incoherent from the quantum correction to the Drell-Yan process.

\paragraph{The medium-induced collinear divergence.}
The next two terms $\mathcal{J}_{q/q,F}^{(1),\rm coll}\otimes\Sigma_{FT}^{(0)} \otimes \mathcal{N}_T^{(0)}$ and $\mathcal{J}_{q/q,A}^{(1),\rm coll}\otimes\Sigma_{AT}^{(0)} \otimes \mathcal{N}_T^{(0)}$ are terms in the NLO collinear function that contain collinear divergence. At leading order, the Glauber gluon can only couple to collinear quark which is in the fundamental representation. At NLO, this is more complicated because the quark can radiate a gluon. $\mathcal{J}_{q/q, R}^{(1),\rm coll}$ with subscripts $R=A, F$ collect terms from the collinear sector that couple to the Glauber gluon in the adjoint ($\propto C_A$) or the fundamental representation ($\propto C_F$). The medium-induced collinear divergence has been recently studied within SCET$_{\rm G}$ in~\cite{Ke:2023ixa}. It was shown that it can be canceled by a set of in-medium counter terms, and lead to matter related renormalization of the parton density $f_{q/p}$. The corresponding RG equations encode the energy loss effect for collinear partons. In section~\ref{sec:TMD-NLO-matching:collinear-div}, we outline the calculation in either scenario shown in figure~\ref{fig:TMD-DY-pA-modes}.

\paragraph{The medium-induced rapidity divergence.}
Terms in the third line of Eq.~(\ref{eq:NLO-decomposition-q2q}) all contain rapidity divergence and an infrared divergence.
The first term comes from gluon forward scattering in the NLO collinear function, the second term is the NLO soft correction to the Glauber exchange, and the third term contains the NLO correction to the target sector.
We will show by explicit calculations that the rapidity divergences cancel among the three contributions, and verify that the resulting rapidity renormalization group equation is the BFKL equation. The infrared divergences come from the $\bfq\rightarrow 0$ region of the Glauber gluon, which will be regulated by the effective mass $\xi$.

The last line of Eq. (\ref{eq:NLO-decomposition-q2q}) are finite leftovers, i.e., fixed order terms. They are not enhanced by any large logarithmic when $\mu$ and $\nu$ take the natural scale of each sector.
Finally, Eq. (\ref{eq:NLO-decomposition-g2q}) are also contributions from the gluon to quark conversion. It only contains a factorized scattering term and medium-induced collinear divergences.

\begin{figure}[b!]
    \centering
    \includegraphics[width=\textwidth]{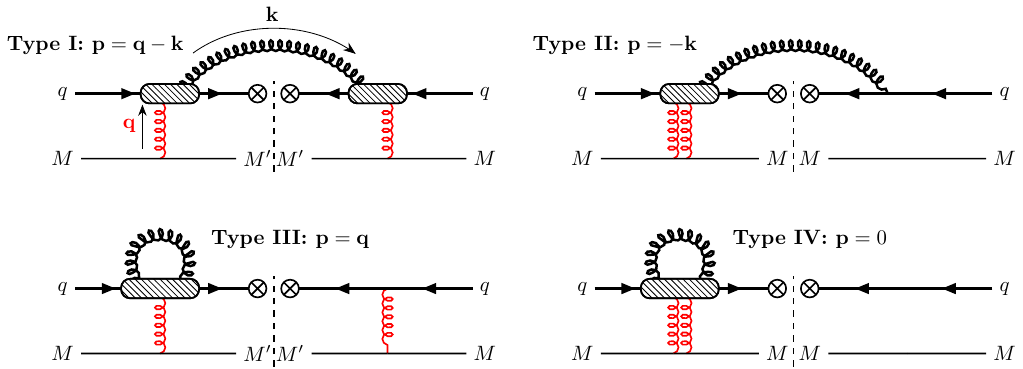}
    \caption{Four types of contributions to the NLO collinear matching coefficient. The crossed circle is the hard vertex. The dotted line represents the final-state cuts. The red vertical lines are the Glauber gluons. The shaded blob denotes the summation of all possible attachments between collinear partons and the Glauber gluons at this order.}
    \label{fig:four-types-of-diagrams}
\end{figure}

\section{The NLO collinear function $\mathcal{J}_{q/q,R}^{(1)}$ to first order in opacity}
\label{sec:TMD-NLO-matching}
\subsection{Example calculations}
In this section, we use the background field method to extract the ${\cal O}(\chi^1)$ collinear function at NLO. This is equivalent to calculating the $x$ and $k_T$ differential splitting function at first order in opacity. In the existing literature, only the real emission contributions have been computed explicitly~\cite{Ovanesyan:2011xy,Ovanesyan:2015dop} from SCET$_{\rm G}$. This is sufficient for studying collinear hadron production with transverse momentum integrated out, where the only effect of virtual corrections is to ensure flavor and energy-momentum sum rules. 
For TMD observables, both the real emission diagrams and virtual diagrams are needed because a diagram containing a collinear loop can also receive transverse momentum recoil from the Glauber gluon. Depending on the recoils received by the collinear quark, the NLO calculation can be broken down into four types of diagrams, as shown in figure~\ref{fig:four-types-of-diagrams}. 
In type-I diagrams, the quark receives transverse momentum recoil from both the Glauber gluon and the collinear radiation. In type-II diagrams, the double Glauber exchange in the contact limit transfers zero net transverse momentum to the quark, and the recoil is given by the radiated gluon. Type-III diagrams involve interference between the collinear loop with the single-Glauber exchange diagram. The recoil comes from the Glauber exchange. Finally, type-IV diagrams give zero change in the transverse momentum of the quark. 
In this section, we will only demonstrate the evaluation of two example diagrams that have not been calculated explicitly in Refs.~\cite{Ovanesyan:2011xy,Ovanesyan:2015dop}. The complete calculation of these diagrams can be found in  appendix~\ref{sec:app:collinear_sector}. 

\begin{figure}[hb!]
    \centering
    \includegraphics[scale=1.2]{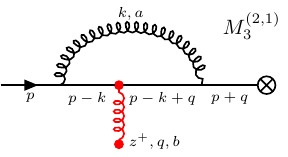}~~~
    \includegraphics[scale=1.2]{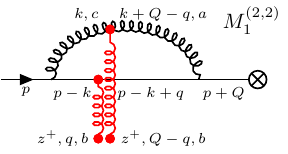}
    \caption{Left: an amplitude that enters the calculation of Type-III diagrams in figure~\ref{fig:four-types-of-diagrams}. Right: an amplitude that enters the calculation of Type-IV diagrams in figure~\ref{fig:four-types-of-diagrams}.}
    \label{fig:example-diagrams}
\end{figure}

\paragraph{Example A: loop correction to a diagram with single-Glauber exchange.} 
The left panel of figure~\ref{fig:example-diagrams} shows one such diagram. It has a collinear loop with one Glauber exchange with a medium color source at $z^+$. It is labeled as $M_3^{(2,1)}$ in the full calculations in Appendix~\ref{sec:app:collinear_sector}. 
Note that if we remove the Glauber gluon, the corresponding vacuum diagram vanishes because it is scaleless in the limit $b^2 \Lambda_{\rm QCD}^2 \ll 1$. The interaction with the Glauber gluon introduces three scales to $M_3^{(2,1)}$: 1) the impact parameter via the phase factor $e^{i\bfq\cdot\bfb}$, 2) $\mu_E$ that enters from the LPM phase, and 3) the inverse range of the interaction $\xi$.  
Using the effective Feynman rules for SCET$_{\rm G}$ in the light-cone gauge (provided in appendix~\ref{sec:app:SCETG_feynman_rule}), the amplitude for this diagram is
\begin{align}
M_3^{(2,1)} &= A_G^-(\bfq)\int \frac{dq^-}{2\pi} e^{iq^-z^+/2} \int \frac{dk^+  d^2\bfk}{2(2\pi)^3} \int \frac{dk^-}{2\pi}
 \frac{i}{k^2+i\eta}\left[\sum_{\lambda}\epsilon_\lambda^\mu\epsilon_\lambda^\nu-\frac{4k^2}{(k^+)^2}\nbar^\mu\nbar^\nu\right]
\nonumber\\
& \hspace*{.4cm}\times  \frac{\sln}{2}\frac{i}{p^-+q^--\frac{(\bfp+\bfq)^2}{p^+} + \frac{i\eta}{p^+}} \nonumber\\
& \hspace*{.4cm}\times  igt^a\frac{\SP_\mu^*(\bfQ_2)}{k^+(p^+-k^+)/p^+} \frac{\slnbar}{2}\frac{\sln}{2}\frac{i}{p^-+q^--k^--\frac{(\bfp+\bfq-\bfk)^2}{p^+-k^+}+\frac{i\eta}{p^+-k^+}}\nonumber\\
& \hspace*{.4cm}\times  igt^b\frac{\slnbar}{2}  \frac{\sln}{2}\frac{i}{p^--k^--\frac{(\bfp-\bfk)^2}{p^+-k^+}+\frac{i\eta}{p^+-k^+}}  igt^a\frac{\SP_\nu(\bfQ_1)}{k^+(p^+-k^+)/p^+} \frac{\slnbar}{2} \frac{\sln}{2}\chi_n(p).
\end{align}
$\chi_n(p)$ is the spinor state of the incoming quark. $igt^a\frac{\SP_\nu(\bfQ_1)}{k^+(p^+-k^+)/p^+}$ is the splitting amplitude that we defined in the appendix~\ref{sec:app:SCETG_feynman_rule}. 
$\bfQ_1 = x\bfk-(1-x)(\bfp-\bfk)$ and $\bfQ_2 = x\bfk-(1-x)(\bfp-\bfk+\bfq)$ are the relative transverse momentum of the two daughter partons for each amplitude.
The interaction with the background field $A_G^-(\bfq)$ transfers momentum $q=(0, q^-,\bfq)$ to the collinear parton. From Eq. (\ref{eq:factorize-medium-correlation-contact-limit}) we have also included the $e^{iq^-z^+/2}$ phase factor and the integral over the minus component of the Glauber momentum.

To complete the loop integral, one first notices that the gauge part $\frac{4k^2}{(k^+)^2}\nbar^\mu\nbar^\nu$ of the light-cone-gauge gluon propagator does not contribute to the final result.
This is because it cancels the $k^2$ in the denominator of the gluon propagator. The remaining $k^-$ contour integration always has two poles on the same side of the real axis, which gives zero.
For the remaining terms, since two of the three $k^-$ poles have imaginary part $\frac{\eta}{p^+-k^+}$, in order to get non-zero contribution from the $k^-$ contour integration, the third pole from the gluon propagator must have an imaginary part $\frac{\eta}{k^+}$ on the opposite side of the real axis from $\frac{\eta}{p^+-k^+}$. This requires $0<k^+<p^+$. Then, we can define $x=k^+/p^+$ that takes value from $0$ to $1$.
Finally, in performing the $q^-$ integration, the phase factor $e^{iq^-z^+/2}$ requires us to close the contour integration in the upper (lower) plain if $z^+>0$ ($z^+<0$). It turns out that the only $q^-$ pole is in the lower-half plain, so the whole expression is only non-zero for $z^+<0$. This is expected since the multiple collisions of the quark take place in the initial state. Finally, we get
\begin{align}
M_3^{(2,1)} &=  A_G^-(\bfq)\Theta(-z^+) ig_s^3 t^a t^b t^a  \int_0^1 \frac{dx}{(1-x)}  \nonumber\\
& \hspace*{.4cm}\times  \int \frac{d^2\bfk}{2(2\pi)^3} \left(e^{i\omega_2z^+/2}-e^{i\omega_1 z^+/2}\right)
\frac{\sum_\lambda \SP_\lambda^*(\bfQ_2)  \SP_\lambda(\bfQ_1)}{\bfQ_2^2 \bfQ_1^2 }\frac{\sln}{2}\chi_{n}(p).
\end{align}
The frequencies are $\omega_1 = \frac{\bfk^2}{k^+} + \frac{(\bfp-\bfk+\bfq)^2}{p^+-k^+} - \frac{\bfp^2}{p^+}$ and $\omega_2 = \frac{(\bfp+\bfq)^2}{p^+} - \frac{\bfp^2}{p^+}$. They are also defined in appendix~\ref{sec:app:collinear_sector}. $\mathcal{P}_\lambda(\bfQ_n) = \epsilon_{\lambda}^{\mu}\mathcal{P}_\mu(\bfQ_n)$ is the contraction of the polarization vector with the splitting amplitude.

\paragraph{Example B: loop correction to a diagram with double-Glauber exchange.} 
In this example (the right panel of figure~\ref{fig:example-diagrams}, labeled as $M^{(2,2)}_1$), both the collinear gluon and collinear quark lines interacts with the same color source at location $z^+$. Note that the former is proportional to the transverse part of the metric tensor $g_T^{\alpha\beta}$. The contraction of $g_T^{\alpha\beta}$ with the gauge part of the gluon propagator vanishes, $g_{T,\alpha\mu} \frac{-4k^2}{(k^+)^2}\nbar^\mu\nbar^\nu = 0$.
From Eq.~(\ref{eq:factorize-medium-correlation-contact-limit}), we can see that in the contact limit, the colors of the two Glauber exchanges are the same, with opposite transverse momenta. We make a transformation of variables to use $q=(0,q^-,\bfq)$ as the momentum exchange in the first Glauber exchange, and $\ell=(0, \ell^-, 0_\perp)$ as the total momentum exchange from the two Glauber interactions.
Then, we can write down
\begin{align}
M_1^{(2,2)} &=  \int\frac{dk^+d^2\bfk}{2(2\pi)^3} \int\frac{dk^-}{2\pi} \int\frac{d\ell^-}{2\pi}e^{i\ell^-z^+/2}\int\frac{dq^-}{2\pi} \frac{\sln}{2}\frac{i}{p^-+\ell^--\frac{\bfp^2}{p^+}+\frac{i\eta}{p^+}} \nonumber\\
& \hspace*{.4cm}\times  igt^a\frac{\SP^*_\mu(\bfQ_3)}{k^+(p^+-k^+)/p^+}\frac{\slnbar}{2} \frac{\sln}{2}\frac{i}{p^-+q^--k^--\frac{(\bfp+\bfq-\bfk)^2}{p^+-k^+} + \frac{i\eta}{p^+-k^+}} \nonumber\\
& \hspace*{.4cm}\times igt^b A^-(\bfq) \frac{\slnbar}{2}  \frac{\sln}{2}\frac{i}{p^--k^--\frac{(\bfp-\bfk)^2}{p^+-k^+} + \frac{i\eta}{p^+-k^+}}  igt^c\frac{\SP_\nu(\bfQ_1)}{k^+(p^+-k^+)/p^+}\frac{\slnbar}{2} \frac{\sln}{2}\chi_{n}(p) \nonumber\\
& \hspace*{.4cm}\times \frac{i\sum_\lambda \e_\lambda^\mu(k+\ell-q)\e_\lambda^{\mu'}(k+\ell-q)}{(k+\ell-q)^2+i\eta} gf^{abc} A^{-*}(\bfq) k^+ g_{T, \mu'\nu'} \frac{i\sum_\theta \e_\theta^\nu(k)\e_\theta^{\nu'}(k)}{k^2+i\eta} , 
\end{align}
where $\bfQ_3 = x(\bfk-\bfq)-(1-x)(\bfp-\bfk+\bfq)$.
In accordance with the contact limit, it is important to perform $q^-$ integration before the $k^-$ integration. Then, it is straightforward to get
\begin{align}
M_1^{(2,2)} &=  \Theta(-z^+)|A^-(\bfq)|^2 (-i) g_s^4 f^{abc} t^a t^b t^c \int_0^1\frac{dx}{1-x}\nonumber\\
&  \hspace*{.4cm}\times  \int\frac{d^2\bfk}{2(2\pi)^3} \left[1-e^{i\omega_6 z^+/2}\right] \frac{\sum_\lambda \SP^*_\lambda(\bfQ_3) \SP_\lambda(\bfQ_1)}{\bfQ_3^2 \bfQ_1^2} \frac{\sln}{2}\chi_{n}(p),
\end{align}
with $\omega_6 = \frac{(\bfk-\bfq)^2}{k^+} + \frac{(\bfp-\bfk+\bfq)^2}{p^+-k^+} - \frac{\bfp^2}{p^+}$.

\subsection{The full collinear function} Adding up the contributions  of type-I, II, II, and IV, at the cross section level  we get the expression for the NLO collinear matching coefficient at first order in opacity:
\begin{align}
\label{eq:NLO-collinear-matching-coeff-full}
&\mathcal{J}_{q/q,R}^{(1)}(x,\bfp,\bfq) \nonumber\\
&= \frac{g_s^2C_F}{2\pi} P_{qq}(x) \int d^2 \bfk\left[\delta^{(2)}(\bfp-\bfq+\bfk) \mathcal{I}_{\rm I,R}(x,\bfk,\bfq) + \delta^{(2)}(\bfp+\bfk) \mathcal{I}_{\rm II,R}(x,\bfk,\bfq) \right] \\
& + \frac{g_s^2C_F}{2\pi}\delta(1-x)\int_0^1 dx' P_{qq}(x') \int d^2 \bfk
 \left[\delta^{(2)}(\bfp-\bfq)\mathcal{I}_{\rm III,R}(x',\bfk,\bfq) + \delta^{(2)}(\bfp)\mathcal{I}_{\rm IV,R}(x',\bfk,\bfq) \right] . \nonumber
\end{align}
The delta function reflects the transverse momentum recoil from each type of contribution.

Expressions for $\mathcal{I}_{K,R}(x, \bfk, \bfq)$ for $K=$ I, II, III, IV and $R=F,A$ has been collected in table~\ref{tab:NLO-collinear}. For compactness, we have defined 
\begin{align}
\bfQ_1 &= x\bfk-(1-x)(\bfp_0-\bfk),\\
\bfQ_2 &= x\bfk-(1-x)(\bfp_0-\bfk+\bfq), \\
\bfQ_3 &= x(\bfk-\bfq)-(1-x)(\bfp_0-\bfk+\bfq),\\
\bfQ_4 &= x(\bfk+\bfq)-(1-x)(\bfp_0-\bfk),
\end{align}
where $\bfp_0$ is the initial-state transverse momentum of the quark and $\bfk,\bfq$ are the transverse momentum flows in the collinear gluon and Glauber gluon lines. The path-length averaged LPM interference phase factors $\Phi_n$ are 
\begin{align}
\Phi_n = 1-\frac{\sin{\left(\frac{\bfQ_n^2 L^+}{2x(1-x)p_1^+}\right)}}{\frac{\bfQ_n^2 L^+}{2x(1-x)p_1^+}}.
\end{align}
The collinear function depends on both the transverse momentum $|\bfp|\sim \mu_b$ and the semi-hard scale $\mu_E = p_1^+/L^+$ through the LPM phase factor $\phi_n$. In the limit $\mu_b\gg \mu_E$ or $\mu_b\ll \mu_E$, we will further reduce it to a single-scale function by power expansion.

\begin{table}[ht!]
\centering
\renewcommand{\arraystretch}{1.5}
\begin{tabular}{c|c|c}
\hline
Type $K$ & $\mathcal{I}_{K,F}(x, \bfk, \bfq)$ & $\mathcal{I}_{K,A}(x, \bfk, \bfq)$\\
\hline
I & $\frac{1}{\bfQ_1^2}+2\frac{\bfQ_2}{\bfQ_2^2}\cdot\left(\frac{\bfQ_2}{\bfQ_2^2}-\frac{\bfQ_1}{\bfQ_1^2}\right)\phi_2$  & $\frac{1}{\bfQ_3^2}-\frac{\bfQ_1}{\bfQ_1^2}\cdot\frac{\bfQ_3}{\bfQ_3^2}+\frac{\bfQ_2}{\bfQ_2^2}\cdot\left(\frac{\bfQ_1}{\bfQ_1^2}-\frac{\bfQ_3}{\bfQ_3^2}\right)\phi_2$ \\
\hline
II  & $-\frac{1}{\bfQ_1^2}$ & $\frac{\bfQ_1}{\bfQ_1^2}\cdot\left(\frac{\bfQ_1}{\bfQ_1^2}-\frac{\bfQ_3}{\bfQ_3^2}\right)(\phi_1-1)$\\ 
\hline
III  & $-2\frac{\bfQ_2}{\bfQ_2^2}\cdot\left(\frac{\bfQ_2}{\bfQ_2^2}-\frac{\bfQ_1}{\bfQ_1^2}\right)\phi_2 $ & $-\frac{\bfQ_1\cdot\bfQ_2}{\bfQ_1^2\bfQ_2^2}\phi_2 + \frac{\bfQ_1}{\bfQ_1^2}\cdot \frac{\bfQ_4}{\bfQ_4^2} \phi_4 $\\
\hline
IV & $0$ &
$-\frac{1}{\bfQ_1^2}\phi_1+\frac{\bfQ_1}{\bfQ_1^2}\cdot\frac{\bfQ_3}{\bfQ_3^2}\phi_3$\\
\hline
\end{tabular}  
\caption{$\mathcal{I}_{K,F}$ and $\mathcal{I}_{K,A}$ functions ($K = $ I, II, III, IV) that show up in Eq.~(\ref{eq:NLO-collinear-matching-coeff-full}).}
\label{tab:NLO-collinear}
\end{table}
As a cross check, we have verified that the sum of type-I and II contributions reproduces the initial-state splitting functions obtained in Ref.~\cite{Ovanesyan:2015dop}. The type-III and IV contributions in Eq. (\ref{eq:NLO-collinear-matching-coeff-full}) are obtained in this work.
To properly define the integrals under power expansion, we will regulate the transverse momentum ones by dimensional regularization (DR) with $d=4-2\epsilon$ (leaving $2-2\epsilon$ in transverse space). The rapidity divergences are regulated by the $\eta$-regulator~\cite{Chiu:2012ir}, which for the collinear sector is
\begin{align}
\eta(x) = \left(\frac{(1-x)p_1^+}{\nu}\right)^{-\tau}.
\end{align}
Under dimensional regularization, it can be shown that the flavor sum rule is also satisfied when summing over the results from all four types of diagrams, 
\begin{align}
\int_0^1 dx \int d^2\bfp \int d^2\bfq\mathcal{J}_{q/q,R}^{(1)}(x,\bfp,\bfq) = 0
\end{align}
which is a consistency check for the type III and IV results.

Because the initial transverse momentum $\bfp_0$ is of non-perturbative origin, we will set $\bfp_0=0$ when extracting the perturbative collinear function and define
\begin{align}
&\bfA \equiv \left.\bfQ_1\right|_{\bfp_0=0} = \bfk,\\
&\bfB \equiv \left.\bfQ_2\right|_{\bfp_0=0} = \bfk-(1-x)\bfq,\\
&\bfC \equiv \left.\bfQ_3\right|_{\bfp_0=0} = \bfk-\bfq, \\
&\bfD \equiv \left.\bfQ_4\right|_{\bfp_0=0} = \bfk+x\bfq,
\end{align}
for later conveniences.

\subsection{The factorized scattering of the quark}
The rapidity divergence of $\mathcal{J}_{q/q,F}$ comes from the two terms proportional to $1/\bfQ_1^2$. Taking the Fourier transform to the impact parameter space, we find
\begin{align}
\left(\mathcal{J}_{q/q,F}^{(1), \rm rap}\otimes \Sigma_{FT}^{(0)}\otimes \mathcal{N}_{j,T}^{(0)}\right)&(x,b, \mu, \zeta_1/\nu^2)= \left[\Sigma_{FT}^{(0)}(\bfb)-\Sigma_{FT}^{(0)}(0)\right]\nonumber\\
& \times \alpha_s^{(0)}\frac{C_F}{2\pi^2} \left(\frac{p^+}{\nu}\right)^{-\tau} \frac{1+x^2}{(1-x)^{1+\tau}} \int \frac{d^{2-2\epsilon}\bfk}{(2\pi)^{2-2\epsilon}}\frac{e^{i\bfb\cdot\bfk}}{\bfk^2},
\end{align}
This is precisely the product of the LO forward scattering cross-section (including the unitary correction) times the NLO vacuum matching coefficient. 
In the perturbative region, after we power expand in the small number $\xi^2b^2$, the unitary correction term $\Sigma_{FT}(0)$ is scaleless. So we will drop it for the rest of the section.

If we sum over the collinear matching coefficient in the vacuum at LO ($C_{q/q}^{(0)} = \delta(1-x)$) and NLO $C_{q/q}^{(1)}$, the LO matching coefficient at ${\cal O}(\chi^1)$, and the factorized scattering contribution of the NLO matching coefficient at ${\cal O}(\chi^1)$, we find 
\begin{align}
\mathcal{B}_{q/q,0}+\chi \mathcal{B}_{q/q,1}  \supset \left[\delta(1-x)+C_{q/q}^{(1)}\left(x,b,\mu,\frac{\zeta_1}{\nu^2}\right)\right]\left[1+\rho_0^-L^+ f_T \Sigma_{FT}(b)\right].
\end{align}
The physical meaning of this subset of corrections is clear: the collisional recoil is independent from the radiative correction.

\subsection{The medium-induced collinear divergence} 
\label{sec:TMD-NLO-matching:collinear-div}
After taking out the factorized scattering terms, the remaining part of $\mathcal{J}_{q/q,F}^{(1)}$ only contains collinear divergence,
\begin{align}
&\left(\mathcal{J}_{q/q,F}^{(1),\rm coll}\otimes \Sigma_{FT}^{(0)}\otimes \mathcal{N}_{j,T}^{(0)}\right)(x, b, \mu, \mu_E) =   \int \frac{d^{2-2\epsilon}\bfq}{(2\pi)^{2-2\epsilon}} 
 g_s^2\frac{C_T}{d_A}\frac{e^{-i\bfb\cdot \bfq}}{\bfq^4} g_s^2 C_F \int \frac{d^{2-2\epsilon}\bfk}{(2\pi)^{2-2\epsilon}} g_s^2\frac{C_F}{2\pi} \nonumber\\
& \left\{ e^{i\bfb\cdot \bfk}P_{qq}(x) 2\frac{\bfB}{\bfB^2}\cdot\left(\frac{\bfB}{\bfB^2}-\frac{\bfA}{\bfA^2}\right)\Phi_B - \delta(1-x)  \int dx' P_{qq}(x') 2\frac{\bfB'}{{\bfB'}^2}\cdot\left(\frac{\bfB'}{{\bfB'}^2}-\frac{\bfA'}{{\bfA'}^2}\right)\Phi_{B'} \right\} ,
\label{eq:coll-div-F}
\end{align}
where we have already dropped several scaleless integrals. Quantities with a prime indicates that their $x$ dependence is replaced by $x'$, a dummy integration variable.
Because $\bfB = \bfA$ in the limit $x=1$, the factor $(\bfA/\bfA^2-\bfB/\bfB^2)$ cancels the $x=1$ pole in $P_{qq}(x)$. Therefore, it does not contain the rapidity divergence, so we have taken $\tau=0$ in the rapidity regulator.

The rest of the collinear divergences come from $\mathcal{J}_{q/q,A}^{(1)}$. To decouple this collinear divergence from the rapidity divergence near $x=1$ is subtle. The detailed separation procedure is given in the appendix~\ref{app:separation_Cnu_from_Cmu}.
We write down the final result for the separated collinear divergence,
\begin{align}
 &\mathcal{J}_{q/q,A}^{(1),\rm coll}\otimes \Sigma_{AT}^{(0)}\otimes \mathcal{N}_{j,T}^{(0)} = \int \frac{d^{2-2\epsilon}\bfq}{(2\pi)^{2-2\epsilon}} 
 g_s^2\frac{C_T}{d_A} \frac{e^{-i\bfb\cdot \bfq}}{\bfq^4}g_s^2 C_A\nonumber\\
 & \times \left[g_s^2\frac{C_F}{2\pi}P_{qq}(x) \int \frac{d^{2-2\epsilon}\bfk}{(2\pi)^{2-2\epsilon}}  e^{i\bfb\cdot \bfk} \left\{\frac{\bfB}{\bfB^2}\cdot\left(\frac{\bfA}{\bfA^2}-\frac{\bfC}{\bfC^2}\right)\Phi_B + \frac{\bfC}{\bfC^2}\cdot\left(\frac{\bfC}{\bfC^2}-\frac{\bfA}{\bfA^2}\right)\Phi_C \right\}\right]_+. \label{eq:coll-div-A}
 \end{align}
The plus prescription acts on the whole expression because the $\bfA, \bfB, \bfC$ and the phase $\Phi$ all depends on $x$,
\begin{align}
  [f(x)]_+ = f(x) - \delta(1-x)\int_0^1 f(x') dx'.
 \end{align}
The summation of Eqs. (\ref{eq:coll-div-F}) and (\ref{eq:coll-div-A}) give the full collinear divergence of Eq. (\ref{eq:NLO-collinear-matching-coeff-full}). They still contains two scale $\mu_b$ and $\mu_E$. From the discussion in section~\ref{sec:TMD-med-scales}, we can reduce the problem into a single-scale problem when there is a large separation of scales.

First, consider the limit $\mu_b\gg \mu_E$, corresponding to the scenario shown on the right panel of figure~\ref{fig:TMD-DY-pA-modes}. We can further expand the calculation in $\mu_E^2/\mu_b^2 \sim b^2 p_1^+/L^+$. To leading power, all the TMD phases in the above integrals become unity, as the typical medium-induced radiation has scale $p^2\sim p_1^+/L^+$. Then, the sum of all the collinear divergences takes a familiar form that has been studied in our previous paper~\cite{Ke:2023ixa}
\begin{align}
 &\mathcal{J}_{q/q,F}^{(1),\rm coll}\otimes \Sigma_{FT}^{(0)}\otimes \mathcal{N}_{j,T}^{(0)}+\mathcal{J}_{q/q,A}^{(1),\rm coll}\otimes \Sigma_{AT}^{(0)}\otimes \mathcal{N}_{j,T}^{(0)} \nonumber\\
 &= \int \frac{d^{2-2\epsilon}\bfq}{(2\pi)^{2-2\epsilon}} 
\frac{g_s^2 C_A g_s^2C_T}{d_A} \frac{1}{\bfq^4} 
 \int \frac{d^{2-2\epsilon}\bfk}{(2\pi)^{2-2\epsilon}} g_s^2\frac{C_F}{2\pi} \left[P_{qq}(x)\left\{\frac{\bfB}{\bfB^2}\cdot\left(\frac{\bfB}{\bfB^2}-\frac{\bfC}{\bfC^2}\right)\Phi_B \right.\right. \nonumber\\
 & \left.\left.+ \frac{\bfC}{\bfC^2}\cdot\left(\frac{\bfC}{\bfC^2}-\frac{\bfA}{\bfA^2}\right)\Phi_C + \left(\frac{2C_F}{C_A}-1\right)\frac{\bfB}{\bfB^2}\cdot\left(\frac{\bfB}{\bfB^2}-\frac{\bfA}{\bfA^2}\right)\Phi_B \right\}\right]_+ + \mathcal{O}\left(\frac{\mu_E^2}{\mu_b^2}\right). 
 \end{align}
Using the techniques developed in Ref.~\cite{Ke:2023ixa}, the convolution of the matching coefficient with the collinear PDF gives
 \begin{align}
&  \sum_{T,j} xf_{q/p}(x) \otimes \mathcal{J}_{q/q,F}^{(1),\rm coll}\otimes \Sigma_{FT}^{(0)}\otimes \mathcal{N}_{j,T}^{(0)} \otimes f_{j/N} \rho_0^-L^+   \nonumber\\
& = \frac{\alpha_s^2(\mu^2) \rho_G^- L^+}{8\mu_E^2} B\left(w\right)  \left(\frac{1}{2\epsilon}+\ln\frac{\mu^2}{\gamma(w) \mu_E^2}\right)2C_F\left(\frac{2C_A+C_F}{x} - 2 C_A \frac{d}{dx}\right)\left[xf_{q/a}(x)\right] . \label{eq:coll-div-convolve}
 \end{align}
Here $w=\mu_b^2/\mu_E^2$, and $B(w)$ and $\gamma(w)$ are two coefficient functions defined as
\begin{align}
B(w) &= \frac{4}{\pi}\int_0^{w} \Phi(x) \frac{dx}{x^2}, \label{eq:Bw}\\
\gamma(w) &= 2\exp\left\{\frac{1}{B(w)}\frac{4}{\pi}\int_0^{w} \Phi(x) \ln(x) \frac{dx}{x^2} \right\}. \label{eq:chiw}
\end{align}
In the considered limit $\mu_b^2\gg \mu_E^2$, $B(w)$ approaches $B(\infty)=1$ while $\gamma(w)$ approaches $\gamma(\infty)=2e^{3/2-\gamma_E}\approx 5.03$.
The $\rho_G^-$ in Eq.~(\ref{eq:coll-div-convolve}) is an effective density defined as
\begin{align}
\rho_G^- = \rho_0^- g_s^2 \sum_j\frac{C_j}{d_A} \int dx_t f_{j/N}(x_t) 
\label{eq:rhoG}
\end{align}
This expression only applies when the medium is consists of weakly-coupled partons where it is meaningful to take about the distribution of partons in the medium. For a nuclear matter at zero temperature, we will treat $\rho_G^-$ as a non-perturbative input parameter.

In Eq.~(\ref{eq:coll-div-convolve}), the $1/\epsilon$ pole is of infrared origin since we have dropped the NP effective mass $\xi$ from the calculation. Following Ref.~\cite{Ke:2023ixa}, it is cancelled by an in-medium counter terms the parton density $F(x)\equiv x f(x)$ is renormalized. It leads to the following evolution equations that resum medium-induced collinear radiations,
\begin{align}
\frac{\partial F_{q-\bar{q}}}{\partial \tau} &= \left( 4C_FC_A   \frac{\partial}{\partial x} - \frac{4C_FC_A+2C_F^2}{x} \right) F_{q-\bar{q}}~,\label{eq:collinear_evolution-1}\\
\frac{\partial F_{q+\bar{q}}}{\partial\tau} &= \left( 4C_FC_A   \frac{\partial}{\partial x} - \frac{4C_FC_A+2C_F^2}{x} \right)  F_{q+\bar{q}} + C_F \frac{F_g}{x}~, \label{eq:collinear_evolution-2}\\
\frac{\partial F_g}{\partial \tau} &= \left(4C_A^2 \frac{\partial }{\partial x} - \frac{2N_f C_F}{x}\right)F_g  + 2C_F^2  \sum_{q}\frac{F_{q+\bar{q}}}{x}~.
\label{eq:collinear_evolution-3}
\end{align}
$\tau$ is a redefinition of the evolution variable
\begin{align}
\tau(\mu^2) = \frac{ B(w)\rho_G^- L^+}{8p_1^+/L^+}  \frac{4\pi}{\beta_0}\left[\alpha_s(\mu^2) - \alpha_s\left(\frac{\gamma(w) p_1^+}{L^+}\right)\right]. \label{eq:evolution_tau}
\end{align}
It evolves from the natural scale to $\xi^2$. To illustrate the major effect of the in-medium evolution, it is instructive to look at the travelling-wave solution of the flavor non-singlet sector
\begin{align}
F_{q-\bar{q}}(x, \tau) = \frac{F_{q-\bar{q}}(x+4C_FC_A\tau, 0)}{\left(1+4C_FC_A\tau/x\right)^{1+C_F/(2C_A)}}.
\label{eq:travelling-wave}
\end{align}
For a parton spectrum that is decreasing fast with $x$, the most prominent effect of the evolution is the shift of the spectrum $\Delta x = -4C_FC_A\tau$. This is interpreted as the  
energy loss $\Delta p_1^+ = p_1^+ \Delta x$ of the parton in the cold nuclear matter
\begin{align}
\left.\Delta p_1^+\right|_{\mu_E\ll \mu_b} &= \frac{ B(w)\rho_G^- (L^+)^2}{8}  \frac{4\pi}{\beta_0}\left[\alpha_s(\xi^2) - \alpha_s\left(\gamma(w)\mu_E^2\right)\right]. \label{eq:eloss-2}
\end{align}

\begin{figure}[hb!]
\centering
\includegraphics[scale=.8]{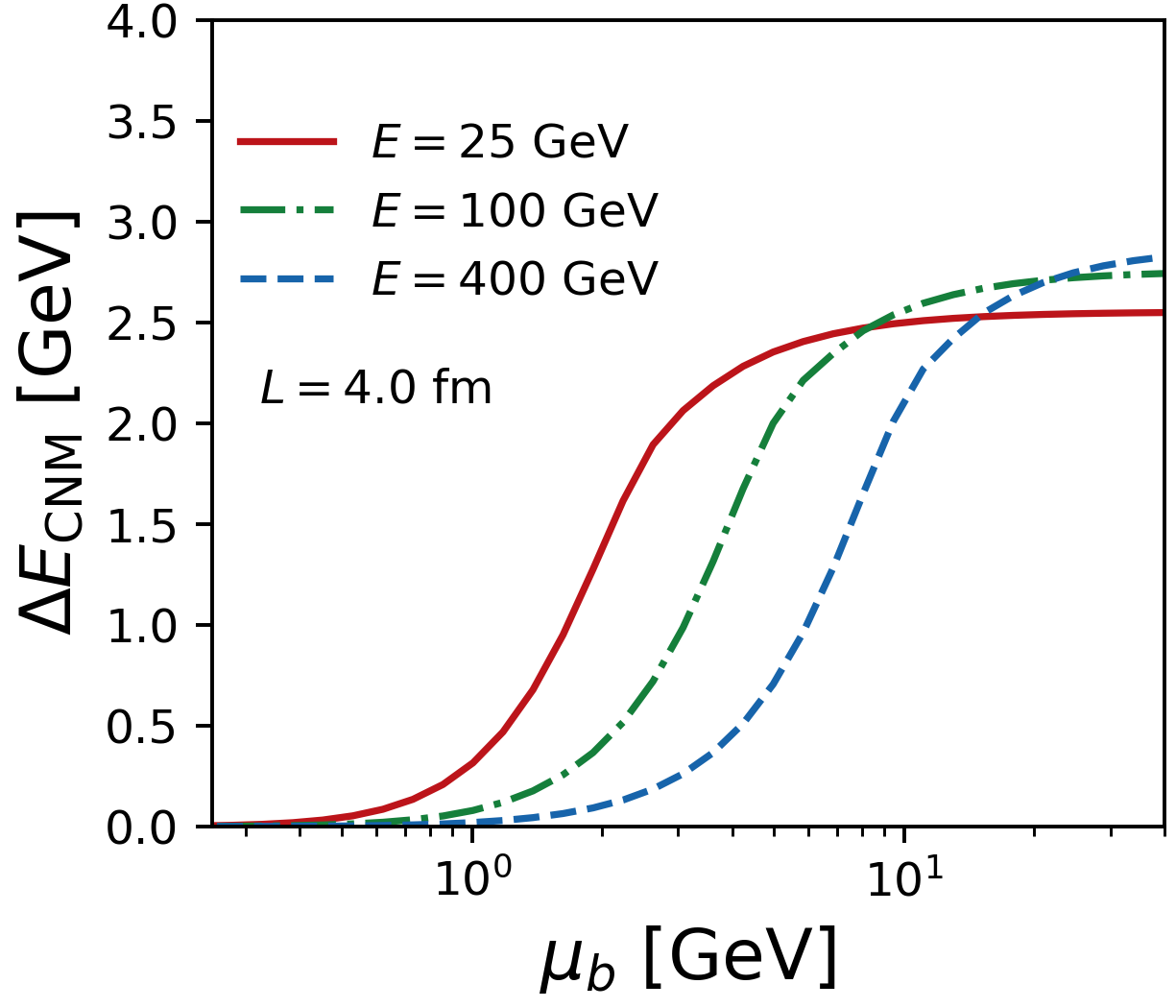}~
\includegraphics[scale=.8]{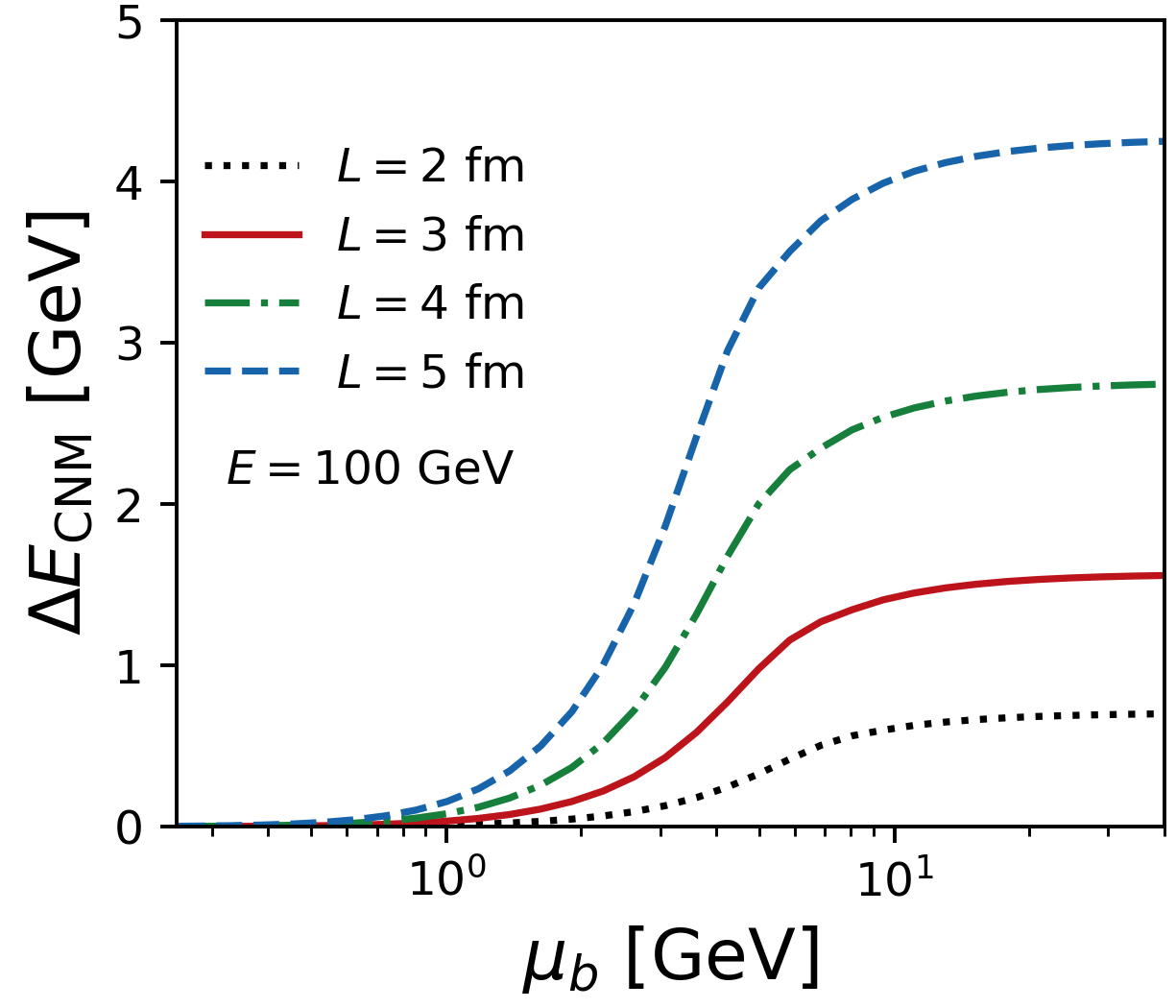}
\caption{The parton energy loss (primary effect of in-medium collinear evolution) a function of $\mu_b$. Left: varying the parton energy $E$ in the nuclear rest frame. Right: varying the path length as seen in the nuclear rest frame.}
\label{fig:eloss-mub}
\end{figure}

Next, we turn to the other limit $\mu_b\ll\mu_E$ shown on the left panel of figure~\ref{fig:TMD-DY-pA-modes}. The transverse momentum integration is limited by the smallness of $\mu_b$, so the amount of energy loss is further power suppressed by $\frac{\mu_b^2}{\mu_E^2}$ and can be neglected. 
For phenomenological applications, we need an approximate formula that smoothly interpolates between the two limiting scenario. We note that the set of equations.~(\ref{eq:Bw}) and
(\ref{eq:chiw}) already contains the desired interpolating behavior. For example, examine the $w = \mu_b^2/\mu_E^2 \ll 1$ limit of the coefficient functions $B(w)$ and $\gamma(w)$,
\begin{align}
& B(w\ll 1) \approx \frac{2}{3\pi}\frac{\mu_b^2}{\mu_E^2},\quad \gamma(w\ll 1) \approx \frac{2}{e} \frac{\mu_b^2}{\mu_E^2}.
\end{align}
If these values are substituted into the energy loss formula,
\begin{align}
\left.\Delta p_1^+\right|_{\mu_b\ll \mu_E} &\approx  \frac{2}{3\pi}\frac{\mu_b^2}{\mu_E^2}\frac{ \rho_G^- (L^+)^2}{8}  \frac{4\pi}{\beta_0}\left[\alpha_s(\xi^2) - \alpha_s\left(2e^{-1} \mu_b^2\right)\right], \label{eq:eloss}
\end{align}
we find that the $\Delta p_1^+$ is indeed further power suppressed by $\mu_b^2/\mu_E^2$. Furthermore, the natural scale of the running coupling is replaced by $2 e^{-1}\mu_b^2$. 
Because of this interpolating feature, we will use Eqs. (\ref{eq:Bw}), (\ref{eq:chiw}) and the evolution Eqs. (\ref{eq:collinear_evolution-1}), (\ref{eq:collinear_evolution-2}), and (\ref{eq:collinear_evolution-3}) for phenomenological applications at all $\mu_b$ values. This is the first transverse-momentum (or impact-parameter) dependent parton energy loss formula.

In figure~\ref{fig:eloss-mub}, we plot the initial-state quark energy loss in the rest frame of the cold nuclear matter. The left panel shows the energy loss as a function of $\mu_b$ for different parton energy. The energy loss is almost zero at small $\mu_b$, then rapidity increases when $\mu_b\sim \mu_E$, and eventually saturates for $\mu_E \gg \mu_b$.
This $b$ dependence of the energy loss is a reflection of the ``survival'' bias of parton propagation in the medium: partons that carry large longitudinal momentum tend to have smaller transverse momentum recoils from the medium. Because typical medium-induced collinear radiation causes momentum recoil of order $\sqrt{p_1^+/L^+}$, partons produced with small transverse momentum cannot undergo such radiations and, as a result, lose less energy.

Finally, we discuss the nature of the medium-induced collinear evolution in Eqs. (\ref{eq:collinear_evolution-1}), (\ref{eq:collinear_evolution-2}), and (\ref{eq:collinear_evolution-3}).
Going back to the travelling wave solution in Eq.~(\ref{eq:travelling-wave}), it can be Taylor-expanded in $\tau$. For a fast-changing $F_{q-\bar{q}}$, focus on the most important terms in the expansion  involves the gradients  $\partial^n F_{q-\bar{q}}$,
\begin{align}
F_{q-\bar{q}}(x, \tau) \approx  \frac{1}{\left(1+4C_FC_A\tau/x\right)^{1+C_F/(2C_A)}} \sum_{n=0}^\infty \frac{(4 C_F C_A \tau)^n}{n!} \frac{\partial^{n} F_{q-\bar{q}}(x, 0)}{\partial x^n}  \, .
\end{align}
Because $\tau$ is proportional to the medium opacity $\tau \propto \rho_G^- L^+ \propto \chi$, it means that in-medium collinear evolution equations sums terms from higher orders in opacity.
More specifically, for the energy loss effect the evolution equation sums the leading gradient and leading $\ln\left(\frac{\gamma(w)\mu_E^2}{\xi^2}\right)$ terms from each order in opacity. 
This is very different from the nature of the rapidity evolution that will be discussed later, which is a renormalization of the forward cross-section at fixed orders of opacity.

\subsection{The rapidity divergence from gluon rescattering}
\label{sec:TMD-NLO-matching:rapidity-div}
We now switch to the term that contains the rapidity divergences from the gluon rescattering
\begin{align}
& \mathcal{J}_{q/q,A}^{(1),\rm rap}\otimes\Sigma_{AT}^{(0)} \otimes \mathcal{N}_T^{(0)} \nonumber \\
&= -\frac{2}{\tau}\delta(1-x) g_s^2\frac{C_F}{2\pi}  \int \frac{d^{2-2\epsilon}\bfk}{(2\pi)^{2-2\epsilon}} \int \frac{d^{2-2\epsilon}\bfq}{(2\pi)^{2-2\epsilon}} 
 g_s^2 \frac{C_T}{d_A} \frac{1}{\bfq^4} g_s^2 C_A  \nonumber\\
 & \times \left\{ \left(e^{i\bfb\cdot(\bfk-\bfq) }- e^{-i\bfb\cdot\bfq}\right) \left(\frac{1}{\bfk^2}-\frac{\bfk\cdot(\bfk-\bfq)}{\bfk^2(\bfk-\bfq)^2}\right) \left[\frac{\min\{p_1^+, e^{\gamma_E-1}\bfk^2 L^+/2\}}{\nu}\right]^{-\tau} \right.\nonumber\\
 &\left.+ \left(e^{i\bfb\cdot(\bfk-\bfq)}-1\right)  \left(\frac{1}{(\bfk-\bfq)^2}-\frac{\bfk\cdot(\bfk-\bfq)}{\bfk^2(\bfk-\bfq)^2}\right) \left[\frac{\min\{p^+, e^{\gamma_E-1}(\bfk-\bfq)^2 L^+/2\}}{\nu}\right]^{-\tau} \right\}.
 \label{eq:Arap}
\end{align}
From appendix~\ref{app:separation_Cnu_from_Cmu}, the choice of the Collins-Soper scale already takes into account both scenarios $\mu_b\ll \mu_E$ and $\mu_b \gg \mu_E$.
With a large transverse momentum scale as shown on the right panel of figure~\ref{fig:TMD-DY-pA-modes}, the phase space region that contributes to the rapidity logarithm is unaffected by the LPM effect. Therefore, the CS scale is the same as the one in the vacuum, $\sqrt{\zeta_1}$. 
When the transverse momentum scale is small (left panel of figure~\ref{fig:TMD-DY-pA-modes}), the rapidity logarithm is cut off by the LPM effect at the critical line, which results in a new CS scale of order $\mu_b^2 L^+/2$, which we referred as the LPM CS scale. This is directly demonstrated in appendix~\ref{app:separation_Cnu_from_Cmu}. 
The use of the minimum function is an approximate way to interpolate between the two limiting scenarios. 

One technical issue is that, for the scenario $\mu_b\ll \mu_E$, the transverse momentum scales in the rapidity regulator are different in each term. Nevertheless, so long as we work in the impact parameter space, both $\bfk^2 L^+/2$ and $(\bfk-\bfq)^2 L^+/2$ in Eq.~(\ref{eq:Arap}) will all be converted to $\mu_b^2 L^2/2$.
To show this, we perform the integral  for $\mu_b \ll \mu_E$ and find that
\begin{align}
\mathcal{J}_{q/q,A}^{(1),\rm rap}\otimes\Sigma_{AT}^{(0)}&  \otimes \mathcal{N}_T^{(0)} = \delta(1-x) \frac{1}{2\pi} \alpha_s(\mu^2)C_F \alpha_s(\mu^2)C_A \frac{g_s^2C_T}{d_A}\frac{ b^2}{4}     \left(\frac{\mu^2 b^2}{4e^{-2\gamma_E}}\right)^{2\epsilon} \nonumber\\
&\times \left( -\frac{2}{\tau} \right)\left(\frac{2L^+/\nu}{b^2}\right)^{-\tau} \left\{\frac{\Gamma(1-2\epsilon-\tau)}{\Gamma(2+\epsilon+\tau)}\frac{\pi\epsilon}{\sin(\pi\epsilon)}\frac{(-2\epsilon)\mathrm{B}(-\epsilon, 1-\epsilon)}{\Gamma(1-\epsilon)}\right.\nonumber\\
&\left.-\frac{\Gamma(-1-2\epsilon-\tau)}{\Gamma(1+\tau)}\frac{\mathrm{B}(1-\epsilon, -\epsilon-\tau)}{1+\epsilon+\tau}+\frac{\Gamma(-1-2\e-\tau)}{\Gamma(1+\tau)}\mathrm{B}(-\e,-\e-\tau)\right\} \nonumber\\
&= \delta(1-x) \alpha_s(\mu^2)C_F\frac{g_s^2C_T}{d_A}\frac{ b^2}{4} \left(\frac{\mu^2 }{\mu_b^2}\right)^{2\epsilon} \left\{-\frac{1}{2\epsilon^2}+\frac{1}{2\epsilon} +\frac{3}{2} -\frac{\pi^2}{12} + \mathcal{O}(\epsilon)\right\} \nonumber\\
&\times\frac{\alpha_s(\mu^2)C_A}{2\pi} \left( -\frac{2}{\tau} \right)\left(\frac{2L^+/b^2}{\nu}\right)^{-\tau}\left[1 + \mathcal{O}(\tau)\right],
\end{align}
where in the second equation we have expanded first in $\tau\rightarrow 0$ and then in $\epsilon\rightarrow 0$.
The main result is that the transverse momentum integrals do not lead to extra poles in $\tau$, and the natural scale for $\nu$ is replaced by $2L/b^2$. If one aims at extracting the leading-log contribution, it is sufficient to replace both $\bfk^2L^+/2$ and $(\bfk-\bfq)^2L^+/2$ 
in Eq.~(\ref{eq:Arap}) by $\mu_b^2L^+/2$ from the very beginning. This way, one can pull the rapidity regulator out of the transverse momentum integral and rewrite the expression in a more familiar form
\begin{align}
\nonumber
 & \mathcal{J}_{q/q,A}^{(1),\rm rap}\otimes\Sigma_{AT}^{(0)} \otimes \mathcal{N}_T^{(0)} \nonumber\\
 &= -\frac{2}{\tau } \left[\frac{\min\{2L^+/b^2, p^+\}}{\nu}\right]^{-\tau} \delta(1-x) \int \frac{d^{2-2\epsilon}\bfq}{(2\pi)^{2-2\epsilon}} 
\frac{d\sigma^{(0)}_{FT}}{d^2\bfq} \int \frac{d^{2-2\epsilon}\bfk}{(2\pi)^{2-2\epsilon}} \frac{g_s^2 C_F }{2\pi}  \nonumber\\
 & \times \left\{e^{i(\bfk-\bfq)\cdot\bfb} \left[\frac{1}{\bfk^2}+\frac{1}{(\bfk-\bfq)^2}-\frac{2\bfk\cdot(\bfk-\bfq)}{\bfk^2(\bfk-\bfq)^2}\right]-e^{-i\bfq\cdot\bfb}\left[\frac{1}{\bfk^2}-\frac{\bfk\cdot(\bfk-\bfq)}{\bfk^2(\bfk-\bfq)^2}\right]\right\} \nonumber \\
& = \delta(1-x)  \left[ -\frac{1}{\tau } + \mathcal{L}_n + \mathcal{O}(\tau)\right]   \int \frac{d^{2-2\epsilon}\bfq}{(2\pi)^{2-2\epsilon}} 
\hat{\mathcal{C}}\left[\frac{e^{-i\bfq\cdot \bfb}}{\bfq^2}\right] \bfq^2\frac{d\sigma_{FT}^{(0)}}{d^2\bfq} \, ,
\label{eq:A-rap-final-form}
\end{align}
where $\mathcal{L}_n=\ln\frac{\min\{2L^+/b^2, p^+\}}{\nu}$ is the rapidity logarithm given by the collinear function. $\hat{\mathcal{C}}$ is the BFKL kernel. Its action on a function $v(\bfq^2)$ is defined as
\begin{align}
\hat{\mathcal{C}}[v(\bfq^2)] = \frac{g_s^2 C_A}{\pi}   \int \frac{d^{2-2\epsilon}\bfk}{(2\pi)^{2-2\epsilon}} \left[\frac{1}{(\bfq-\bfk)^2}v(\bfk^2) - \frac{\bfq^2}{2\bfk^2(\bfq-\bfk)^2}v(\bfq^2)\right].
\label{eq:BFKL-k-space}
\end{align}
To recover the factorization form of Eq. (\ref{eq:Opacity-one-factorization-1}), we can insert the identity 
\begin{align}
1=\int d^2 \bfp\delta^{(2)}(\bfp-\bfq) \int d^2 \bfq\delta^{(2)}(\bfq-\bfq') \, .
\end{align}
Then, the sum of the LO result and the leading-log correction to the collinear function can be compactly written as
\begin{align}
&\mathcal{J}_{q/q,F}^{(0)}\otimes\Sigma_{FT}^{(0)} \otimes \mathcal{N}_T^{(0)} + \mathcal{J}_{q/q,A}^{(1),\rm rap}\otimes\Sigma_{AT}^{(0)} \otimes \mathcal{N}_T^{(0)} \nonumber\\
&=\delta(1-x_1)\int d^{2-2\epsilon}\bfp  e^{-i\bfp\cdot \bfb} \delta^{2-2\epsilon}(\bfp-\bfq) \int \frac{d^{2-2\epsilon}\bfq}{(2\pi)^{2-2\epsilon}} \int \frac{d^{2-2\epsilon}\bfq'}{(2\pi)^{2-2\epsilon}} \nonumber\\
&\times \left[1 +\left( -\frac{1}{\tau } + \mathcal{L}_n\right)\hat{\mathcal{C}}\right]\left[\frac{\delta(1-x)\delta^{(2-2\epsilon)}(\bfp-\bfq)}{\bfq^2}\right] \bfq^2\bfq'^2 \Sigma_{FT}^{(0)} \delta^{(2-2\epsilon)}(\bfq-\bfq') \frac{1}{{\bfq'}^2} \nonumber \\
&=\int \frac{d^{2-2\epsilon}\bfp}{(2\pi)^{-2\epsilon}}  e^{-i\bfp\cdot \bfb} \int \frac{d^{2-2\epsilon}\bfq}{(2\pi)^{2-2\epsilon}} \int \frac{d^{2-2\epsilon}\bfq'}{(2\pi)^{2-2\epsilon}} \nonumber\\
&\times\left[1 +\left( -\frac{1}{\tau } + \mathcal{L}_n\right)\hat{\mathcal{C}}\right]\left[\frac{\mathcal{J}_{q/q,F}^{(0)}}{\bfq^2}\right] \bfq^2{\bfq'}^2 \Sigma_{FT}^{(0)} \frac{\mathcal{N}_T^{(0)}}{{\bfq'}^2}.
\end{align}

\section{Cancellation of the rapidity divergence at first order in opacity}\label{sec:TMD-NLO-soft}
\subsection{The soft and the anti-collinear sectors}
Without the complication from the LPM effects, the calculations for the anti-collinear sector are straightforward and are given in appendix~\ref{app:anticollinear}.
The rapidity divergence from the NLO calculation of the anti-collinear sector is
\begin{align}
&\mathcal{J}_{q/q,F}^{(0)}\otimes\Sigma_{FT}^{(0)} \otimes \mathcal{N}_T^{(1)}  \nonumber\\
&= \int \frac{d^{2-2\epsilon}\bfp}{(2\pi)^{-2\epsilon}}  e^{-i\bfp\cdot \bfb} \int \frac{d^{2-2\epsilon}\bfq}{(2\pi)^{2-2\epsilon}} \int \frac{d^{2-2\epsilon}\bfq'}{(2\pi)^{2-2\epsilon}}
\frac{\mathcal{J}_{q/q,F}^{(0)}}{\bfq^2} \bfq^2{\bfq'}^2 \Sigma_{FT}^{(0)} \left( -\frac{1}{\tau } + \mathcal{L}_{\bar{n}}\right)\hat{\mathcal{C}}
\left[\frac{\mathcal{N}_T^{(0)}}{{\bfq'}^2}\right] \, ,
\end{align}
with $\mathcal{L}_{\bar{n}} = \ln \frac{x_t P^-}{\nu}$. Its Collins-Soper scale is the minus component of the momentum of the target parton. It carries $x_t$ fraction of the momentum of the target nucleon. 

For the soft radiation induced by the Glauber exchange, the rapidity regulator is~\cite{Chiu:2012ir}
\begin{equation}
\left|\frac{k_z}{\nu}\right|^{-\tau/2} = \left|\frac{k^+-k^-}{2\nu}\right|^{-\tau/2}.
\end{equation}
From the derivations provided in appendix~\ref{app:soft}, we have
\begin{align}
&\mathcal{J}_{q/q,F}^{(0)}\otimes\Sigma_{FT}^{(1)} \otimes \mathcal{N}_T^{(0)} \nonumber\\
&= \int \frac{d^{2-2\epsilon}\bfp}{(2\pi)^{-2\epsilon}}  e^{-i\bfp\cdot \bfb} \int \frac{d^{2-2\epsilon}\bfq}{(2\pi)^{2-2\epsilon}} \int \frac{d^{2-2\epsilon}\bfq'}{(2\pi)^{2-2\epsilon}}
\frac{\mathcal{J}_{q/q,F}^{(0)}}{\bfq^2} \left( \frac{2}{\tau } + \mathcal{L}_{s}\right)\hat{\mathcal{C}}
\left[\bfq^2{\bfq'}^2 \Sigma_{FT}^{(0)}\right] \frac{\mathcal{N}_T^{(0)}}{{\bfq'}^2},
\end{align}
where the soft logarithm is $\mathcal{L}_{s} = \ln \frac{\nu^2}{\mu_b^2}$. 

If one substitutes the LO expression for $\mathcal{J}_{q/q,F}^{(0)}$, $\Sigma_{FT}^{(0)}$, and $\mathcal{N}_T^{(0)}$, it is straightforward to show that the rapidity divergence in the form of poles in $1/\tau$ cancels when summing the NLO corrections to all three sectors. The explicit result reads
\begin{align}
&\mathcal{J}_{q/q,F}^{(0)}\otimes\Sigma_{FT}^{(1)}\otimes \mathcal{N}_T^{(0)}+\mathcal{J}_{q/q,A}^{(1), \rm rap}\otimes\Sigma_{AT}^{(0)}\otimes \mathcal{N}_T^{(0)}\nonumber\\
&+\mathcal{J}_{q/q,F}^{(0)}\otimes\Sigma_{FT}^{(1)}\otimes\mathcal{N}_T^{(0)}
+\mathcal{J}_{q/q,F}^{(0)}\otimes\Sigma_{FT}^{(0)} \otimes \mathcal{N}_T^{(1)} \nonumber\\
&= \delta(1-x) \frac{g_s^2 C_F g_s^2 C_T}{d_A} \int \frac{d^{2-2\epsilon}\bfq}{(2\pi)^{2-2\epsilon}} e^{i\bfq\cdot\bfb} \frac{1}{\bfq^2}\left[1+\mathcal{L}_1\hat{\mathcal{C}}\right] \frac{1}{\bfq^2}.
\end{align}
The final logarithmic enhancement factor is also independent of the rapidity renormalization scale $\nu$:
\begin{align}
\mathcal{L}_1 &= \mathcal{L}_n + \mathcal{L}_s + \mathcal{L}_{\bar{n}} \\
&= \ln\frac{\min\{2L^+\mu_b^2, x_1 P_a^+\}}{\nu} + \ln \frac{\nu^2}{\mu_b^2} + \ln \frac{x_t P_b^-}{\nu}. 
\end{align}
Because $L^+ P^-$ is boost invariant, it is easiest to express the product in the rest frame of the nucleus, where $L^+ = 2L$ and $P^- = m_N$. The logarithmic factor is rewritten as
\begin{align}
\mathcal{L}_1 &= \ln\left(\min\left\{ 4 x_t m_N L  , \frac{x_1 x_t s}{\mu_b^2}\right\}\right) \nonumber\\
&= \ln\left(\min\left\{ 4 m_N L  , \frac{x_1 s}{\mu_b^2}\right\}\right) + \ln x_t \label{eq:N=1-log}
\end{align}
where we have used $s=P_a^+P_b^-$.

\subsection{The rapidity evolution equations}
Following the rapidity renormalization, one can write down the BFKL-type rapidity evolution equation for each sector in Eq.~(\ref{eq:Opacity-one-factorization-1}).
For clarity, we distinguish the $\nu$ that separates the collinear radiation from soft radiation and $\nu'$ that separates the soft radiation from the anti-collinear radiation.
The evolution equations are
\begin{align}
\frac{g_s^2}{\bfq^2}\frac{\partial \mathcal{J}_R(x, \bfp, \bfq;\nu)  }{\partial \ln \nu}  &=   -\hat{\mathcal{C}}\left[ \frac{g_s^2}{\bfq^2}\mathcal{J}_R(x, \bfp, \bfq;\nu)\right], \\
\frac{g_s^2}{{\bfq'}^2} \frac{\partial \mathcal{N}_T(\bfq';\nu') }{\partial \ln \nu'}  &=   -\hat{\mathcal{C}}\left[\frac{g_s^2}{ {\bfq'}^2}\mathcal{N}_T(\bfq';\nu') \right],\\
\left(\frac{g_s^2}{{\bfq}^2}\right)^{-1}\left(\frac{g_s^2}{{\bfq'}^2}\right)^{-1} \frac{\partial  \Sigma_{RT}(\bfq, \bfq'; \nu, \nu')}{\partial \ln \nu} &= \hat{\mathcal{C}}\left[ \left(\frac{g_s^2}{{\bfq}^2}\right)^{-1}\left(\frac{g_s^2}{{\bfq'}^2}\right)^{-1}\Sigma_{RT}(\bfq, \bfq';\nu, \nu')\right],
\end{align}
For the evolution of the soft function with respect to $\nu$, the $\hat{\mathcal{C}}$ operator acts on its $\bfq$ dependence. A similar equation for the $\nu'$ dependence with $\hat{\mathcal{C}}'$ acting on $\bfq'$ can be written down, but it is not shown here explicitly.
The initial conditions of each sector come from Eq.~(\ref{eq:LO-Glauber-cross-section}).

The BFKL evolution will match two adjacent sectors at some common rapidity scale $\nu$ and $\nu'$ (or one can simply choose $\nu=\nu'$). For example, we can choose to keep the soft function at its natural scale $\nu=\nu'=\mu_b$ and evolve the collinear (anti-collinear) function from $\nu=\min\{2L^+\mu_b^2,x_1P_a^+\}$ ($\nu'=x_tP_b^-$) to $\nu=\mu_b$ ($\nu'=\mu_b$). For this choice, the soft function is just a delta function plus NLO corrections, if we target the leading logarithmic behavior, the evolution of the collinear and anti-collinear functions can be combined into a single evolution in $\nu$ that resums the entire logarithm, such that $\ln \nu_{\rm max}/\nu_{\rm min} = \mathcal{L}_1$. 

\begin{figure}
    \centering
    \includegraphics[scale=.85]{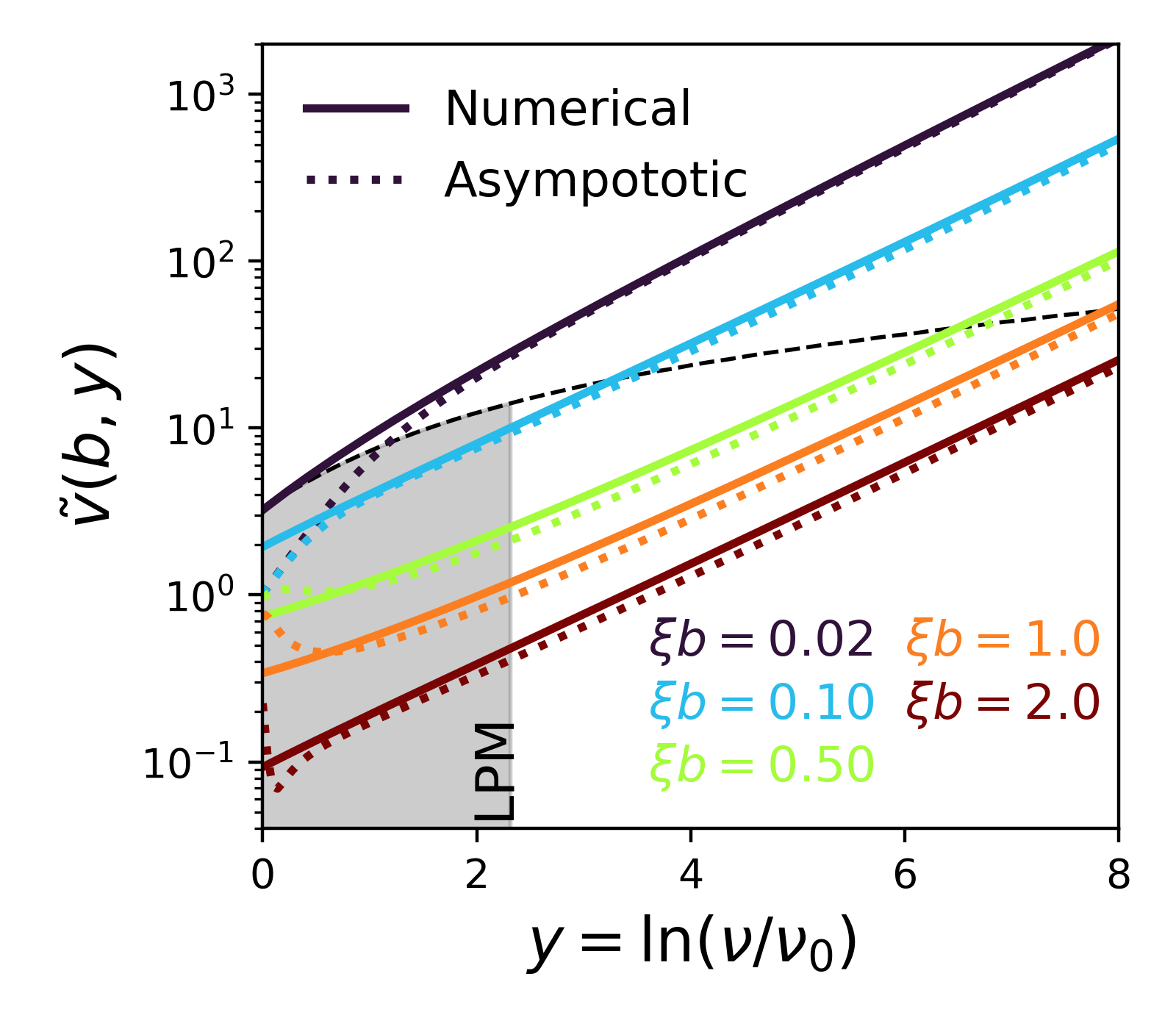}
    \caption{Numerical solution versus the asymptotic behavior of the BFKL solution in the diffusion limit $\xi b \sim 1$. The test uses a fixed coupling $\alpha_{s, \rm fix} = 0.3$. }
    \label{fig:BFKL-numerical}
\end{figure}

\subsection{Numerical solution to the BFKL equation}
The leading order BFKL equation can be solved by the eigenfunction method, and semi-analytic asymptotic solutions can be obtained in certain regions of phase space. However, due to the LPM effect, the rapidity logarithm is not very large. For example, using the average path length $\langle L \rangle=\frac{3}{4}r_0 A^{1/3}$  in this estimate and if the typical $x_t \sim 0.1$, then $\ln(4x_t m_N \langle L \rangle)\approx 2.3$ for a lead nucleus ($A=208$). Consequently, the region of validity of the asymptotic solutions may not be reached for realistic nuclei.

For phenomenological applications, we use a numerical solver for the BFKL equation in the impact parameter ($b$) space.
To perform a Fourier transformation to the BFKL equation, we use dimensional regularization to properly regulate the singular behavior of the propagator,
\begin{align}
&\frac{\partial \tilde{v}_\epsilon(\bfb)}{\partial \ln \nu} =   \frac{\alpha_s^{(0)}C_A}{\pi^2}  \int \frac{d^{2-2\epsilon}\bfk }{(2\pi)^{-2\epsilon}} e^{i\bfk\cdot\bfb}\int \frac{d^{2-2\epsilon}\bfq }{(2\pi)^{2-2\epsilon}} \left[ \frac{1}{(\bfk-\bfq)^2} v(\bfq^2)- \frac{\bfk^2}{2\bfq^2(\bfk-\bfq)^2}v(\bfk^2)\right] \nonumber \\
&= \frac{\alpha_s(\mu^2) C_A}{\pi}e^{-\epsilon\gamma_E} \left\{ \tilde{v}_\epsilon(\bfb)\left(\frac{\mu^2}{\mu_{b}^2}\right)^\epsilon \Gamma(-\epsilon) -\frac{B(-\epsilon, -\epsilon)\Gamma(1-2\epsilon)}{2[\Gamma(1-\epsilon)]^2\Gamma(\epsilon)}\frac{\pi\epsilon}{\sin(\pi\epsilon)}\left[\tilde{v}_\epsilon(\bfb)\left(\frac{\mu^2R^2}{\mu_b^2}\right)^{\epsilon}\frac{1}{\epsilon}\right.\right.\nonumber\\
&\left.\left. + \frac{\Gamma(1-\epsilon)}{\pi^{1-\epsilon}}\left(\int_{|\bfb-\bfb'|>R|\bfb|}\frac{d^{2-2\epsilon}\bfb'\tilde{v}_\epsilon(\bfb')}{|\bfb-\bfb'|^{2(1-2\epsilon)}}+\int_{|\bfb-\bfb'|<R|\bfb|}\frac{d^{2-2\epsilon}\bfb'\left(\tilde{v}_\epsilon(\bfb')-\tilde{v}_\epsilon(\bfb)\right)}{|\bfb-\bfb'|^{2(1-2\epsilon)}}\right)\right]\right\}.
\label{eq:BFKL-Fourier}
\end{align}
Here, $R$ is an arbitrary positive number that separates two regions of the $\bfb'$ integration. $\tilde{v}_\epsilon(\bfb)$ is the Fourier transform of $v(\bfq^2)$ under dimensional regularization.
When taking the limit $\epsilon\rightarrow 0$, the explicit $1/\epsilon$ pole in Eq. (\ref{eq:BFKL-Fourier}) cancels. 
Because this is a linear and homogeneous equation, any poles in the $\tilde{v}_\epsilon(\bfb)$ can be absorbed by a separate renormalization of $\lim_{\epsilon\rightarrow 0}Z_\epsilon^{-1}\tilde{v}_\epsilon(\bfb) = \tilde{v}(\bfb,\mu)$. Consequently,  in the $\epsilon\rightarrow 0$ limit, we arrive at 
\begin{align}
\frac{\partial \tilde{v}(\bfb,\mu)}{\partial y}
= \frac{\alpha_s(\mu^2) C_A}{\pi} &\left( \tilde{v}(\bfb,\mu)\ln R^2 + \int_{|\bfb-\bfb'|>R|\bfb|}\frac{d^{2}\bfb'}{\pi}\frac{\tilde{v}(\bfb',\mu)}{|\bfb-\bfb'|^{2}} \right.\nonumber\\
&\left.+ \int_{|\bfb-\bfb'|<R|\bfb|}\frac{d^{2}\bfb'}{\pi}\frac{\tilde{v}(\bfb',\mu)-\tilde{v}(\bfb,\mu)}{|\bfb-\bfb'|^{2}}\right),
\end{align}
with $y = \ln \nu + {\rm const}$.
The equation does not depend on the choice of the separation parameter $R$. 
As a cross-check, if one takes $R\rightarrow\infty$, it is straightforward to show that the eigenvalue of a basis function $\tilde{v}(b)\sim b^{-2\gamma}$ is $\frac{\alpha_{s, \rm fix}C_A}{\pi}[2\Psi(1)-\Psi(\gamma)-\Psi(1-\gamma)]$, with $\Psi$ being the di-gamma function, which is indeed the eigenvalue of the BFKL equation. For a well-defined numerical scheme, we choose the separation parameter $R=1$.

To construct an initial condition for $\tilde{v}(\bfb,\mu)$, we first examine the leading-order expression for the anti-collinear sector
\begin{align}
g_s^2 C_T \int \frac{d^{2-2\epsilon}\bfq}{(2\pi)^{2-2\epsilon}} \frac{e^{-i\bfq\cdot\bfb}}{\bfq^2} = \alpha_s(\mu^2)C_T\left[-\frac{1}{\epsilon}+\ln\frac{\mu_b^2}{\mu^2}+\mathcal{O}(\epsilon)\right].
\end{align}
It reflects the scale dependence of $\tilde{v}$ at small $b$, but neglects the details in the infrared.
We can model the infrared behavior of the anti-collinear function using an in-medium effective mass $\xi$,
\begin{align}
g_s^2 C_T \int \frac{d^2\bfq}{(2\pi)^2} \frac{e^{-i\bfq\cdot\bfb}}{\bfq^2+\xi^2} = 2\pi \frac{\alpha_s C_T}{\pi} K_0(\xi b) \approx \alpha_s C_T\left[\ln\frac{\mu_b^2}{\xi^2}+\mathcal{O}(\xi^2 b^2)\right].
\end{align}
To interpolate the two scenarios, we will take $\mu=\mu_b$ and use the following expression to model the initial condition 
\begin{align}
\tilde{v}_{T}(\bfb; y=0) = 2\pi\frac{\alpha_s\left(\mu_b^2+\xi^2\right) C_T}{\pi} K_0\left(b \sqrt{\mu_b^2+\xi^2}\right).
\end{align}

Using fixed coupling of $\alpha_{s, \rm fix}=0.3$, we tested the numerical solver with results shown in  figure~\ref{fig:BFKL-numerical}. The solid lines show the evolution of $\tilde{v}(\bfb, y)$ with respect to the rapidity $y$ at various impact parameters $b$ measured in units of $\xi^{-1}$. 
After about one to two units of rapidity evolution, the solution looses most of the memory to the initial condition and acquires a power-law dependence on  $\nu$.
This can be compared to the double-log approximation (DLA) solution in~\cite{Kovchegov_Levin_2012}. The DLA solution transformed to the $b$ space reads
\begin{align}
\tilde{v}_T^{\rm DLA}(\bfb; y) = C_0 e^{\left(\alpha_P-1\right)y } \int\frac{d^2\bfq}{(2\pi^2)}\frac{e^{-i\bfb\cdot\bfq}}{2|\bfq| \xi}  \frac{ e^{-\frac{\left[\ln|\bfq|-\ln \xi\right]^2}{2\sigma^2 y}}}{\sqrt{2\pi\sigma^2 y }}  \, ,
\end{align}
where BFKL pomeron intercept is $\alpha_P-1=\frac{\alpha_{s, \rm fix} C_A}{\pi} 4\ln 2$, $\sigma = 7\zeta(3)\frac{\alpha_{s, \rm fix} C_A}{\pi} y$, and the parameter $C_0$ is chosen such that $\tilde{v}_T^{\rm DLA}(\bfb; y=0)$ matches our numerical initial condition at $\xi b=0.35$.
At large rapidity, the numerical solution traces the behavior of the asymptotic double-log solution (dashed lines) very well. 

\section{Effects from higher-orders in opacity}\label{sec:final-formula}
The NLO calculation is now complete to first order in opacity. The rapidity divergence are shown to cancel among the collinear, anti-collinear, and the soft sector. The collinear divergence is canceled by medium counter terms that are regarded as a simple model for the collinear-soft sector. 
Nevertheless, results at a fixed order in opacity are difficult to inverse Fourier transform from the impact parameter space to the transverse momentum space. In this section, we will discuss a partial resummation of terms from higher orders in opacity to arrive at a result that shows improved behavior and can be applied to phenomenology.

From the end of section~\ref{sec:TMD-NLO-matching:collinear-div}, we recall that the in-medium collinear evolution equations already sum a subset of terms from higher orders in opacity. For example, the in-medium parton energy loss reflected in the modification of the parton spectrum $F_{i/p}(z)=z f_{i/p}(z)$ emerges from the summation of terms of order $\left(\chi \mathcal{L}_E\right)^n \frac{\partial^n}{\partial z^n} F_{i/p}(z)$ at each order in opacity with $\mathcal{L}_E = \ln\left(\frac{\gamma(w)\mu_E^2}{\xi^2}\right)$. 

As for the momentum broadening effect, the summation of all orders in opacity at leading order accuracy is well-known~\cite{Gyulassy:2002yv} and leads to an exponential broadening factor
\begin{align}
1 + \chi\mathcal{B}_{q/i,1}^{(0)}+ \cdots =\exp\left\{\sum_j\rho_0^-L^+ \int dx_t f_{j/N}(x_t) \left[\tilde{\Sigma}_{ij}^{(0)}(b)-\tilde{\Sigma}_{ij}^{(0)}(0)\right]\right\}.
\label{eq:LO-broadening}
\end{align}
The subtraction term $\tilde{\Sigma}(0)$ comes from the unitary correction. If one uses a dimensional regularization, it is scaleless and $\tilde{\Sigma}(0)=0$. For other regularizations, such as the use of an effective mass mass in Ref.~\cite{Gyulassy:2002yv} and in 
Eq.~(\ref{eq:LO-Glauber-cross-section}), $\tilde{\Sigma}(0)$ is non-zero and has to be subtracted. The subtraction guarantees that the broadening factor is unity at $b=0$, reflecting conservation of probability. 
At NLO, the forward scattering cross-section is dressed by soft gluon emissions. The rapidity RG evolution equation renormalizes the forward scattering between the projectile parton with a single parton from the target. Thus, soft radiations do not alter the structure of the opacity summation in this framework. The only change is replacing $\tilde{\Sigma}_{ij}^{(0)}$ by the renormalized quantity, 
\begin{align}
\tilde{\Sigma}_{ij}^{(0)}(b) \Longrightarrow \tilde{\Sigma}_{ij}(b,y) \, ,  
\end{align}
in Eq.~(\ref{eq:LO-broadening}). $\tilde{\Sigma}_{ij}(b,y)$ is obtained from the solution of $\tilde{v}_T(\bfb,y)$
\begin{align}
\tilde{\Sigma}_{ij}(b,y) = \int d^2\bfb'\tilde{v}_i(\bfb-\bfb',y) \tilde{v}_j(\bfb',0).
\end{align}
As pointed out in Ref.~\cite{Vaidya:2021vxu}, the validity of such an exponentiation beyond LO holds when the medium is dilute. When the system is dense, contributions from soft radiations interacting with two or more collision centers coherently can become important, consequently, the renoramlization is no longer local to a single collision center. In the very dense limit, this can be analyzed by summing an infinite number of sources that coherently interact with the soft radiation~\cite{Caucal:2021lgf,Caucal:2022fhc,Caucal:2022mpp}. For nuclear tomography and Drell-Yan process at not very small $x$, we consider the nuclear matter to still be dilute.

\begin{figure}
    \centering
    \includegraphics[scale=.95]{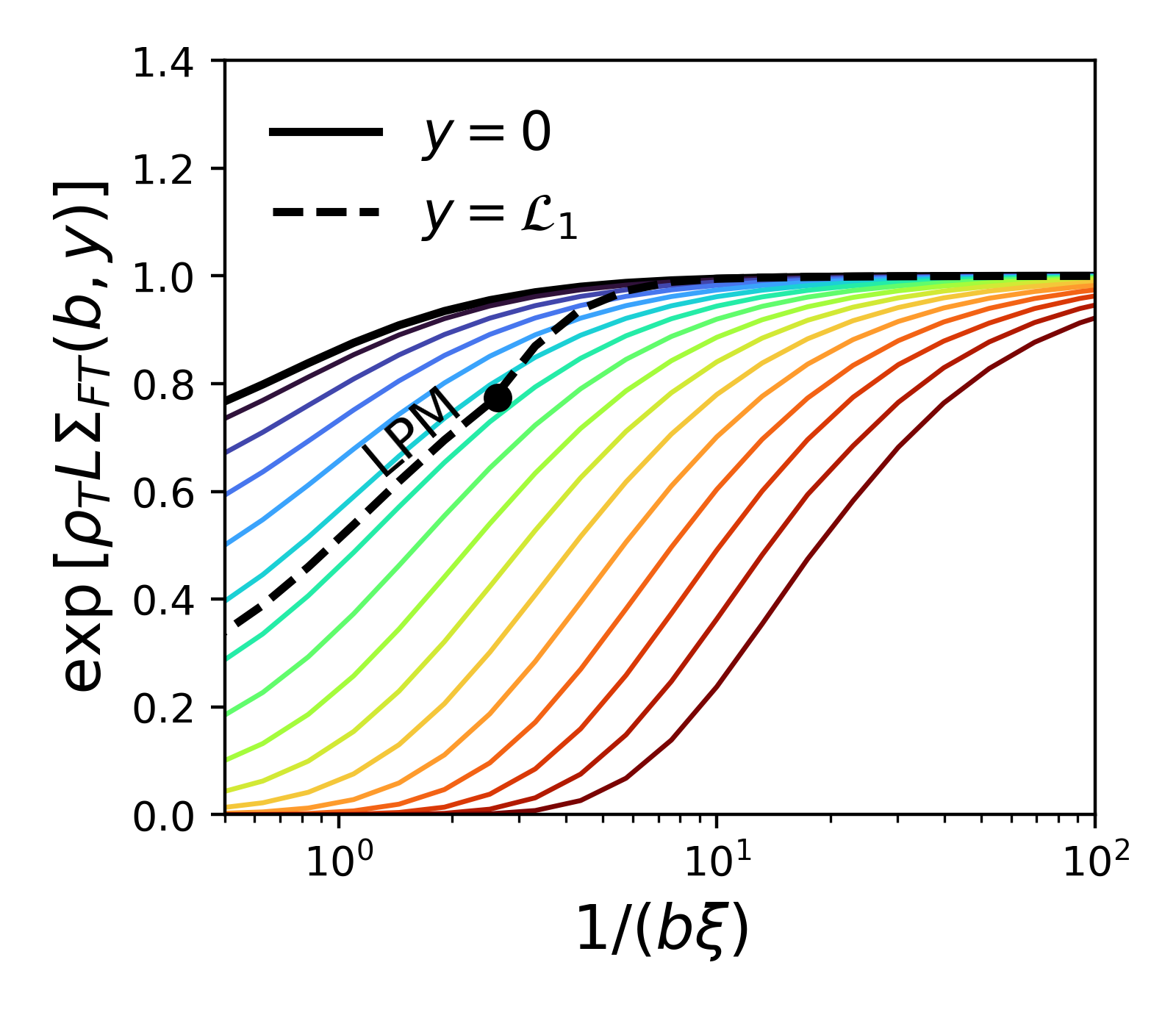}
    \caption{The exponentiation of the transverse momentum broadening factor. The initial condition is shown as the black solid line. The evolution towards a larger rapidity is shown by the colored lines. The physical boundary that takes into account the LPM effect is shown as the black dashed line.}
    \label{fig:Broadening-factor-bspace}
\end{figure}

In figure~\ref{fig:Broadening-factor-bspace}, we show the broadening factor exponentiation as a function of $1/b$. The back solid line is the LO result as the initial condition of the evolution. The colored lines from purple to red show the evolution towards larger rapidity $y$. The numerical results are obtained using a fixed coupling $\alpha_s=0.3$. The physical boundary of the evolution, shown as the black dashed line, is given by Eq.~(\ref{eq:N=1-log}) that takes into account the LPM effect. 
The kink labeled by the black dot is the critical value of $\mu_b$, where $\mu_b  = \sqrt{\frac{x_1 s}{2 P_b^-L^+}}$. At $\mu_b$ larger than this critical value, the range of the BFKL evolution is unaffected by the LPM effect $y=\ln \frac{x_1 x_t s}{\mu_b^2}$. Note that this range of evolution decreases at larger $\mu_b$, therefore the correction relative to the LO result strongly depends on $\mu_b$ in this region. 
At smaller $\mu_b$, the LPM effect takes effect and the evolution boundary is $\mu_b$ independent and is a function of $A^{1/3}$. This is why the dashed line follows a curve of a fixed value of $y$ below the critical point.

Comparing the dashed and the solid line in figure~\ref{fig:Broadening-factor-bspace}, we conclude that the rapidity renormalization has a significant impact on the size of the transverse momentum broadening. In the small $\mu_b$ (or $p_T$) region, the evolution leads to an almost three times larger effect than the leading order calculation.  
We finally note that even in the absence of matter (the LPM) there are kinematic effects and higher order corrections that slow down BFKL evolution~\cite{Deak:2019wms,Ciafaloni:2003rd}, which is empirically too fast. It might be interesting to investigate the interplay of such effects in the future.

\subsection{The projectile beam function evolving in cold nuclear matter}
Combining the higher-order opacity partial summation from both collinear evolution of the parton density and the transverse momentum broadening, we arrive at the final formula for the beam function of projectile ``$a$'' with cold nuclear matter effects
\begin{align}
\mathcal{B}_{q/a}^{\rm CNM}\left(x_1, b, \mu, \frac{\zeta_1}{\nu^2}; \mu_E, \mathcal{L}_1\right) = &\sum_i \int_{x_1}^1 \frac{dx}{x} f_{i/a}\left(\frac{x_1}{x}, \mu_b^*, \mu_E\right) C_{q/i}\left(x, b, \mu_b^*, \frac{\zeta_1}{\nu^2}\right) e^{-S_{\rm NP}^f(b, \zeta_1)} \nonumber\\
& \times \exp\left\{\rho_0^-L^+ \sum_j \int dx_t f_{j/N}(x_t) \left[\tilde{\Sigma}_{ij}(b, \mathcal{L}_1)-\tilde{\Sigma}_{ij}(0, \mathcal{L}_1)\right]\right\}\nonumber\\
& \times \left(1 + \rho_0^-L^+ \sum_j \int dx_t f_{j/N}(x_t) \Delta \sigma_{ij\rightarrow q}^{\rm NLO} \right).
\label{eq:Beam-Function-in-CNM}
\end{align}
The parton density $f_{i/a}\left(\frac{x_1}{x}, \mu_b, \mu_E\right)$ is evolved by the vacuum PDF evolution to scale $\mu_b$ and then includes the cold nuclear matter effects using the in-medium evolution equations in Eqs. (\ref{eq:collinear_evolution-1}), (\ref{eq:collinear_evolution-2}), and (\ref{eq:collinear_evolution-3}).
$C_{q/i}$ is the TMD matching coefficient in the vacuum, and $S_{\rm NP}$ is a non-perturbative function. The second line of Eq. (\ref{eq:Beam-Function-in-CNM}) is the exponentiation of the transverse momentum broadening factor with the renormalized partonic forward scattering cross-sections. $y$ is evolved from 0 to $\mathcal{L}_1$ in Eq. (\ref{eq:N=1-log}).
The last line contains the leftover of the NLO correction at opacity order one.
This result is accurate to NLO+NNLL for elementary collisions. For the CNM effects, it is accurate to NLO at first order in opacity and sums leading-log contribution to the parton energy loss and momentum broadening effects from higher orders in opacity.

The introduction of $S_{\rm NP}$ and $b^*$ (and $\mu_b^*$) is to model the intrinsic non-perturbative TMD parton distribution inside the nucleon and then interpolate to the perturbative calculations. A common parametrization of $S_{\rm NP}$ is taken from Refs.~\cite{Sun:2014dqm,Kang:2015msa},
\begin{align}
S_{\rm NP}^f(b, \zeta) &= \frac{g_2}{2} \ln\frac{b}{b^*} \ln\frac{\sqrt{\zeta}}{\sqrt{\zeta_0}} + g_1^f b^2 \, .
\end{align}
Here, $g_2 = 0.84$, $g_1^f = 0.106$ GeV$^2$, and the reference scale $\sqrt{\zeta_0}=Q_0 = \sqrt{2.4}$~GeV are taken from Ref.~\cite{Echevarria:2020hpy,Alrashed:2021csd}.
The $b^*$ prescription provides a soft cut-off on $b$,
\begin{align}
b^* &= \frac{b}{\sqrt{1+b^2/b^2_{\rm max}}}, \quad \mu_b^* = 2e^{-\gamma_E}/b^* \, , 
\end{align}
ensures that $b^*\approx b$ at perturbative values and saturates at $b_{\rm max}=1.5$~GeV$^{-1}$~\cite{Echevarria:2020hpy}.
Ref.~\cite{Alrashed:2021csd} also provides a phenomenological extraction of the $g_1^f$ in the nucleus with nuclear mass number dependence. Nevertheless, we want to examine to what extent dynamical effects can explain the modifications, so the same $g_1^f$ as that of a free proton will be used. 

In the calculations, quantities $x_1$, $s$, $\mu_b\sim 1/P_T$, $r_0 A^{1/3}$ can be controlled by experimental kinematics and the choice of different nuclear targets. This is not the case for $x_t$, because one cannot precisely determine the momentum fraction of every medium constituent in the multiple collisions. In principle, $x_t$ should be averaged over the distribution $f_{j/N}(x_t)$ of color charge $j$ in the medium. It is not the goal of this paper to construct a detailed model for $f_{j/N}(x_t)$, instead, we try to relate this quantity to the cold nuclear matter parameter $\rho_G$ in calculating the parton energy loss. We know that, without the rapidity evolution, the forward scattering cross section is the same as that used in the energy loss calculation. Therefore, we approximate $y$ and replace the medium color charge density using the definition of $\rho_G$ in Eq.~(\ref{eq:rhoG})
\begin{align}
&\rho_0^- L^+ \sum_j\int_{0}^1 dx_t f_{j/N}(x_t) \left[\Sigma_{ij}(b,\mathcal{L}_1)-\Sigma_{ij}(0,\mathcal{L}_1)\right] \nonumber\\ 
& \hspace*{.4cm}\approx \rho_G^- L^+ \sum_j \frac{d_A}{g_s^2 C_j}\left[\tilde{\Sigma}_{ij}(b,\bar{\mathcal{L}}_1)-\tilde{\Sigma}_{ij}(0,\bar{\mathcal{L}}_1)\right]\,.
\end{align}
Then, we use an averaged $\bar{x}_t$ to represent the typical longitudinal momentum fraction of medium color source, which defines $\bar{\mathcal{L}}_1$ through Eq.~(\ref{eq:N=1-log}). In the following calculations, we will use $\bar{x}_t=0.05$ for an estimation.

\begin{figure}
\centering
\includegraphics[scale=.85]{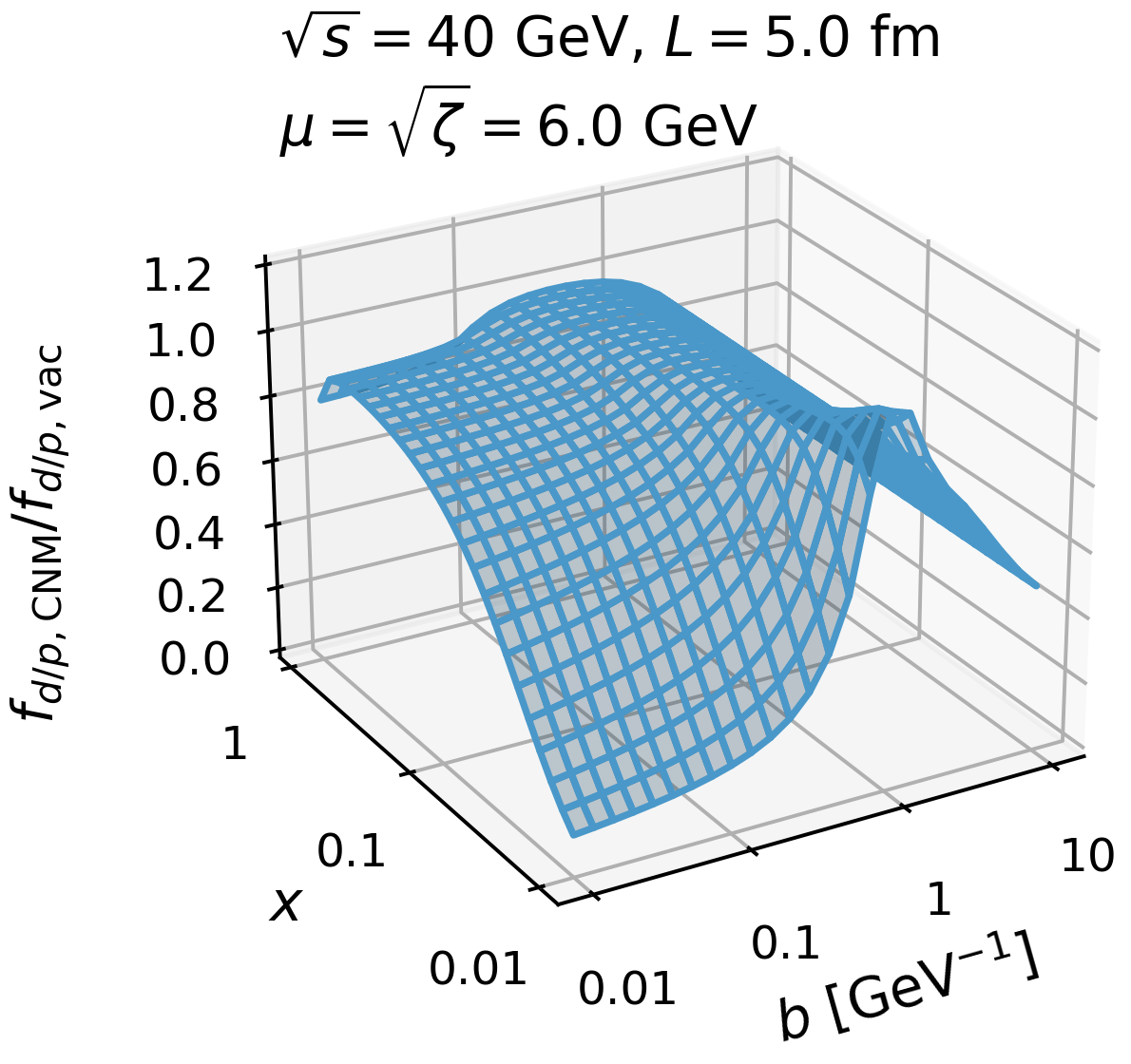}~~
\includegraphics[scale=.85]{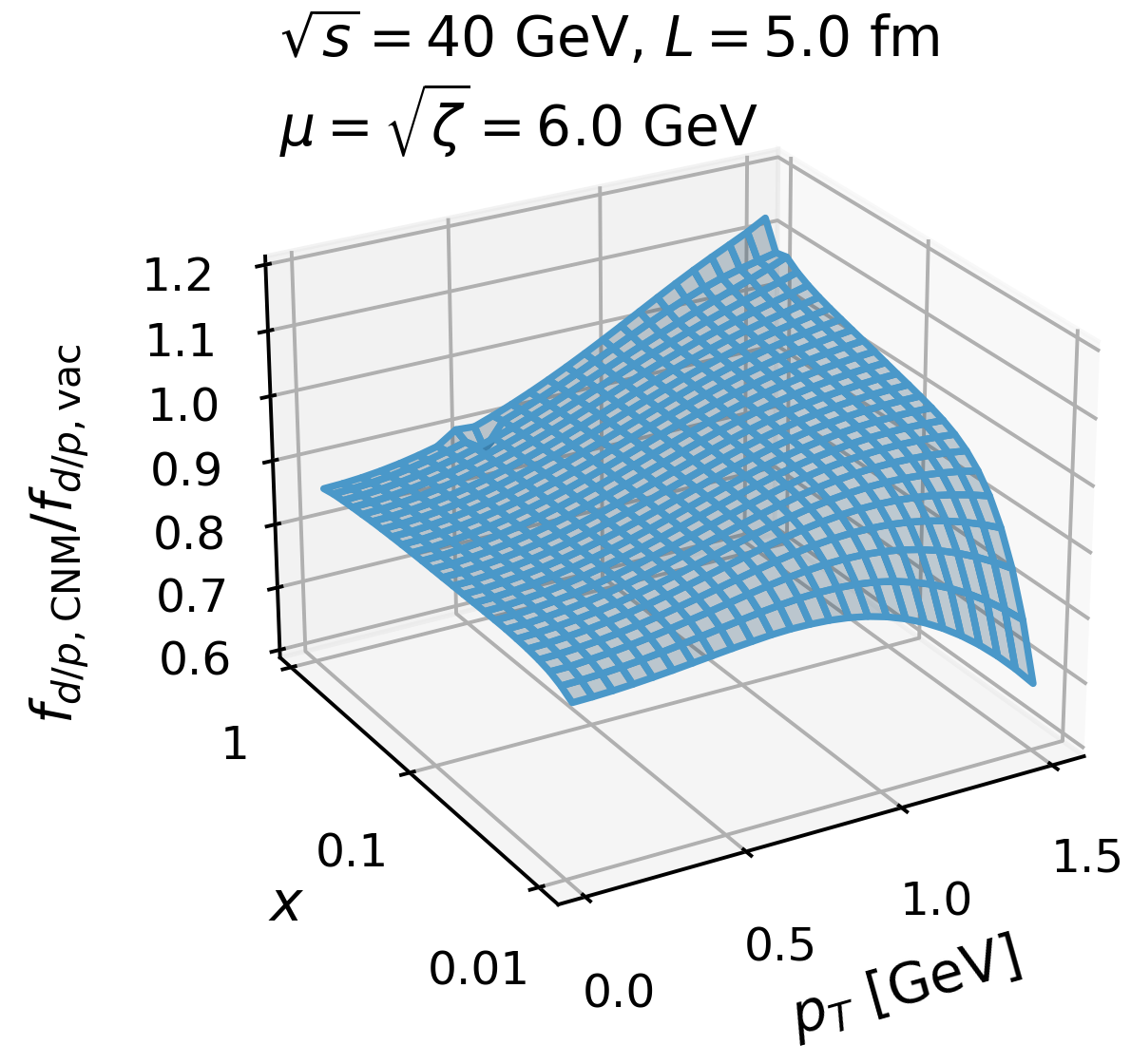}
\caption{The ratio between the proton TMD parton density evolved in the cold nuclear matter to that evolved in the vacuum. Left: the ratio in the impact parameter space. Right: the ratio in the transverse momentum space. }
\label{fig:TMDdensity}
\end{figure}

In the left panel of figure~\ref{fig:TMDdensity}, we compute the ratio of the proton TMD PDF evolved in cold nuclear matter of size $L=5.0$~fm to that evolved in the vacuum.
The TMD PDF has been evolved to $\mu=\sqrt{\zeta}=6.0$ GeV. The energy of the parton in the rest frame of the nuclear medium is $E=x s/(2m_N)$ with $\sqrt{s}=40$ GeV.
Thus, the semi-hard scale $\mu_E=\sqrt{E/L}$ varies from 0.6 GeV to 6 GeV for $0.01<x<1$. 
The ratio as a function of $x$ and $b$ displays quite a nontrivial structure. To understand the physics it contains, we transform $b$ back to the transverse momentum space, which is shown on the right panel of figure~\ref{fig:TMDdensity}.
At large $x$ the parton is energetic in the rest frame of the nucleus and the fractional energy loss is negligible. The ratio reflects the $p_T$ broadening effect, i.e., there is a depletion at low $p_T$ and an enhancement at high $p_T$.
At small $x$ the fractional energy loss is significant, so the ratio decreases as at small $x$. Furthermore, we have seen that the energy loss is correlated with $b$ such that parton with a higher $p_T$ loses more energy. Therefore, the $p_T$ broadening effect is eventually overcome by the energy loss effect at small $x$, which is exemplified by the ratio decreasing with $p_T$ at $x= 0.01$.

\subsection{Qualitative comparison to other approaches}
In this section, we discuss how this formula is related and compared to other approaches used in the field of jet tomography in characterizing the properties of the cold nuclear matter and the quark-gluon plasma (hot and dense matter). The transverse momentum dependence of medium correction is important for understanding observables such as angular correlations between di-jets and $\gamma/Z$/hadron-hadron/jet correlations, as well as jet substructure observables. 
The problem faced in such calculations is similar to our current work, where one aims to include momentum broadening and radiative corrections in the medium.
For example, in the limit of dense medium and large number of scatterings, it was found that multiple soft emissions that are strongly ordered in both formation time and transverse momentum renormalize the parton momentum broadening~\cite{Wu:2011kc,Liou:2013qya,Blaizot:2014bha} and lead to the so-called anomalous diffusion in the transverse direction~\cite{Caucal:2022fhc}.  An evolution equation is written down for the broadening $\langle\Delta p_T^2\rangle$ and the asymptotic solution is obtained in the double logarithmic limit.
The renormalization can be traced back to the BFKL evolution of the dipole cross section in this formalism and has been recently improved to next-to-leading-logarithm accuracy~\cite{Caucal:2022fhc}.
Combined with TMD factorization in the vacuum, such formula has been used to study the broadening of di-hadron, hadron-jet, and di-jet correlations~\cite{Chen:2016vem, Jia:2019qbl}.
Another example~\cite{Vaidya:2020lih,Vaidya:2021vxu} takes the opacity expansion approach and treats the interaction between the jet and the plasma using collinear-soft Glauber exchanges. The factorization formula is developed with both RG and RRG evolution equation established and can be used to study jet substructures in the medium.

The feature that is common among these calculation, including our work, is the importance of dressing the Glauber gluon exchange with soft emissions. Furthermore, the soft emission phase space is limited by the LPM effect (from the formulation-time considerations). In a dilute medium treated under the opacity expansion, the phase-space is limited by the comparison of emission formulation time with the medium size. In a dense medium treated using the multiple-soft approach, it is limited by the time scale on which multiple collision center contribute coherently to the soft emission.
Our work focuses on parton-medium interactions in the initial state of the hard scattering. Nevertheless, the physics of the forward scattering is similar to those happening in the final state. The problem is further set up in the collider frame where the parton and the medium have a large momentum pointing in opposite light-cone direction. The reason behind this choice is to relate the medium information to parton distribution function at small $b$. Based on previous studies, we identify a novel set of collinear evolution that encode parton energy loss and its correlation with the impact parameter. This allows a three-dimensional description of parton propagation in matter. Importantly, the framework we develop allows us to write down expressions at the cross section level and treat explicitly and self-consistently scale and rapidity evolution. As we will show below, these features enable theoretical predictions that can be compared directly to experimental measurements.

\section{Phenomenology}
\label{sec:result}

\begin{figure}
    \centering
    \includegraphics[scale=.85]{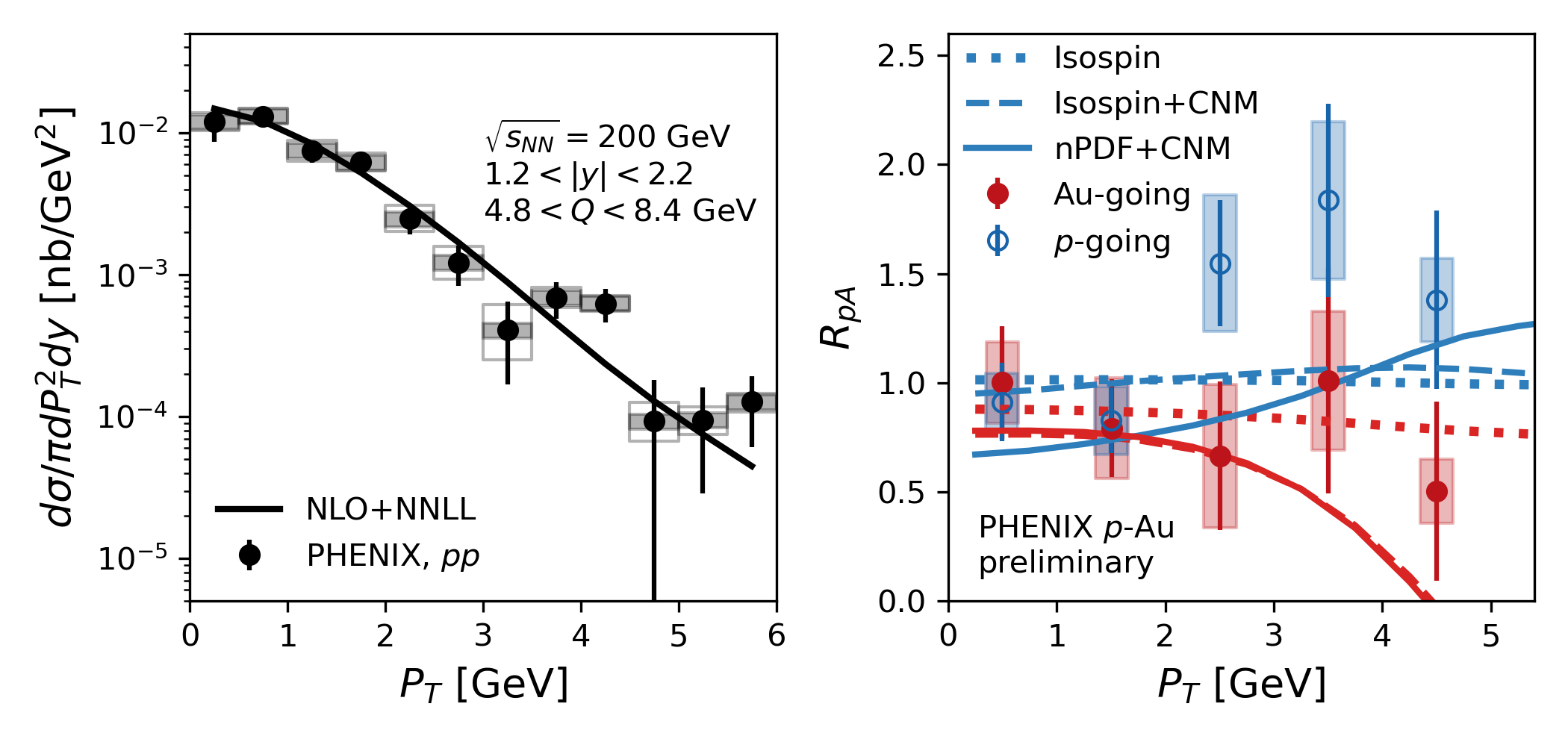}
    \caption{A comparison of the NLO+NNLL calculation and the $P_T$ spectrum as measured by the PHENIX experiments~\cite{PHENIX:2018dwt,Leung:2019wd}. $p$+$p$ collisions are shown in the left panel and $p$+Au collisions are shown in the right panel.}
    \label{fig:TMD-DY-pp-pA}
\end{figure}

\paragraph{Drell-Yan measurement from PHENIX} In figure~\ref{fig:TMD-DY-pp-pA}, we compare a calculation of the DY cross section in the TMD region to the PHENIX experimental measurement at $\sqrt{s_{NN}}$ = 200 GeV. The baseline computation, in the absence of CNM effects, is performed using a code at NLO+NNLL accuracy. For the $p$+$p$ baseline shown in the left panel of figure~\ref{fig:TMD-DY-pp-pA}, we used the CT18nlo proton PDF~\cite{Hou:2019efy} with non-perturbative functions and parameters from Ref.~\cite{Sun:2014dqm}. The calculation agrees well with the data at $P_T<3$ GeV. There are sizable deviations at $P_T>3$ GeV, which may be because we have not included the Y term for the matching to fixed-order calculations. Despite this caveat, the low $P_T$ region is adequate to test the new formalism for in-medium corrections.

In $p$+Au collisions, the nuclear modification factor is defined as the ratio of TMD cross sections between $p$+Au and $p$-$p$, normalized by the number of binary collisions, $A=197$:
\begin{align}
R_{pA} = \frac{1}{A}\frac{d\sigma_{pA}/d\mathcal{PS}}{d\sigma_{pp}/d\mathcal{PS}}\,.
\end{align}
The $p$+Au data is taken in both the Au-going side ($-2.2<y<-1.2$) and the $p$-going side ($1.2<y<2.2$). The Au-going side is sensitive to the small $x$ region of the proton PDF ($0.0027<x_1<0.013$) and the large $x$ region of the nuclear PDF ($0.08<x_2<0.38$). In the $p$-going side, the probed regions of $x_1$ and $x_2$ are flipped. The empirical nuclear PDF is modified non-trivially in different $x_2$ regions. To isolate the dynamical cold nuclear matter effects, we first construct a simple nuclear PDF including only the isospin effects:
\begin{align}
f_{i/A}^{\rm iso}(x,\mu) = \frac{Z}{A}f_{i/p}(x,\mu) + \frac{A-Z}{A}f_{i/n}(x,\mu),
\end{align}
where the neutron PDF is obtained from the free proton PDF using isospin symmetry. The calculations using only the ``isospin" PDF and without dynamical CNM effects are shown as dotted lines in the Au-going (red) and $p$-going (blue) sides. The ratio is consistent with unity on the $p$-going side and is below unity on the Au-going side with a weak $P_T$ dependence.

The calculations with the full dynamical CNM effects are shown in dashed lines, labeled ``Isospin+CNM".
The nuclear medium parameters are $\rho_G=0.4$~fm$^{-3}$ and $\xi^2=0.12$ GeV$^2$, which are found to provide a reasonable description of the nuclear modification factor of the collinear fragmentation function in $e$+$A$ semi-inclusive DIS in our previous work~\cite{Ke:2023ixa}. 
In the Au-going side, the parton is relatively less energetic ($57<E<270$ GeV) in the rest frame of the nucleus. From the discussion of figure~\ref{fig:TMDdensity}, we know that the fractional energy loss can be sizable and increase with $P_T$. This explains the further suppression of $R_{pA}$ relative to the ``Isopsin'' case and its decreasing trend with $P_T$.
The suppression is so large that the cross-section becomes negative for $P_T>4.5$ GeV, nevertheless, this is already outside of the domain of the TMD factorization.
On the $p$-going side, the fractional parton energy loss is negligible. The ratio slightly increases with $P_T$, which is due to momentum broadening.

Finally, we uses the empertical collinear nuclear PDF (nPDF) provided by the EPPS21 parametrization~\cite{Eskola:2021nhw}. Calculations are shown in solid lines, labeled ``nPDF+CNM'.
Compared to the ``Isospin+CNM' calculation, the difference in the Au-going side is tiny. However, in the $p$-going side, which corresponds to the small-$x$ region of the nuclear PDF, the use of EPPS21 nPDF further increases the $P_T$ slope of $R_{pA}$ and improves the comparison to data in terms of shape, though we recognize that the error bars of the measurements are quite large to make a stronger statement. The reason that the nPDF makes such a huge difference in the $p$-going region is traced back to the slightly different $\mu=\mu_b$ dependence of the EPPS21 nPDF and the isopsin-averaged PDF.

\begin{figure}[hb!]
\centering
\includegraphics[scale=0.84]{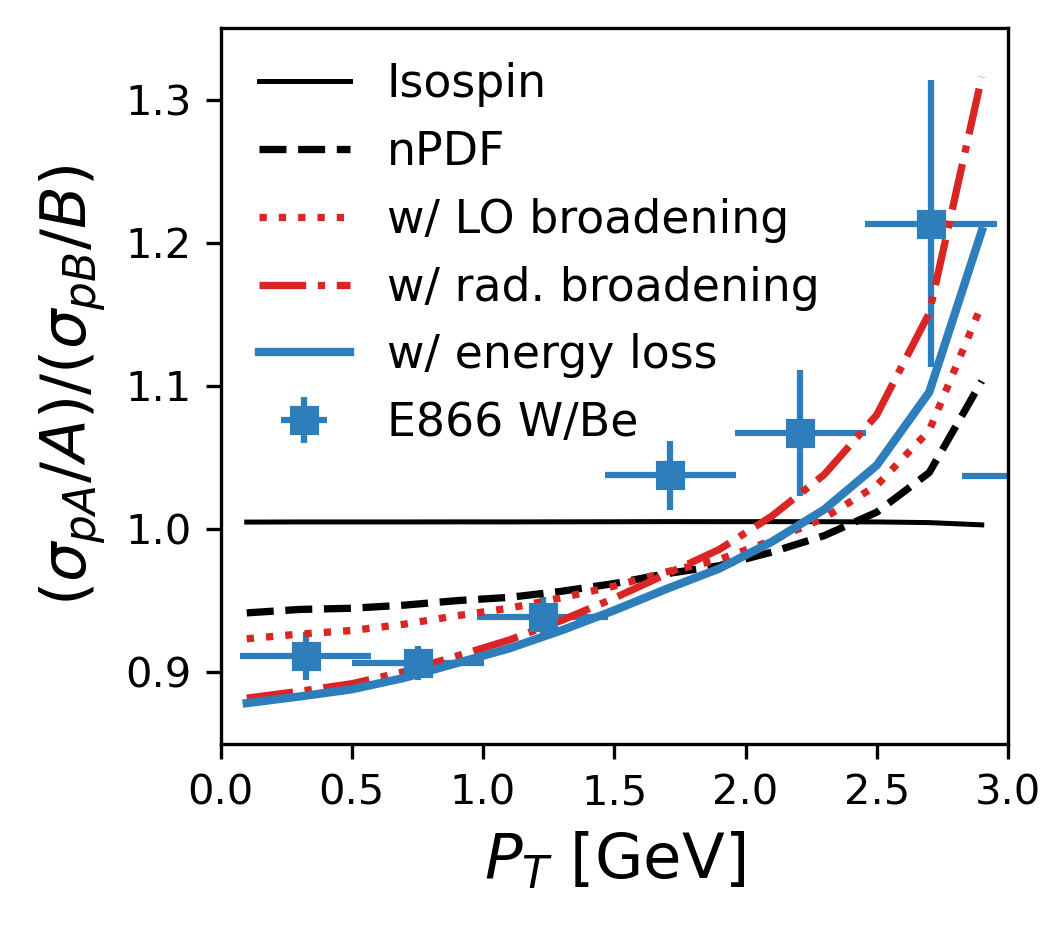}
\includegraphics[scale=0.84]{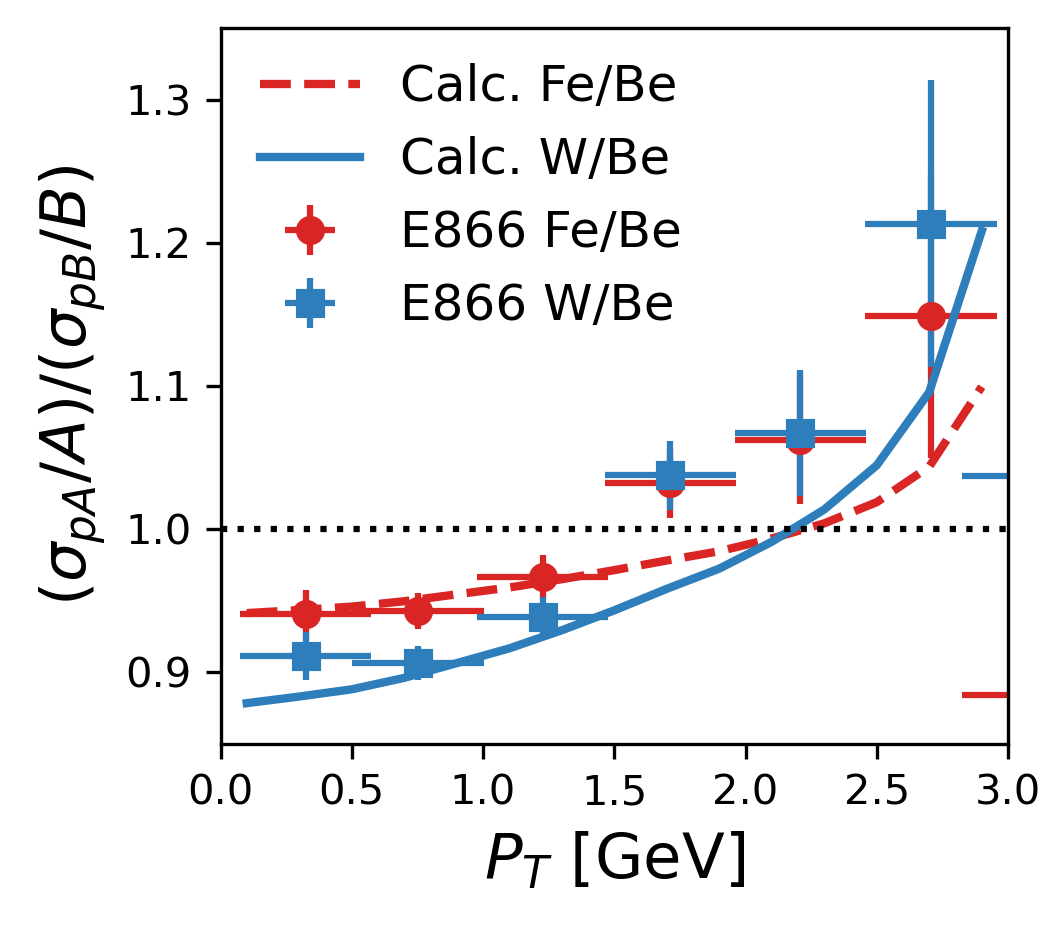}
\caption{Left: progresssively including different nuclear effects in the calculation of the nuclear modification factor of W relative to Be. Right: the nuclear modification factors of Fe and W relative to Be. Calculations are compared to data from the E866 experiment~\cite{NuSea:1999egr}.
}
\label{fig:SIDIS-DY}
\end{figure}

\paragraph{Fixed target measurement from E866} In search of more precise $p$+$A$ data in the TMD region, we finally compare our calculation to the E866 experimental measurements with Be, Fe, and W ($A\; =\; $9, 56, and 184) targets~\cite{NuSea:1999egr}.
The proton beam has an energy of $E=800$ GeV. The measurement uses dilepton paris within the region $4.0<Q<8.4$ GeV, $0.21<x_1<0.95$, $0.01<x_2<0.12$, and $0.13<x_F<0.93$ with $x_F=x_1-x_2$. These kinematic cuts are implemented in the calculations.

In the left panel of figure~\ref{fig:SIDIS-DY}, we take the ratio of W/Be and examine the role of different nuclear modifications in our calculation.
The thin black solid line is obtained using the vacuum TMD factorization formula with the isospin construction of the nuclear PDF, and it is practically consistent with unity.
Including the EPPS21 nuclear PDF, the back dashed line increases with $P_T$, but still underestimates the slope in the data.
With leading order collisional broadening (the red dotted line), we observe only a marginal improvement.
The red dashed line includes the BFKL evolution of the forward cross-section that resums radiative broadening. It significantly increases the $P_T$ slope, bringing it much closer to the data.
Finally, the inclusion of radiative energy loss (the blue splid line) slightly reduces the ratio at higher $P_T$.
Because the minimum mass of the dilepton pair is only 4.0 GeV, we consider a proper TMD limit to be $P_T$ less than 1.5 GeV. In this region, the calculation nicely describes the ratio for W/Be. For larger $P_T$, even though the TMD factorization starts to break down, but the calculation still captures the overall increasing trend until $P_T=3$ GeV.
In the right panel of figure~\ref{fig:SIDIS-DY}, we compare the full calculation to the nuclear modification factors of W and Fe relative to Be, and there is an overall agreement with the data. It means that the calculation captures the correction mass number dependence of the modifications. 

These phenomenological applications demonstrate that it is possible to use the same set of cold nuclear matter parameters ($\rho_G$, $\xi^2$) to largely explain both the nuclear modification to collinear hadron production in $e$+$A$  SIDIS~\cite{Ke:2023ixa} and the TMD Drell-Yan production in $p$+$A$ shown in this paper. Such CNM corrections can be included in the future global determination of the intrinsic non-perturbative nuclear TMD PDFs and TMD fragmentation functions.

\section{Summary}\label{sec:summary}
In this paper, we have considered the nuclear medium-induced corrections to transverse momentum dependent Drell-Yan pair production in $p$+$A$ at first order in opacity and up to NLO. Physically, this includes the momentum broadening and energy loss effects of the proton-collinear parton, and the calculation accounts for the leading correlation between the hard production vertex and the scattering with medium partons. We find that up to this order we need to consider the collinear sector, collinear-soft sector, anti-collinear sector, and soft radiation from the Glauber exchange. Even though SCET$_{\rm G}$, where the medium is treated as a background field, is sufficient to handle the collinear and collinear-soft sectors, one needs to go beyond this approach and include quantum fluctuations in the anti-collinear and soft sectors, which are essential to understanding the TMD observable. Such a set of ingredients has already been pointed out in~\cite{Rothstein:2016bsq,Vaidya:2021vxu}, where a forward scattering EFT was built. In the cases we consider, the collinear parton enters a hard process, which leads to LPM interference effects at finite opacity.

The calculation of each sector exhibits both collinear and rapidity divergences. The collinear divergences cancel among the collinear and collinear-soft sectors, leading to RG evolution equations that resum leading terms from higher-order opacity and encode the parton energy loss in cold nuclear matter, similar to those derived in semi-inclusive DIS in~\cite{Ke:2023ixa} but with transverse momentum dependence. The rapidity divergences, on the other hand, cancel among the collinear, soft, and collinear-soft sectors. This is a property guaranteed by RG consistency~\cite{Rothstein:2016bsq,Vaidya:2021vxu}, but we demonstrate it explicitly in this paper in the presence of the LPM effect. The RG evolution equation has a BFKL form and resum the effect of fast, soft radiation into the forward scattering cross section. The calculation exponentiates the multiple forward scatterings in the medium that lie on the jet's path.

The theoretical formalism developed here was applied to QCD phenomenology. We find that with a small set of cold nuclear matter parameters deduced from earlier studies of the SIDIS process in $e$+$A$, the TMD Drell-Yan pair production in $p$+$A$ can be successfully understood. This paves the way for future work on a systematic extraction of the cold nuclear matter parameters using both SIDIS and Drell-Yan data. Furthermore, in addition to structure modifications that might be parameterized via traditional (TMD) nuclear PDFs, such dynamically cold nuclear matter effects modify the rapidity and transverse momentum dependence of the Drell-Yan pair production in ways that exhibit different behavior as functions of kinematics and nuclear size. Such differences merit further investigation, and in the long term, it will be very useful to separate them from non-perturbative parameterizations in future global analyses.

\section*{Acknowledgements}
The authors thank the Institute for Nuclear Theory at the University of Washington for their hospitality. The work of J.T. and I.V. is supported by the US Department of Energy through the Los Alamos National Laboratory. Los Alamos National Laboratory is operated by Triad National Security, LLC, for the National Nuclear Security Administration of the U.S. Department of Energy (Contract No. 89233218CNA000001). This research is funded by LANL’s Laboratory Directed Research and Development (LDRD) program under project numbers 20220715PRD1 and 20230787ER. The work of W.K. is supported by the LDRD program at LANL.

\bibliographystyle{JHEP}
\bibliography{ref}

\newpage
\appendix
\section{Perturbative expressions in elementary collisions}\label{app:vac-matching}

\paragraph{Hard matching function} The explicit expression for the Drell-Yan hard matching coefficient is given by
\begin{align}
H(Q, \mu) = 1 + \frac{\alpha_s C_F}{2\pi}\left(3L_Q^2-L_Q-8+\frac{7\pi^2}{6}\right) + \mathcal{O}\left(\alpha_s^2\right)\, . \label{eq:hard-coeff}
\end{align}
with the logarithm being $L_Q=\ln(Q^2/\mu^2)$.

\paragraph{Collinear matching function} The collinear matching coefficients are presented as a perturbative expansion in $\alpha_s$
\begin{align}
     C_{i\leftarrow j}(x,\mu_i,\mu,\zeta) = \sum_n \left(\frac{\alpha_s}{4\pi}\right)^n C_{i\leftarrow j}^{(n)}(x,\mu_i,\mu,\zeta)\, , \label{eq:coll-coeff-1}
\end{align}
At leading order, the non-zero coefficient that is relevant for TMD Drell-Yan process is
\begin{align}
    C_{q\leftarrow q}^{(0)}(x,\mu_i,\mu,\zeta) & = \delta\left(1-x\right).  \label{eq:coll-coeff-2}
\end{align}
The one-loop corrections are~\cite{Echevarria:2016scs}
\begin{align}
    C_{q\leftarrow q}^{(1)}(x,\mu_i,\mu,\zeta) & = C_F\left[-2 L_\mu p_{qq}(x)+2(1-x)+\delta\left(1-x\right)\left(-L_\mu^2+2L_\mu L_\zeta-\frac{\pi^2}{6}\right)\right]\,,  \label{eq:coll-coeff-3}\\
    C_{q\leftarrow g}^{(1)}(x,\mu_i,\mu,\zeta) & = T_R\left[-2L_\mu p_{gq}(x)+4 x(1-x)\right]\,,   \label{eq:coll-coeff-4}
\end{align}
with $L_\mu = \log\left(\frac{\mu_i^2}{\mu_b^2}\right)$
and $L_\zeta = \log\left(\frac{\mu_i^2}{\zeta}\right)$.
The collinear splitting functions in the vacuum ($P_{ji}(x)$) and the corresponding real-emission contributions ($p_{ji}(x)$) are
\begin{align}
    P_{qq}(x) &= \frac{1+x^2}{(1-x)_+}+\frac{3}{2}\delta\left(1-x\right) \equiv p_{qq}(x)  +\frac{3}{2}\delta\left(1-x\right) \,, \\
    P_{qg}(x) &= 1-2x(1-x) \equiv p_{qg}(x)  \,.
\end{align}

\paragraph{TMD anomalous dimensions} The anomalous dimensions for the hard, beam, and soft function are given by
\begin{align}
\gamma_\mu^H\left(\mu, \frac{Q}{\mu}\right) &= 2C_F \gamma^{\rm cusp}(\alpha_s(\mu)) \ln \frac{Q^2}{\mu^2} + 4\gamma_q(\alpha_s(\mu)), \\
\gamma_\mu^B\left(\mu, \frac{\zeta}{\nu^2}\right) &= -C_F \gamma^{\rm cusp}(\alpha_s(\mu)) \ln\frac{\zeta}{\nu^2} - 2\gamma_q(\alpha_s(\mu)), \\
\gamma_\mu^S\left(\mu, \frac{\mu}{\nu}\right) &= 2C_F \gamma^{\rm cusp}(\alpha_s(\mu)) \ln\frac{\mu^2}{\nu^2}, \\
\gamma_\nu^B(\mu,b) = - \frac{1}{2}\gamma_\nu^S(\mu,b) &= -2C_F\int_{\mu_b}^\mu \frac{d\mu'}{\mu'}\gamma^{\rm cusp}(\alpha_s(\mu'))  - C_F\gamma^r(\alpha_s(\mu_b))\, .
\end{align}
The cusp and non-cusp dimensions can be expanded in perturbative series
\begin{align}
\gamma^{\rm cusp}(\alpha_s) &= \sum_n \left(\frac{\alpha_s}{4\pi}\right)^{n+1} \gamma^{\rm cusp}_n, \\
\gamma^{q}(\alpha_s) &= \sum_n \left(\frac{\alpha_s}{4\pi}\right)^{n+1} \gamma^{q}_n, \\
\gamma^{r}(\alpha_s) &= \sum_n \left(\frac{\alpha_s}{4\pi}\right)^{n+1} \gamma^{r}_n\, .
\end{align}
The perturbative expansion of the QCD $\beta$ function is
\begin{align}
\beta(\alpha_s) = -2\alpha_s\sum_n\left(\frac{\alpha_s}{4\pi}\right)^n \beta_n.
\end{align}
According to Ref.~\cite{Duhr:2022yyp}, to achieve the desired order of accuracy NLO+NNLL, we use the one-loop expression for the hard and collinear matching coefficients given in Eqs.~(\ref{eq:hard-coeff}) and (\ref{eq:coll-coeff-1}) to (\ref{eq:coll-coeff-4}). The $\beta$ function and $\gamma^{\rm cusp}$ are kept to the three-loop order while the non-cusp dimensions are kept to two-loop order~\cite{Korchemsky:1987wg,Moch:2004pa,Moch:2005tm, Moch:2005id, Idilbi:2005ni,Idilbi:2006dg,Becher:2006mr}. Here we quote these coefficients for completeness. The $\beta$-function coefficients are
\begin{align}
\beta_0 &= \frac{11}{3}C_A - \frac{4}{3}T_R N_f\, ,\\
\beta_1 &= \frac{34}{3}C_A^2 - \frac{20}{3}C_A T_R N_f - 4C_F T_R N_f\, ,\\
\beta_2 &= \frac{2857}{54}C_A^3 + \left(2C_F^2 - \frac{205}{9}C_F C_A - \frac{1415}{27}C_A^2\right)T_R N_f \nonumber\\
      &+ \left(\frac{44}{9}C_F + \frac{158}{27}CA\right)T_R^2 N_f^2 \, .
\end{align}
The cusp coefficients to three loops are
\begin{align}
\gamma^{\rm cusp}_0 &= 4\, ,\\
\gamma^{\rm cusp}_1 &= C_A \left[\frac{268}{9} -8\zeta(2) -\frac{40}{9}N_f\right]\, \\
\gamma^{\rm cusp}_2 &= 
C_A^2\left[-\frac{1072\zeta(2)}{9} + \frac{88\zeta(3)}{3} + 88\zeta(4) + \frac{490}{3}\right]\, , \nonumber\\
&+ C_A N_f\left[\frac{160\zeta(2)}{9} - \frac{112\zeta(3)}{3} - \frac{836}{27}\right] \nonumber\\
&+ C_F N_f \left[32\zeta(3)-\frac{110}{3}\right] - \frac{16}{27}N_f^2\, .
\end{align}
Finally, the non-cusp coefficients to two loops are
\begin{align}
\gamma_0^q &= - 3 C_F\, , \\
\gamma_1^q &= C_A C_F\left[-11\zeta(2)+26\zeta(3)-\frac{961}{54}\right]\, ,\nonumber\\
        & + C_F^2\left[12\zeta(2)-24\zeta(3)-\frac{3}{2}\right]
         + C_F N_f \left[2\zeta(2)+\frac{65}{27}\right] \, ,\\
\gamma_0^r &= 0\, ,\\
\gamma_1^r &= C_A\left[\frac{22}{3}\zeta(2)+28\zeta(3)-\frac{808}{27}\right] + N_f\left[\frac{112}{27} - \frac{4\zeta(2)}{3}\right] - 2\zeta(2)\beta_0\, .
\end{align}

\section{SCET$_{\rm G}$ Feynman rules and elementary splitting amplitude}
\label{sec:app:SCETG_feynman_rule}

\begin{figure}[ht!]
\centering
\includegraphics[scale=1]{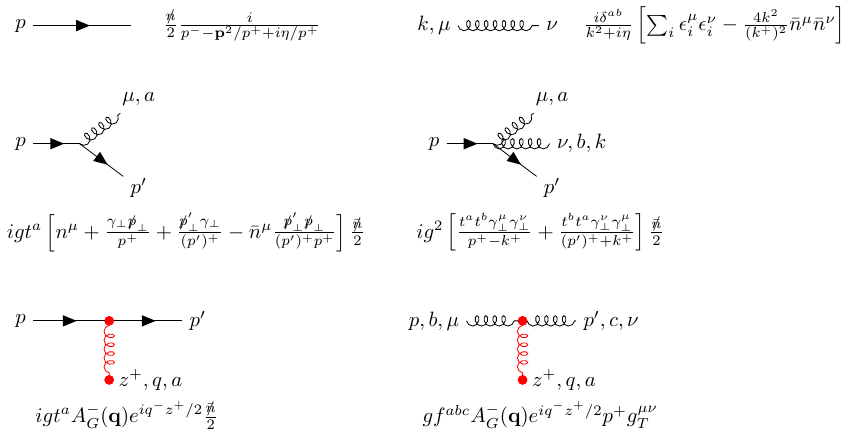}
\caption{Feynman rules for SCET$_{\rm G}$ in the hybrid gauge.}
\label{fig:RR123}
\end{figure}

The Feynman rules for SCET$_{\rm G}$ were derived in ~\cite{Ovanesyan:2011xy} and used to calculate the medium-induced splitting kernels  for massless partons~\cite{Ovanesyan:2011kn,Fickinger:2013xwa,Ovanesyan:2015dop} and  heavy quarks~\cite{Kang:2016ofv}.
We follow the hydrid gauge used in~\cite{Ovanesyan:2011kn}, where the collinear sector uses the light-cone gauge $\nbar \cdot A=0$, while the Glauber gluon and the soft sector uses the covariant gauge. The relevant rules in the hybrid gauge are given in figure~\ref{fig:RR123}, where we use the convention that $p$ and $p'$ denote quark momenta while $k$ denote gluon momenta. The two basis polarization vectors for the collinear gluon are 
\begin{align}
\e_\lambda^\mu(k) = \left[0, \frac{2\bfe_\lambda\cdot \bfk}{k^+}, \bfe_\lambda\right],
\end{align}
with $\bfe_1=\hat{x}$ and $\bfe_2=\hat{y}$. Since for our choice the basis vectors are real, we will not distinguish $e_\lambda^\mu$ and $(e_\lambda^\mu)^*$ hereafter.

In the $q(p+k, E^+)\rightarrow q(p,xE^+)+g(k,(1-x)E^+)$ vertex, so long as the gluon line is on shell or it ends in a Glauber interaction vertex, we can always contract it with a polarization vector and obtain
\begin{align}
 & igt^a\left[n\cdot \epsilon_\lambda+\frac{\slashed{\e}_{\lambda,\perp} (\slp_\perp+\slk_\perp)}{p^++k^+}+\frac{\slp_\perp\slashed{\e}_{\lambda,\perp}}{p^+}\right]\frac{\slnbar}{2} \nonumber\\
 & =\frac{igt^a}{x(1-x)E^+}\left[(1+x)\bfe_\lambda\cdot\bfQ \Gamma_S-i(1-x)\Gamma_T^{ij} \bfe_\lambda^i\bfQ^j\right]\frac{\slnbar}{2} \nonumber \\
  & \equiv \frac{igt^a}{x(1-x)E^+} P_\lambda(\bfQ) \frac{\slnbar}{2}  ,
  \label{eq:splitting}.
\end{align}
where $\bfQ = x\bfk-(1-x)\bfp$, $\Gamma_S = \I_{4\times 4}$ and $\Gamma_T^{ij} = \frac{i}{2}[\gamma^i, \gamma^j]$.
$P_\lambda(\bfQ)=\e_{\lambda,\alpha}(k) P^\alpha(\bfQ)$ is a short-hand notation with
\begin{align}
P^\alpha(\bfQ) = -(1+x)Q_\perp^\alpha \Gamma_S-i(1-x)\Gamma_T^{\alpha \beta} Q_{\perp,\beta} \,.
\end{align}
If we reverse the momentum flow and let $k$ and $p-k$ merge into a final-state quark of momentum $p$, $P(\bfQ)$ is replaced by $P(\bfQ)^*$ in Eq. (\ref{eq:splitting}).

In appendix~\ref{sec:app:collinear_sector} we will encounter the interferences between different elementary splitting amplitudes. Using the facts that 1) $\sln$ commutes with $\Gamma_S$ and $\Gamma_T^{ij}, i,j \in \{1,2\}$, 2) $(\Gamma_T^{ij})^2=\I$, and 3) $\Tr\left\{\Gamma_T^{ij}\frac{\sln}{2}\bar{J}J\right\}=0$, a generic interference term goes like
\begin{align}
(\chibar_{n,p} P_\lambda(\bfQ) J)^\dagger(\chibar_{n,p} P_\lambda(\bfQ') J) 
= p^+&\left[(1+x)^2(\bfe_\lambda\cdot\bfQ)(\bfe_\lambda\cdot\bfQ')  \right.\nonumber \\ 
& \left. + (1-x)^2(\bfe_\lambda\times\bfQ)\cdot(\bfe_\lambda\times\bfQ')  \right]\Tr\left\{ \frac{\sln}{2} \bar{J}J\right\}  .
\end{align}
If one sum over the two polarizations $\bfe_{1,2}$, then
\begin{align}
\sum_{\lambda} (\chibar_{n,p} P(\bfQ) J)^\dagger(\chibar_{n,p} P(\bfQ') J) =  2p^+\left[(1+x^2)-\epsilon(1-x)^2\right]\bfQ\cdot \bfQ'\Tr\left\{ \frac{\sln}{2} \bar{J}J\right\}. 
\end{align}
Note that in $d=4-2\epsilon$ dimensions the transverse vectors are in a $2-2\epsilon$ dimensional subspace, so $\sum_\lambda(\bfe_\lambda\cdot \bfQ)(\bfe_\lambda\cdot \bfQ') = \bfQ\cdot \bfQ'$ and $\sum_\lambda(\bfe_\lambda\times \bfQ)(\bfe_\lambda \times \bfQ') = (1-2\epsilon)\bfQ\cdot \bfQ'$. We can understand the  latter relation as follows: first choose one of the basis vectors $\bfe_1$ to be parallel to $\bfQ$, 
then  decompose $\bfQ'$ into a component parallel to $\bfQ$ and an orthogonal component. The orthogonal component gives zero, i.e. $\sum_\lambda(\bfe_\lambda\times \bfQ)(\bfe_\lambda \times \bfQ_\perp') = 0$.
Because the dimension of the subspace that is orthogonal to $\bfQ$ is $(1-2\epsilon)$, the parallel component gives $\sum_\lambda(\bfe_\lambda\cdot\bfQ)(\bfe_\lambda\cdot\bfQ_{\|}') = (1-2\epsilon)\bfQ\cdot \bfQ'$. 

\section{Medium ensemble average and the contact limit}
\label{sec:app:ensemble-avg}
In the covariant gauge, the background vector potential mediated by Glauber gluons is generated by the color current $J^\mu$ collinear to the motion of the medium
\begin{align}
  & A_G^{a,-}(q^+, q^-, \bfq) =  ig_s \int \frac{dx^+ dx^-}{2} d^2\bfx \frac{i  e^{iq^+x^-/2}e^{iq^- x^+/2} e^{-i\bfq\cdot\bfx}}{\bfq^2+\xi^2} \langle M_F|j_{\nbar}^{a\,-}(x)|M_I\rangle.   \label{eq:Glauber-1}
\end{align}
Only the ``minus'' component of $A_G$ is retained from power counting. When Glauber gluons are coupled to the collinear sector, $q^+\ll p^+$ will be neglected everywhere except for the phase factor in Eq.~(\ref{eq:Glauber-1}). The integration over $q^+$ can then be carried out
\begin{align}
A_G^{-,a}(q^-, \bfq) &=  ig_s\int \frac{dq^+}{2\pi} A_G^{-,a}(q) \nonumber\\
&= ig_s \int dx^+ d^2\bfx \frac{i e^{iq^- x^+/2} e^{-i\bfq\cdot\bfx}}{\bfq^2+\xi^2} \langle M_F|j_{\nbar}^{a\,-}(x^-=0, x^+, \bfx)|M_I\rangle.
\end{align}

Going back to Eq.~(\ref{eq:factorize-medium-correlation}), we now show the procedure to sum over the medium ensemble and how to obtain the contact limit for double Glauber interactions. Transforming from coordinate space to  momentum space ($x\rightarrow q$, $y\rightarrow k$) and performing integration over the plus component of the Glauber momentum
\begin{align}
\chi\mathcal{B}_{q/q,1} =& \sumint_J e^{-i\bfp_J\cdot\bfb} \delta(\nbar\cdot p_J-(1-z)\nbar\cdot p_q) \int \frac{d k^- d^2\bfk}{2(2\pi)^3} \int \frac{dq^- d^2\bfq}{2(2\pi)^3} \nonumber\\
& \times D^{ab}(-k^-, -\bfk, -q^-, -\bfq) W^{ab}(k^-, \bfk, q^-, \bfq) + \cdots\, .
\end{align}
The Fourier-transformed medium correlator is
\begin{align}
W^{ab}(k^-, \bfk, q^-, \bfq)
=& g_s^2  \int dx^+ d^2\bfx \int dy^+ d^2\bfy  \frac{i e^{ik^- x^+/2-iq^- y^+/2} e^{-i\bfk\cdot\bfx+i\bfq\cdot\bfy}}{(\bfk^2+\xi^2)(\bfq^2+\xi^2)}
\nonumber \\
& \times \left\{\mathrm{Tr}\left\{j_{\nbar}^{a\,-}(x^+, 0, \bfx) j_{\nbar}^{b\, -}(y^+, 0, \bfy) \hat{\rho}_i\right\}\right\} \, , 
\end{align}
where we have applied the completeness relations and denote the initial-state medium density operator by $\hat{\rho}_i$.
$\hat{\rho}_i$ could be a complicated many-body density matrix that describes the distribution of nucleons within the nucleus and the distribution of color sources within the nucleons.
However, due to confinement, the color correlation can only exist within a single nucleon, i.e. the
range of color correlation is smaller than the intra-nucleon distances.
Thus, one can reduce the correlation function to the product of the single nucleon distribution function and the correlation within one nucleon:
\begin{align}
& \mathrm{Tr}\left\{j_{\nbar}^{a\,-}(x^+, \bfx) j_{\nbar}^{b\, -}(y^+, \bfy)  \rho_i \right\}\nonumber\\
&\qquad = \delta^{ab} \int ds^+ d\bfs \rho^-(s^+, \bfs) \langle N(s^+, \bfs)|j_{\nbar}^{a\,-}(x^+, \bfx) j_{\nbar}^{b\,-}(y^+, \bfy) |N(s^+, \bfs)\rangle,
\end{align}
where the nucleon states are normalized by $\langle N|N \rangle = 1$. The label $s^+, \bfs$ means the nucleon state is translated to this location and contains a translation phase factor.
Because the initial and final state are the same, we must have $a=b$.
With a transformation of the integration variables
\begin{align}
&X^+ = \frac{x^++y^+}{2}, ~ \delta x^+ = x^+-y^+, ~ \bfX = \frac{\bfx+\bfy}{2}, ~ \delta \bfx = \bfx - \bfy,
\end{align}
and  a shift of the nucleon state to $(0^+, \mathbf{0})$, the correlator becomes
\begin{align}
W^{ab}(k^-, \bfk, q^-, \bfq)
=&  \delta^{ab} g_s^2 \int ds^+ d^2\bfs \rho^-(s^+, \bfs)  \int dX^+ d^2\bfX e^{i(k^--q^-)X^+/2} e^{-i(\bfk-\bfq)\cdot\bfX}  \nonumber\\
& \times \int d\delta x^+ d^2\delta \bfx  e^{i\frac{k^-+q^-}{2}\cdot \delta x^+/2}e^{-i\frac{\bfk+\bfq}{2}\cdot \delta \bfx}\frac{1}{(\bfk^2+\xi^2)(\bfq^2+\xi^2)}\nonumber\\
& \times \langle N(0)|j_{\nbar}^{a\,-}\left(X^+-s^++\frac{\delta x^+}{2}, \bfX-\bfs+\frac{\delta\bfx}{2}\right)  \nonumber\\
& \hspace*{1.2cm}\times  j_{\nbar}^{a\,-}\left(X^+-s^+-\frac{\delta x^+}{2}, \bfX-\bfs-\frac{\delta\bfx}{2}\right) |N(0)\rangle.
\end{align}
Because the nucleon density distribution $\rho(s^+, \bfs)$ is a slowly varying function compared to the correlation range within a single nucleon, and the later is sharply peaked around the source location, we expand 
\begin{align}
&\langle N(0)|j_{\nbar}^{a\,-}\left(X^+-s^++\frac{\delta x^+}{2}, \bfX-\bfs+\frac{\delta\bfx}{2}\right) j_{\nbar}^{a\,-}\left(X^+-s^+-\frac{\delta x^+}{2}, \bfX-\bfs-\frac{\delta\bfx}{2}\right) |N(0)\rangle  \nonumber\\
&\approx \langle N(0)|j_{\nbar}^{a\,-}\left(\frac{\delta x^+}{2}, \frac{\delta\bfx}{2}\right) j_{\nbar}^{a\,-}\left(-\frac{\delta x^+}{2}, -\frac{\delta\bfx}{2}\right) |N(0)\rangle \delta(X^+-s^+)\delta^{(2)}(\bfX-\bfs).
\end{align}
Therefore,
\begin{align}
W^{ab}(k^-, \bfk, q^-, \bfq)
=& \delta^{ab} g_s^2 \int ds^+ d^2\bfs \rho^-(s^+, \bfs)  e^{i(k^--q^-)s^+/2} e^{-i(\bfk-\bfq)\cdot\bfs}  \nonumber\\
& \times \int d\delta x^+ d^2\delta \bfx  e^{i\frac{k^-+q^-}{2}\cdot \delta x^+/2}e^{-i\frac{\bfk+\bfq}{2}\cdot \delta \bfx}\frac{1}{(\bfk^2+\xi^2)(\bfq^2+\xi^2)}\nonumber\\
& \ \times \langle N(0)|j_{\nbar}^{a\,-}\left(\frac{\delta x^+}{2}, \frac{\delta\bfx}{2}\right) j_{\nbar}^{a\,-}\left(-\frac{\delta x^+}{2}, -\frac{\delta\bfx}{2}\right) |N(0)\rangle .
\end{align}
Finally, from the power counting of $\bfq,\bfk$ and $q^-,k^-$, $(k^--q^-)s^+/2$ is an order one quantity; $(q^- - k^-)\delta x^+$ is much smaller than one, while $(\bfq-\bfk)\cdot \bfs$ is fast oscillating. We can take the $\bfs$ integration while neglecting the change in the nucleon density around the impact parameter $\bfb$ such that $\rho(s^+,\bfs)=\rho(s^+,\bfb)+(\bfs-\bfb)\cdot\nabla_\perp \rho(X^+, \bfb)+\cdots$. To the leading order in the gradient expansion,
\begin{align}
W^{ab}(k^-, \bfk, q^-, \bfq)
=& \delta^{ab} (2\pi)^2 \delta^{(2)}(\bfk-\bfq) \frac{g_s^2 \rho_0^- L^+ }{(\bfq^2+\xi^2)^2} \nonumber\\
&\sum_T C_T n_T(\bfq) \int \frac{ds^+}{L^+}\frac{\rho_N^-(s^+, \bfb)}{\rho_0^-}  e^{i(k^--q^-)s^+/2} , 
\end{align}
where $n_T(\bfq)$ is the probability density to find a color source of representation $T$.
\begin{align}
\sum_T C_T n_T(\bfq) = \int d\delta x^+ d^2\delta \bfx  e^{-i\bfq\cdot \delta \bfx} \langle N|j_{\nbar}^{a\,-}\left(\frac{\delta x^+}{2}, \frac{\delta\bfx}{2}\right) j_{\nbar}^{a\,-}\left(-\frac{\delta x^+}{2}, -\frac{\delta\bfx}{2}\right) |N\rangle .
\end{align}
At sufficiently large $\bfq$ and consider the medium consists of weakly coupled partons, we can match on the medium parton distributions 
\begin{align}
n_T(\bfq) = \mathcal{N}_{j,T}(\bfq) \int dx f_{j/N}(x_t)\, .
\end{align}

In summary,  the correlator of the two fields in the limits that we consider becomes a contact correlation in the $x^+$ direction. Furthermore, it scales as the opacity parameter after integrating over $\bfq$ and $\bfk$,
\begin{align}
\int \frac{d^2\bfk}{(2\pi)^2} \int \frac{d^2\bfq}{(2\pi)^2} 
W^{ab}(k^-, \bfk, q^-, \bfq) \propto \delta^{ab} \rho_0^- \sigma L^+ \propto \delta^{ab}\chi .
\end{align}

\section{Collinear matching coefficients up to NLO at first order in opacity }
\label{sec:app:collinear_sector}
\begin{figure}[ht!]
\centering
\includegraphics[width=.35\textwidth]{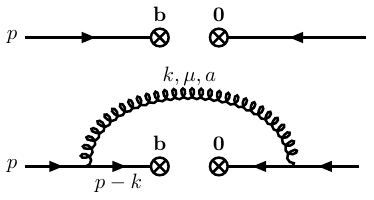}\hskip2em\includegraphics[width=.4\textwidth]{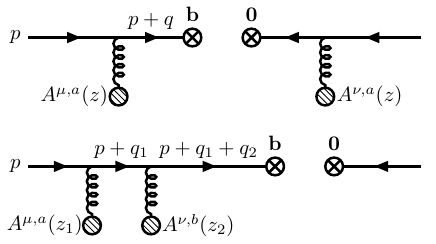}
\caption{Left: diagrams for the LO and NLO contributions at opacity zero. Right: the LO contribution at first order in opacity.}
\label{fig:vac-LO-NLO-and-opacity-one-LO}
\end{figure}

The expressions for the matching coefficients for the simplest contributions to the cross section can be obtained from the diagrams shown in figure~\ref{fig:vac-LO-NLO-and-opacity-one-LO}. 
The fully regulated vacuum matching coefficients for quarks are already in Eqs.~(\ref{eq:coll-coeff-2}) and~(\ref{eq:coll-coeff-3}). For the discussion in this section, we rewrite them in the following forms and keep a finite non-perturbative transverse momentum $\bfp$ of the initial parton 
\begin{align}
C_{q\leftarrow q}^{(0)}(x,b) &= \delta(1-x)e^{-i\bfb\cdot \bfp}\, ,\\
C_{q\leftarrow q}^{(1)}(x,b)  &= \frac{\alpha_s^{(0)}C_F}{2\pi^2}P_{qq}(x) \int \frac{d^{2-2\epsilon}\bfk}{(2\pi)^{-2\epsilon}}\frac{e^{-i\bfb\cdot (\bfp-\bfk)}}{\bfk^2}\, .
\end{align}
The LO expression of beam function in the first order in opacity is given by
\begin{align}
 &\sum_T \rho_T L^+ \mathcal{J}_F^{(0)}\otimes \Sigma_{FT}^{(0)} \otimes \mathcal{N}_{T}^{(0)} \nonumber\\
 &= \delta(1-x)\frac{\alpha_s^{(0)}C_F}{\pi} \int_{-\infty}^0\rho_G(z^+) dz^+ \int \frac{d^{2-2\epsilon}\bfq}{(2\pi)^{-2\epsilon}}\frac{1}{(\bfq^2+\xi^2)^2} \left(e^{-i\bfb\cdot \bfq} - 1\right)e^{-i\bfb\cdot \bfp}.
\end{align}

\section{NLO collinear matching coefficient at the first order in opacity}
To streamline the calculations, it is useful to introduce more compact notation. We defined kinematic variables $\bfQ_i$ in the transverse direction as follows:
\begin{align}
\bfQ_1 &= x\bfk-(1-x)(\bfp-\bfk), \\
\bfQ_2 &= x\bfk-(1-x)(\bfp-\bfk+\bfq), \\
\bfQ_3 &= x(\bfk-\bfq)-(1-x)(\bfp-\bfk+\bfq), \\
\bfQ_4 &= x(\bfk+\bfq)-(1-x)(\bfp-\bfk),\\
\bfQ_3 &= x(\bfk-\bfq)-(1-x)(\bfp+\bfq-\bfk).
\end{align}
These are linear combinations of the transverse momentum of the radiated parton $\bfk$ with momentum fraction $1-x$, the incoming parton $\bfp$, and the Glauber gluon $\bfq$. The Landau-Pomeranchuk-Migdal (LPM) frequencies, the virtuality of the branching processes, at the amplitude level are defined as 
\begin{align}
\omega_1 &= \frac{\bfk^2}{k^+} + \frac{(\bfp-\bfk+\bfq)^2}{p^+-k^+} - \frac{\bfp^2}{p^+},\\
\omega_2 &= \frac{(\bfp+\bfq)^2}{p^+} - \frac{\bfp^2}{p^+},\\
\omega_3 &= \frac{\bfk^2}{k^+} - \frac{(\bfk-\bfq)^2}{k^+},\\
\omega_4 &= \frac{\bfk^2}{k^+} + \frac{(\bfp-\bfk)^2}{p^+-k^+} - \frac{\bfp^2}{p^+},\\
\omega_5 &= \frac{(\bfk+\bfq)^2}{k^+} + \frac{(\bfp-\bfk)^2}{p^+-k^+} - \frac{\bfp^2}{p^+},\\
\omega_6 &= \frac{(\bfk-\bfq)^2}{k^+} + \frac{(\bfp-\bfk+\bfq)^2}{p^+-k^+} - \frac{\bfp^2}{p^+}.
\end{align}
At the level of squared-amplitudes the relevant LPM frequencies  are
\begin{align}
\Omega_1 &= \omega_4 = \frac{\bfQ_1^2}{x(1-x)p^+},\\
\Omega_2 &= \omega_1-\omega_2 = \frac{\bfQ_2^2}{x(1-x)p^+},\\
\Omega_3 &= \omega_1 - \omega_3 = \omega_6 = \frac{\bfQ_3^2}{x(1-x)p^+},\\
\Omega_4 &= \Omega_3 - \omega_2 = \frac{\bfQ_4^2}{x(1-x)p^+},
\end{align}
The LPM interference factor at first order in opacity has a universal form
\begin{align}
\phi_n \equiv 1-\cos(\Omega_n z^+), 
\end{align}
and if we consider cold  nuclear matter of  length $L$ and constant density, it is also useful to define an $z^+$-averaged LPM interference factor
\begin{align}
\Phi_n \equiv \frac{1}{L^+}\int_{-L^+}^0 \phi_n dz^+ = 1 - \frac{\sin \Omega_n L^+}{\Omega_n L^+} .
\end{align}

We now discuss the various contributions to the medium-induced cross sections. 
\paragraph{Type I: real emission with real (one) Glauber interaction.}
\begin{figure}[ht!]
\centering
\includegraphics[scale=1]{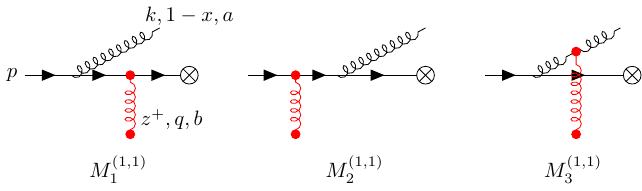}
\caption{Amplitudes with one real emission and one Glauber gluon interaction.}
\label{fig:RR-all}
\end{figure}
The diagrams with one Glauber gluon exchange and radiation are shown in figure~\ref{fig:RR-all} and given by 

\begin{align}
M^{(1,1)}_1 &= \int \frac{dq^-}{2\pi}\frac{\sln}{2} \frac{i}{p^--k^-+q^--\frac{(\bfp-\bfk+\bfq)^2}{p^+-k^+}+\frac{i\eta}{p^+-k^+}} igt^b A_G^-(\bfq)e^{iq^-z^+/2}\frac{\slnbar}{2} \nonumber\\
& \  \  \times \frac{\sln}{2}\frac{i}{p^--k^--\frac{(\bfp-\bfk)^2}{p^+-k^+}+\frac{i\eta}{p^+-k^+}} \frac{ig t^a}{x(1-x)p^+}\SP_\lambda(\bfQ_1) \frac{\slnbar}{2} \frac{\sln}{2}\chi_n(p)  \nonumber \\
&= i g_s^2 t^b t^a A^-_G(\bfq)\Theta(-z^+)e^{i\omega_1z_+/2} \frac{\SP_\lambda(\bfQ_1)}{\bfQ_1^2}\frac{\sln}{2}\chi_{n}(p) ,
\end{align}

\begin{align}
M^{(1,1)}_2 &= \int \frac{dq^-}{2\pi}\frac{\sln}{2} \frac{i}{p^--k^-+q^--\frac{(\bfp-\bfk+\bfq)^2}{p^+-k^+}+\frac{i\eta}{p^+-k^+}}  \frac{ig t^a}{x(1-x)p^+}\SP_\lambda(\bfQ_2)\frac{\slnbar}{2} \nonumber\\
&  \ \ \times \frac{\sln}{2}\frac{i}{p^-+q^--\frac{(\bfp+\bfq)^2}{p^+}+\frac{i\eta}{p^+}} igt^b A_G^-(\bfq)e^{iq^-z^+/2}\frac{\slnbar}{2} \frac{\sln}{2}\chi_n(p)  \nonumber \\
&= -i g_s^2 t^a t^b A^-_G(\bfq)\Theta(-z^+)\left[e^{i\omega_1z_+/2}-e^{i\omega_2z_+/2}\right] \frac{\SP_\lambda(\bfQ_2)}{\bfQ_2^2}\frac{\sln}{2}\chi_{n}(p) ,
\end{align}

\begin{align}
M^{(1,1)}_3 &= \int \frac{dq^-}{2\pi}\frac{\sln}{2} \frac{i}{p^--k^-+q^--\frac{(\bfp-\bfk+\bfq)^2}{p^+-k^+}+\frac{i\eta}{p^+-k^+}}  \frac{ig t^c}{x(1-x)p^+}\SP^\rho(\bfQ_3)\frac{\slnbar}{2}\frac{\sln}{2}\chi_{n}(p)  \nonumber\\
& \ \ \times \e_{\lambda}(k) gf^{abc}A^-_G(\bfq)e^{iq^-z^+/2}k^+ g_T^{\mu\nu} \frac{i}{(k-q)^2+i\eta}\sum_{\theta}\e_{\theta,\nu}(k-q)\e_{\rho, \theta}(k-q) \nonumber  \\
&= ig_s^2\left(t^at^b-t^bt^a\right)A^-_G(\bfq)\left[\Theta(-z^+)e^{i\omega_1 z^+/2} + \Theta(z^+)e^{i\omega_3 z^+/2}\right]\frac{\SP_\lambda(\bfQ_3)}{\bfQ_3^2}\frac{\sln}{2}\chi_{n}(p).    
\end{align}
Concerning the third diagram: first, the contact term in the gluon propagator vanishes when it contracted with $g_T^{\mu\nu}$ in the Glauber vertex. Second, the gluon propagator can give a contribution where the scattering happens after the hard scattering.

Summing over final-state color, spin and polarization, while averaged over initial state spin and color ($2N_c$), we obtain
\begin{align}
 &\overline{|M^{(1,1)}|^2} =\sum_{\lambda}\frac{1}{2N_c}\mathrm{Tr}\left\{|M^{(1,1)}_1+M^{(1,1)}_2+M^{(1,1)}_3|^2\right\} \nonumber \\
 &= \sum_\lambda g_s^4 |A^-_G(\bfq)|^2\Theta(-z^+)\frac{1}{2}\frac{1}{N_c}\textrm{Tr}\left\{ \left| t^bt^a\left[e^{i\omega_1z^+/2}\frac{\SP_\lambda(\bfQ_1)}{\bfQ_1^2}-e^{i\omega_1z^+/2}\frac{\SP_\lambda(\bfQ_3)}{\bfQ_3^2}\right]\right. \right.  \nonumber \\
&\hspace*{.4cm}\left.+t^at^b\left[-e^{i\omega_1z^+/2}\frac{\SP_\lambda(\bfQ_2)}{\bfQ_2^2}+e^{i\omega_2z^+/2}\frac{\SP_\lambda(\bfQ_2)}{\bfQ_2^2}+e^{i\omega_3z^+/2}\frac{\SP_\lambda(\bfQ_3)}{\bfQ_3^2}\right]\right|^2J_{p-k+q} \bar{J}_{p-k+q}\frac{\sln}{2}p^+\left.\right\}\nonumber\\
&= g_s^4 |A^-_G(\bfq)|^2\Theta(-z^+) \nonumber\\
&\hspace*{.4cm}\times \frac{1}{2}\mathrm{Tr}\left\{\sum_\lambda \mathfrak{Re}\left\{C_F^2\left[
\frac{\SP_\lambda^\dagger(\bfQ_1)\SP_\lambda(\bfQ_1)}{\bfQ_1^2\bfQ_1^2}
+2\frac{\SP_\lambda^\dagger(\bfQ_3)\SP_\lambda(\bfQ_3)}{\bfQ_3^2\bfQ_3^2}
-2\frac{\SP_\lambda^\dagger(\bfQ_1)\SP_\lambda(\bfQ_3)}{\bfQ_1^2\bfQ_3^2}\right.\right.\right. \nonumber\\
&\hspace*{.4cm}\left.+2\frac{\SP_\lambda^\dagger(\bfQ_2)\SP_\lambda(\bfQ_2)}{\bfQ_2^2\bfQ_2^2}\left(1-\cos\left(\Omega_2\frac{z^+}{2}\right)\right)
-2\frac{\SP_\lambda^\dagger(\bfQ_2)\SP_\lambda(\bfQ_3)}{\bfQ_2^2\bfQ_3^2}\left(1-\cos\left(\Omega_2\frac{z^+}{2}\right)\right)
\right] \nonumber\\
&\hspace*{.4cm}+C_F(C_F-C_A/2)\left[
2\frac{\SP_\lambda^\dagger(\bfQ_1)\SP_\lambda(\bfQ_3)}{\bfQ_1^2\bfQ_3^2}
-2\frac{\SP_\lambda^\dagger(\bfQ_3)\SP_\lambda(\bfQ_3)}{\bfQ_3^2\bfQ_3^2} 
\right.\nonumber\\
&\hspace*{.4cm}\left.\left. 
+2\frac{\SP_\lambda^\dagger(\bfQ_3)\SP_\lambda(\bfQ_3)}{\bfQ_3^2\bfQ_3^2} \left(1-\cos\left(\Omega_2\frac{z^+}{2}\right)\right)
-2\frac{\SP_\lambda^\dagger(\bfQ_1)\SP_\lambda(\bfQ_2)}{\bfQ_1^2\bfQ_2^2} \left(1-\cos\left(\Omega_2\frac{z^+}{2}\right)\right)
\right]\right\}  \nonumber\\
&\hspace*{.4cm}\left. \times J_{p-k+q} \bar{J}_{p-k+q}\frac{\sln}{2}p^+\right\} .
\end{align}
Performing the polarization sum (with initial spin average)
\begin{align}
\frac{1}{2}\sum_{\lambda} \mathrm{Tr}\left\{\SP_\lambda^\dagger(\bfQ) \SP_\lambda(\bfQ') J \bar{J}\frac{\sln}{2}p^+\right\} =  2p^+\left[(1+x^2)-\epsilon(1-x)^2\right]\bfQ\cdot \bfQ'\frac{\Tr\{\bar{J}J \frac{\sln}{2} \}}{2} .
\end{align}

Relating the LO spin-averaged TMD parton density to the sources $D_q = \Tr\{\bar{J}J \frac{\sln}{2}\}/2$, we get
\begin{align}
\overline{|M^{(1,1)}|^2} 
&= 2p^+ \left[(1+x)^2-\epsilon (1-x)^2\right] g_s^4 |A^-_G(\bfq)|^2\Theta(-z^+) D_q(p-k+q) \nonumber\\
&  \  \  \times \left\{
2C_F^2\left[\frac{1}{2}\frac{1}{\bfQ_1^2} + \frac{1}{\bfQ_3^2}-\frac{\bfQ_1}{\bfQ_1^2}\cdot\frac{\bfQ_3}{\bfQ_3^2} + \frac{1}{\bfQ_2^2}\phi_2 - \frac{\bfQ_2}{\bfQ_2^2}\cdot \frac{\bfQ_3}{\bfQ_3^2}\phi_2\right]\right.\nonumber\\
&\hspace*{1cm} +\left. (2C_F^2-C_FC_A)
\left[-\frac{1}{\bfQ_3^2}+\frac{\bfQ_1}{\bfQ_1^2}\cdot\frac{\bfQ_3}{\bfQ_3^2} - \frac{\bfQ_1}{\bfQ_1^2}\cdot \frac{\bfQ_2}{\bfQ_2^2}\phi_2 + \frac{\bfQ_2}{\bfQ_2^2}\cdot \frac{\bfQ_3}{\bfQ_3^2}\phi_2\right]
\right\} \nonumber \\
&= 2p^+ C_F\left[(1+x)^2-\epsilon (1-x)^2\right] g_s^4 |A^-_G(\bfq)|^2\Theta(-z^+) D_q(p-k+q) \nonumber\\
&\hspace*{.4cm}\times \left\{
C_F\left[\frac{1}{\bfQ_1^2} + 2\frac{\bfQ_2}{\bfQ_2^2}\cdot\left[\frac{\bfQ_2}{\bfQ_2^2}-\frac{\bfQ_1}{\bfQ_1^2}\right]\phi_2\right]  \right. \nonumber\\ 
&\hspace*{1cm}\left. + C_A
\left[\frac{1}{\bfQ_3^2}-\frac{\bfQ_1}{\bfQ_1^2}\cdot\frac{\bfQ_3}{\bfQ_3^2} 
+ \frac{\bfQ_1}{\bfQ_1^2}\cdot \frac{\bfQ_2}{\bfQ_2^2}\phi_2 
- \frac{\bfQ_2}{\bfQ_2^2}\cdot \frac{\bfQ_3}{\bfQ_3^2}\phi_2\right]
\right\} .
\end{align}
where we have defined the LPM interference factors as
\begin{align}
\phi_n = 1-\cos\left(\Omega_n\frac{z^+}{2}\right)\, .
\end{align}

Finally, including the phase space integration over the radiated parton, and the integration and average over the medium, the contribution to the NLO TMD parton density is 
\begin{align}
& \int dz^+ \int\frac{d^2\bfq}{(2\pi)^2}\left\langle\int\frac{d^2\bfk}{(2\pi)^3}\frac{1}{2(1-x)p^+}\overline{|M^{(1,1)}|^2} e^{i\bfb\cdot(\bfp-\bfk+\bfq)} \right\rangle_{\rm med} \nonumber  \\
&= \int_{-\infty }^0 dz^+ \rho(z^+) \int\frac{d^2\bfk}{(2\pi)^3} g_s^2 P_{qq}(x, \epsilon) \rho_G(z^+)\int\frac{d^2\bfq}{(2\pi)^2} \frac{g_s^2}{\bfq^4} e^{i\bfb\cdot(\bfp-\bfk+\bfq)} \nonumber\\
&\hspace*{.4cm}\times \left\{
C_F\left[\frac{1}{\bfQ_1^2} + 2\frac{\bfQ_2}{\bfQ_2^2}\cdot\left[\frac{\bfQ_2}{\bfQ_2^2}-\frac{\bfQ_1}{\bfQ_1^2}\right]\phi_2\right]\right.\nonumber\\
&\hspace*{1cm}\left.+ C_A
\left[\frac{1}{\bfQ_3^2}-\frac{\bfQ_1}{\bfQ_1^2}\cdot\frac{\bfQ_3}{\bfQ_3^2} 
+ \frac{\bfQ_1}{\bfQ_1^2}\cdot \frac{\bfQ_2}{\bfQ_2^2}\phi_2 
- \frac{\bfQ_2}{\bfQ_2^2}\cdot \frac{\bfQ_3}{\bfQ_3^2}\phi_2\right]
\right\} .
\end{align}

\paragraph{Type II: real emission with virtual  (double) Glauber gluon exchange.}
\begin{figure}
\centering
\includegraphics[scale=1]{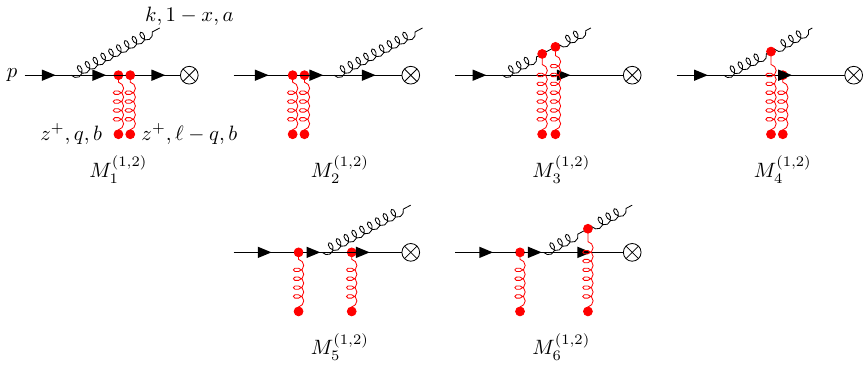}
\caption{Amplitudes with one real emission and two contact Glauber gluon interactions. The 5th and 6th diagrams are zero because the phase factor does not depend on $q^-$ in the contact limit, and the $q^-$ integration involves two poles on the same side of the real axis. }
\label{fig:RV-all}
\end{figure}

Diagrams with two Galuber gluon exchanges and a real emission are shown in figure~\ref{fig:RV-all}.
We parametrize the momentum of the two Glauber exchanges by $q$ and $\ell-q$, such that the total momentum transfer to the jet system is $\ell$.
 We will take the contact limit first, setting the location of the two Glauber exchanges to be the same $z^+$, $\bfl=0$. However, $\ell^-$ is non-zero and will need to be integrated out.
In the contact limit, the two phases associated with the two Glauber exchanges can be combined into one
\begin{align}
e^{iq^-z^+/2}e^{i(\ell^--q^-)z^+/2} = e^{i\ell^-z^+/2}.
\end{align}
The expressions for the non-vanishing diagrams are:

\begin{align}
M^{(1,2)}_1 &= \int\frac{d\ell^-}{2\pi}\int \frac{dq^-}{2\pi}\frac{\sln}{2} \frac{i}{p^--k^-+\ell^--\frac{(\bfp-\bfk)^2}{p^+-k^+}+\frac{i\eta}{p^+-k^+}} igt^b A_G^{-*}(\bfq)\frac{\slnbar}{2} \nonumber\\
& \hspace*{0.4cm} \times \frac{\sln}{2}\frac{i}{p^--k^-+q^--\frac{(\bfp-\bfk+\bfq)^2}{p^+-k^+}+\frac{i\eta}{p^+-k^+}} igt^b A_G^{-}(\bfq)\frac{\slnbar}{2} \nonumber\\
& \hspace*{0.4cm} \times \frac{\sln}{2}\frac{i}{p^--k^--\frac{(\bfp-\bfk)^2}{p^+-k^+} +\frac{i\eta}{p^+-k^+}}
\frac{ig t^a}{x(1-x)p^+}\SP(\bfQ_1) \frac{\slnbar}{2} \frac{\sln}{2} \chi_n(p) e^{i\ell^-z^+/2} \nonumber \\
&= -\frac{1}{2}g_s^3 t^b t^b t^a |A^-_G(\bfq)|^2 \Theta(-z^+) e^{i\omega_4 z^+/2}  \frac{\SP(\bfQ_1)}{\bfQ_1^2} \frac{\sln}{2}\chi_n(p),
\end{align}

\begin{align}
M^{(1,2)}_2 &= \int\frac{d\ell^-}{2\pi}\int \frac{dq^-}{2\pi}\frac{\sln}{2} \frac{i}{p^--k^-+\ell^--\frac{(\bfp-\bfk)^2}{p^+-k^+}+\frac{i\eta}{p^+-k^+}}\frac{ig t^a}{x(1-x)p^+}\SP(\bfQ_1)\frac{\slnbar}{2} \nonumber\\
& \hspace*{0.4cm} \times \frac{\sln}{2}\frac{i}{p^-+\ell^--\frac{\bfp^2}{p^+}+\frac{i\eta}{p^+}} igt^b A_G^{-*}(\bfq)\frac{\slnbar}{2} \nonumber\\
& \hspace*{0.4cm} \times \frac{\sln}{2}\frac{i}{p^-+q^--\frac{(\bfp+\bfq)^2}{p^+} +\frac{i\eta}{p^+}}
  igt^b A_G^{-}(\bfq)\frac{\slnbar}{2} \frac{\sln}{2} \chi_n(p) e^{i\ell^-z^+/2} \nonumber \\
&= \frac{1}{2}g_s^3 t^a t^b t^b |A^-_G(\bfq)|^2 \Theta(-z^+) \left[e^{i\omega_4 z^+/2} -1\right] \frac{\SP(\bfQ_1)}{\bfQ_1^2} \frac{\sln}{2}\chi_n(p),
\end{align}

\begin{align}
M^{(1,2)}_3 &= \int\frac{d\ell^-}{2\pi}\int \frac{dq^-}{2\pi}\frac{\sln}{2} \frac{i}{p^--k^-+\ell^--\frac{(\bfp-\bfk)^2}{p^+-k^+}+\frac{i\eta}{p^+-k^+}}\nonumber\\
&\hspace*{0.4cm} \times \frac{ig t^d}{x(1-x)p^+}\SP(\bfQ_1)\frac{\slnbar}{2} \frac{\sln}{2} \chi_n(p) e^{i\ell^-z^+/2} \nonumber\\
& \hspace*{0.4cm} \times \e_{\lambda}^\mu(k) gf^{abc} A^{-}_G(\bfq) k^+ g_{T,\mu\nu} \frac{i}{(\bfk-\bfq)^2+i\eta} \sum_\theta\e_\theta^\nu(k-q)\e_\theta^{\nu'}(k-q) \nonumber\\
& \hspace*{0.4cm} \times gf^{cbd} A^{-*}_G(\bfq) k^+ g_{T, \nu', \rho'}\frac{i}{(k-Q)^2+i\eta} \sum_{\phi}\e_{\phi}^{\rho'}(k-Q)\e_{\phi}^\rho(k-Q) \nonumber \\
&= \frac{1}{2}g_s^3 f^{abc}f^{cbd}t^d |A^-_G(\bfq)|^2 \left[\Theta(-z^+)e^{i\omega_4 z^+} + \Theta(z^+) \right] \frac{\SP(\bfQ_1)}{\bfQ_1^2} \frac{\sln}{2}\chi_n(p).
\end{align}
We note that the gauge term in the gluon propagator $(\propto \nbar^\mu\nbar^\nu)$ vanishes when contracted with the Glauber vertex ($\propto g_T^{\mu\nu}$). Again, the gluon double scattering can also happen after the hard vertex. The last diagram is
\begin{align}
M^{(1,2)}_4 &= \int\frac{d\ell^-}{2\pi}\int \frac{dq^-}{2\pi}\frac{\sln}{2} \frac{i}{p^--k^-+\ell^--\frac{(\bfp-\bfk)^2}{p^+-k^+}+\frac{i\eta}{p^+-k^+}} ig t^b A^{-*}_G(\bfq) \frac{\slnbar}{2} \nonumber\\
& \hspace*{0.4cm} \times \frac{\sln}{2} \frac{i}{p^--k^-+q^--\frac{(\bfp-\bfk+\bfq)^2}{p^+-k^+}+\frac{i\eta}{p^+-k^+}} \frac{ig t^c}{x(1-x)p^+}\SP(\bfQ_3)\frac{\slnbar}{2} \frac{\sln}{2} \chi_n(p) e^{i\ell^-z^+/2} \nonumber\\
& \hspace*{0.4cm} \times \e_{\lambda}^\mu(k)g f^{abc}A^-_G(\bfq) k^+ g_{T, \mu\nu} \frac{i}{(k-q)^2+i\eta}\sum_\theta\e_\theta^\nu(k-1)\e_\theta^{\nu'}(k-1) \nonumber \\
&= -ig_s^3f^{abc}t^bt^c |A^-_G(\bfq)|^2  \Theta(-z^+) e^{i\omega_4 z^+/2} \frac{\SP(\bfQ_3)}{\bfQ_3^2} \frac{\sln}{2}\chi_n(p) .
\end{align}

The diagrams computed above interfere with $M^{(1,0)}$, yielding the following expressions: 
\begin{align}
& \overline{2\mathfrak{Re}\left\{M^{(1,0)*}M^{(1,2)}_1\right\}} \nonumber \\
&= g_s^4|A_G^-(\bfq)|^2\Theta(-z^+)\sum_{\lambda}\frac{\mathrm{Tr}\left\{t^a t^b t^b t^a\right\}}{N_c}\frac{1}{2} 
2\mathfrak{Re}\mathrm{Tr}\left\{
\frac{-1}{2}\frac{\SP_\lambda^\dagger}{\bfQ_1^2}\frac{\SP_\lambda}{\bfQ_1^2}J_{p-k}\bar{J}_{p-k}\frac{\sln}{2}p^+ e^{i\omega_4 z^+/2}
\right\}  \nonumber \\
&= -C_F g_s^4|A_G^-(\bfq)|^2\Theta(-z^+) 2p^+(1-x)P_{qq}(x, \epsilon)
\frac{1}{\bfQ_1^2}\cos\left(\Omega_1\frac{z^+}{2}\right)
\frac{\mathrm{Tr}\left\{J_{p-k}\bar{J}_{p-k}\frac{\sln}{2}\right\}}{2} ,
\end{align}
\begin{align}
& \overline{2\mathfrak{Re}\left\{M^{(1,0)*}M^{(1,2)}_2\right\}} \nonumber \\
&= - C_F g_s^4|A_G^-(\bfq)|^2\Theta(-z^+) 2p^+(1-x)P_{qq}(x, \epsilon)
\frac{1}{\bfQ_1^2}\left[1-\cos\left(\Omega_1\frac{z^+}{2}\right)\right]
\frac{\mathrm{Tr}\left\{J_{p-k}\bar{J}_{p-k}\frac{\sln}{2}\right\}}{2}  , 
\end{align}
\begin{align}
& \overline{2\mathfrak{Re}\left\{M^{(1,0)*}M^{(1,2)}_3\right\}} \nonumber \\
&= -C_A g_s^4|A_G^-(\bfq)|^2\Theta(-z^+)C_F^2 2p^+(1-x)P_{qq}(x, \epsilon)
\frac{1}{\bfQ_1^2}\cos\left(\Omega_1\frac{z^+}{2}\right)
\frac{\mathrm{Tr}\left\{J_{p-k}\bar{J}_{p-k}\frac{\sln}{2}\right\}}{2} , 
\end{align}
\begin{align}
& \overline{2\mathfrak{Re}\left\{M^{(1,0)*}M^{(1,2)}_4\right\}} \nonumber \\
&= C_A g_s^4|A_G^-(\bfq)|^2\Theta(-z^+)C_F^2 2p^+(1-x)P_{qq}(x, \epsilon)
\frac{\bfQ_1}{\bfQ_1^2}\cdot \frac{\bfQ_3}{\bfQ_3^2}\cos\left(\Omega_1\frac{z^+}{2}\right)
\frac{\mathrm{Tr}\left\{J_{p-k}\bar{J}_{p-k}\frac{\sln}{2}\right\}}{2} .
\end{align}

Summing over the four contributions, performing the average in the medium, and taking the phase space integrals, we have
\begin{align}
& \int dz^+ \int\frac{d^2\bfq}{(2\pi)^2}\left\langle\int\frac{d^2\bfk}{(2\pi)^3}\frac{1}{2(1-x)p^+}\sum_{n=1}^4\overline{2\mathfrak{Re}\left\{M^{(1,0)*}M^{(1,2)}_n\right\}} e^{i\bfb\cdot(\bfp-\bfk+\bfq)} \right\rangle_{\rm med} \nonumber \\
&= \int_{-\infty}^0 dz^+ \rho(z^+) \int\frac{d^2\bfk}{(2\pi)^3} g_s^2 P_{qq}(x, \epsilon) \rho_G(z^+)\int\frac{d^2\bfq}{(2\pi)^2} \frac{g_s^2}{\bfq^4} e^{i\bfb\cdot(\bfp-\bfk)} \nonumber\\
& \ \ \times \left\{
-C_F\frac{1}{\bfQ_1^2} - C_A\frac{1}{\bfQ_1^2}\cos\left(\Omega_1\frac{z^+}{2}\right)  + C_A\frac{\bfQ_1}{\bfQ_1^2}\cdot \frac{\bfQ_3}{\bfQ_3^2}\cos\left(\Omega_1\frac{z^+}{2}\right)
\right\} .
\end{align}

\paragraph{Type III: virtual corrections with a real Glauber gluon interaction.}

\begin{figure}[hb!]
\centering
\includegraphics[scale=1]{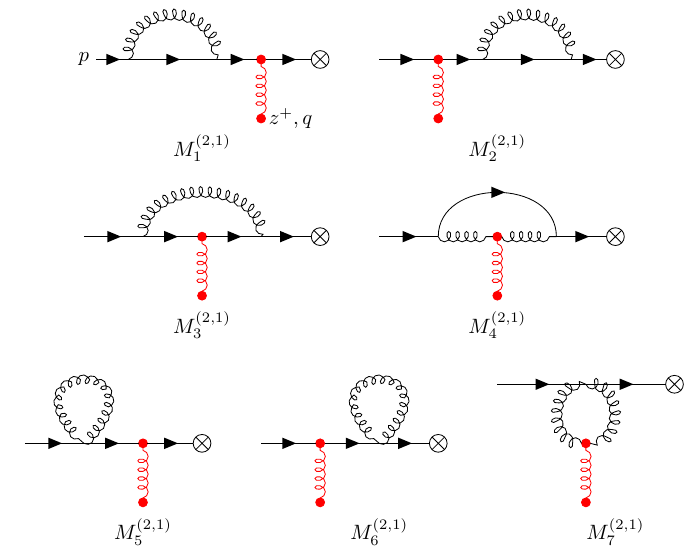}
\caption{One-loop amplitudes with one Glauber gluon interaction that interfere with $M^{0,1}$. The 5th and 6th diagrams are zero, because the $\bfk$ integration is simply scaleless. The 7th diagram is also zero, because the $k^-$ integration involves two poles on the same side of the real axis.}
\label{fig:VR-all}
\end{figure}

Next, we turn to the loop diagrams with one Glauber gluon exchange shown in figure~\ref{fig:VR-all}. We start with
\begin{align}
& M_1^{(2,1)} \nonumber\\
&= \int \frac{dk^+  d^2\bfk}{2(2\pi)^3} \int \frac{dk^-}{2\pi} \int \frac{dq^-}{2\pi} \frac{\sln}{2}\frac{i}{p^-+q^--\frac{(\bfp+\bfq)^2}{p^+}+\frac{i\eta}{p^+}} igt^bA^-(\bfq)e^{iq^-z^+/2}\frac{\slnbar}{2} \nonumber\\
& \hspace*{0.4cm} \times\frac{\sln}{2}\frac{i}{p^--\frac{\bfp^2}{p^+} + \frac{i\eta}{p^+}} igt^a\left[n+\frac{\gamma_\perp(\slp_\perp-\slk_\perp)}{p^+-k^+} + \frac{\slp_\perp\gamma_\perp}{p^+} - \nbar\frac{\slp_\perp(\slp_\perp-\slk_\perp)}{p^+(p^+-k^+)}\right]_\mu\frac{\slnbar}{2} \nonumber\\
& \hspace*{0.4cm} \times\frac{\sln}{2}\frac{i}{p^--k^--\frac{(\bfp-\bfk)^2}{p^+-k^+}+\frac{i\eta}{p^+-k^+}}igt^a\left[n+\frac{\gamma_\perp\slp_\perp}{p^+}+\frac{(\slp_\perp-\slk_\perp)\gamma_\perp}{p^+-k^+}  - \nbar\frac{(\slp_\perp-\slk_\perp)\slp_\perp}{(p^+-k^+)p^+}\right]_\nu\frac{\slnbar}{2} \nonumber\\
& \hspace*{0.4cm} \times\frac{i}{k^2+i\eta}\left[\sum_{\lambda} \e_\lambda^\mu(k)\e_\lambda^\nu(k) -\frac{4k^2}{(k^+)^2}\nbar^\mu\nbar^\nu \right] \frac{\sln}{2}\chi_n(p).
\end{align}
Note that the gluon is off-shell, thus the Lorentz structure contains the tensor $\frac{-4k^2}{(k^+)^2}\nbar^\mu \nbar^\nu$. However, this additional term cancels the pole of the gluon propagator $\frac{i}{k^2+i\eta}$. Then, after the $k^-$ integration, the expression does not contain $\bfk^2$ anymore, which makes the next integration over $\bfk$ scaleless. Therefore, we can drop the $\frac{-4k^2}{(k^+)^2}\nbar^\mu \nbar^\nu$ term from the beginning. Then, we can again use the splitting amplitude defined earlier and rewrite
\begin{align}
M_1^{(2,1)} &= -ig_s^3 t^b t^a t^a A^-(\bfq)\int \frac{dk^+  d^2\bfk}{2(2\pi)^3} \int \frac{dk^-}{2\pi} \int \frac{dq^-}{2\pi}e^{iq^-z^+/2}  \frac{1}{p^-+q^--\frac{(\bfp+\bfq)^2}{p^+}+\frac{i\eta}{p^+}} \nonumber\\
& \hspace*{0.4cm} \times\frac{1}{p^--\frac{\bfp^2}{p^+} + \frac{i\eta}{p^+}}  
\frac{1}{p^--k^--\frac{(\bfp-\bfk)^2}{p^+-k^+}+\frac{i\eta}{p^+-k^+}}    \frac{1}{k^2+i\eta} 
\frac{\sln}{2}\frac{\sum_\lambda \SP_\lambda^*(\bfQ_1) \SP_\lambda(\bfQ_1)}{[k^+(p^+-k^+)/p^+]^2}\chi_n(p) \nonumber \\
&=-ig_s^3 t^b t^a t^a A^-(\bfq)\Theta(-z^+)\int \frac{dk^+  d^2\bfk}{2k^+(2\pi)^3}  (-i)e^{i\left(\frac{(\bfp+\bfq)^2}{p^+}-p^-\right)z^+/2}  \frac{1}{p^--\frac{\bfp^2}{p^+} + \frac{i\eta}{p^+}}  \nonumber\\
& \hspace*{0.4cm} \times \int \frac{dk^-}{2\pi} \frac{1}{p^--k^--\frac{(\bfp-\bfk)^2}{p^+-k^+}+\frac{i\eta}{p^+-k^+}}    \frac{1}{k^--\frac{\bfk^2}{k^+}+\frac{i\eta}{k^+}} 
\frac{\sln}{2}\frac{\sum_\lambda \SP_\lambda^*(\bfQ_1) \SP_\lambda(\bfQ_1)}{[k^+(p^+-k^+)/p^+]^2}\chi_n(p) .
\end{align}
A non-zero $k^-$ integral requires that the two $k^-$  poles are on different sides of the real axis. Therefore,  $k^+(p^+-k^+)>0$ and $0<k^+<p^+$, allowing one to define $x = k^+/p^+$ with $0<x<1$.
\begin{align}
M_1^{(2,1)} &=-ig_s^3 t^b t^a t^a A^-(\bfq)\Theta(-z^+)\int_0^1 \frac{dx}{1-x}\int \frac{d^2\bfk}{2(2\pi)^3}  (-i)e^{i\left(\frac{(\bfp+\bfq)^2}{p^+}-p^-\right)z^+/2}  \frac{1}{p^--\frac{\bfp^2}{p^+} + \frac{i\eta}{p^+}}  \nonumber\\
&\hspace*{0.4cm} \times (i) \frac{1}{\frac{\bfk^2}{k^+}-p^-+\frac{(\bfp-\bfk)^2}{p^+-k^+}-\frac{i\eta}{x(1-x)p^+}}  
\frac{\sln}{2}\frac{\sum_\lambda \SP_\lambda^*(\bfQ_1) \SP_\lambda(\bfQ_1)}{[k^+(p^+-k^+)/p^+]^2}\chi_n(p)  \nonumber \\
&=-ig_s^3 t^b t^a t^a A^-(\bfq)\Theta(-z^+)\int_0^1 \frac{dx}{1-x}\int \frac{d^2\bfk}{2(2\pi)^3}  e^{i\left(\frac{(\bfp+\bfq)^2}{p^+}-p^-\right)z^+/2}   \nonumber\\
&  \hspace*{0.4cm} \times\left[ \frac{1}{p^--\frac{\bfp^2}{p^+}} - \frac{1}{\frac{\bfk^2}{k^+}-p^-+\frac{(\bfp-\bfk)^2}{p^+-k^+}}  \right] \frac{1}{\bfQ_1^2}
\frac{\sln}{2}\frac{\sum_\lambda \SP_\lambda^*(\bfQ_1)  \SP_\lambda(\bfQ_1)}{x(1-x)p^+ }\chi_n(p) .
\end{align}
Recognizing that the $\bfk$ integral over the term containing  $\frac{1}{p^--\frac{\bfp^2}{p^+}}$ is scaleless, the remaining propagator piece gives
\begin{align}
M_1^{(2,1)}
&=ig_s^3 t^b t^a t^a A^-(\bfq)\Theta(-z^+) e^{i\omega_2 z^+/2}\int_0^1 \frac{dx}{1-x}\int \frac{d^2\bfk}{2(2\pi)^3}   \frac{\sum_\lambda \SP_\lambda^*(\bfQ_1)  \SP_\lambda(\bfQ_1)}{\bfQ_1^2 \left[\bfQ_1^2 - x(1-x)p^2\right] }\frac{\sln}{2}\chi_n(p).
\end{align}
Since the off-shellness of the incoming parton ($p^2$) is assumed to be much less than $\bfq^2$, we have approximated
\begin{align}
\frac{(\bfp+\bfq)^2}{p^+}-p^- = \frac{(\bfp+\bfq)^2}{p^+}-\frac{\bfp^2}{p^+} = \omega_2.
\end{align}
Note that it also becomes scaleless at leading power in $\frac{p^2}{\mu_b^2}$, and we will not consider these types of wave function renormalization. In fact, these wave-function renormalization type diagrams will cancel among type-3 and type-4 contributions.

 In the second diagram, for the exact same reasons as above, only the physical polarizations of the gluon propagator contribute to give
\begin{align}
M_2^{(2,1)} &= \int \frac{dk^+  d^2\bfk}{2k^+(2\pi)^3} \int \frac{dk^-}{2\pi} \int \frac{dq^-}{2\pi} \frac{\sln}{2}\frac{i}{p^-+q^--\frac{(\bfp+\bfq)^2}{p^+} + \frac{i\eta}{p^+}} igt^a\frac{\SP^*(\bfQ_2)}{k^+(p^+-k^+)/p^+} \frac{\slnbar}{2}\nonumber\\
& \hspace*{0.4cm} \times\frac{\sln}{2}\frac{i}{p^-+q^--k^--\frac{(\bfp+\bfq-\bfk)^2}{p^+-k^+}+\frac{i\eta}{p^+-k^+}}igt^a\frac{\SP(\bfQ_2)}{k^+(p^+-k^+)/p^+} \frac{\slnbar}{2}\nonumber\\
& \hspace*{0.4cm} \times\frac{\sln}{2}\frac{i}{p^-+q^--\frac{(\bfp+\bfq)^2}{p^+}+\frac{i\eta}{p^+}} igt^bA^-(\bfq)e^{iq^-z^+/2}\frac{\slnbar}{2} \frac{\sln}{2}\chi_n(p) \frac{i}{k^--\frac{\bfk^2}{k^+}+\frac{i\eta}{k^+}} \nonumber \\
&= -ig_s^3 t^a t^a t^b A^-(\bfq)\Theta(-z^+) \int_0^1\frac{dx}{1-x} \int \frac{d^2\bfk}{2(2\pi)^3} \left[e^{i\omega_2z^+/2}-e^{i\omega_1 z^+/2}\right] \nonumber\\
& \hspace*{0.4cm} \times \frac{\sum_\lambda \SP_\lambda^*(\bfQ_2)  \SP_\lambda(\bfQ_2)}{\bfQ_2^2 \bfQ_2^2 }\frac{\sln}{2}\chi_{n}(p) .
\end{align}

For the third amplitude, the $\frac{-4k^2}{(k^+)^2}\nbar^\mu\nbar^\nu$ term of the gluon propagator again does not contribute as it cancels the $\frac{1}{k^2}$ in the gluon propagator. The remaining $k^-$ integral involves two poles on the same side of the real axis, which is zero. This, we can write
\begin{align}
M_3^{(2,1)} &= \int \frac{dk^+  d^2\bfk}{2k^+(2\pi)^3} \int \frac{dk^-}{2\pi} \int \frac{dq^-}{2\pi} \frac{\sln}{2}\frac{i}{p^-+q^--\frac{(\bfp+\bfq)^2}{p^+} + \frac{i\eta}{p^+}} igt^a\frac{\SP^*(\bfQ_2)}{k^+(p^+-k^+)/p^+} \frac{\slnbar}{2}\nonumber\\
& \hspace*{0.4cm} \times\frac{\sln}{2}\frac{i}{p^-+q^--k^--\frac{(\bfp+\bfq-\bfk)^2}{p^+-k^+}+\frac{i\eta}{p^+-k^+}}igt^a\frac{\SP(\bfQ_1)}{k^+(p^+-k^+)/p^+} \frac{\slnbar}{2}\nonumber\\
& \hspace*{0.4cm} \times igt^bA^-(\bfq)e^{iq^-z^+/2}\frac{\slnbar}{2}  \frac{\sln}{2}\frac{i}{p^--k^--\frac{(\bfp-\bfk)^2}{p^+-k^+}+\frac{i\eta}{p^+-k^+}} \frac{\sln}{2}\chi_n(p) \frac{i}{k^--\frac{\bfk^2}{k^+}+\frac{i\eta}{k^+}} \nonumber \\
&= ig_s^3 t^a t^b t^a A^-(\bfq) \Theta(-z^+)\int_0^1\frac{dx}{1-x} \int \frac{d^2\bfk}{2(2\pi)^3} \nonumber\\
& \hspace*{0.4cm}\times \left[e^{i\omega_2z^+/2}-e^{i\omega_1 z^+/2}\right]\frac{\sum_\lambda \SP_\lambda^*(\bfQ_2)  \SP_\lambda(\bfQ_1)}{\bfQ_2^2 \bfQ_1^2 }\frac{\sln}{2}\chi_{n}(p).
\end{align}

For the fourth diagram, the $\frac{-4k^2}{(k^+)^2}\nbar^\mu\nbar^\nu$ terms does not contribute because it is contracted with a Glauber vertex:
\begin{align}
M_4^{(2,1)} &= \int \frac{dk^+  d^2\bfk}{2(2\pi)^3} \int \frac{dk^-}{2\pi} \int \frac{dq^-}{2\pi} \frac{\sln}{2}\frac{i}{p^-+q^--\frac{(\bfp+\bfq)^2}{p^+} + \frac{i\eta}{p^+}} igt^a\frac{\SP_{\mu}^*(\bfQ_4)}{k^+(p^+-k^+)/p^+} \frac{\slnbar}{2}\nonumber\\
& \hspace*{0.4cm} \times \frac{\sln}{2}\frac{i}{p^--k^--\frac{(\bfp-\bfk)^2}{p^+-k^+}+\frac{i\eta}{p^+-k^+}}igt^c\frac{\SP_{\nu}(\bfQ_1)}{k^+(p^+-k^+)/p^+} \frac{\slnbar}{2}\frac{\sln}{2}\chi_n(p)\nonumber\\
& \hspace*{0.4cm} \times  \frac{i\sum_{\lambda}\epsilon_{\lambda}^{\mu}(k+q)\epsilon_{\lambda}^{\mu'}(k+q)}{(k+q)^2+\frac{i\eta}{k^+}} gf^{abc}A^-(\bfq)e^{iq^-z^+/2} k^+ g_{T,\mu'\nu'}  \frac{i\sum_{\theta}\epsilon_{\theta}^{\nu}(k)\epsilon_{\theta}^{\nu'}(k)}{k^2+\frac{i\eta}{k^+}} \nonumber \\
&= -g_s^3 f^{abc} t^a t^c A^-(\bfq)\Theta(-z^+) \int_0^1\frac{dx}{1-x} \int \frac{d^2\bfk}{2(2\pi)^3} \left[e^{i\omega_2z^+/2}-e^{i\Omega_3 z^+/2}\right] \nonumber\\
& \hspace*{0.4cm} \times \frac{\sum_\lambda \SP_{\lambda}^*(\bfQ_4)  \SP_{\lambda}(\bfQ_1)}{\bfQ_4^2 \bfQ_1^2 }\frac{\sln}{2}\chi_{n}(p) .
\end{align}

These diagrams computed above interfere with $M^{(0,1)}$, yielding 
\begin{align}
& \frac{1}{p^+}\int dz^+ \int\frac{d^2\bfq}{(2\pi)^2} \left\langle 2\mathfrak{Re}\left\{M^{(1,0)*}\left(M^{(1,2)}_1+M^{(1,2)}_2+M^{(1,2)}_3+M^{(1,2)}_4\right)\right\}\right\rangle_{\rm med} \nonumber \\
&= \int_{-\infty}^0 dz^+ \rho(z^+) \int\frac{d^2\bfk}{(2\pi)^3} g_s^2 P_{qq}(x, \epsilon) \rho_G(z^+)\int\frac{d^2\bfq}{(2\pi)^2} \frac{g_s^2}{\bfq^4} e^{i\bfb\cdot(\bfp+\bfq)} \nonumber\\
& \hspace*{0.4cm} \times \left\{- 2C_F\frac{1}{\bfQ_2^2}\phi_2 + (2C_F-C_A) \frac{\bfQ_2}{\bfQ_2^2}\cdot \frac{\bfQ_1}{\bfQ_1^2}\phi_2
+ C_A \frac{\bfQ_4}{\bfQ_4^2}\cdot \frac{\bfQ_1}{\bfQ_1^2} \phi_4 
\right\}  . 
\end{align}

\begin{figure}[ht!]
\centering
\includegraphics[scale=1]{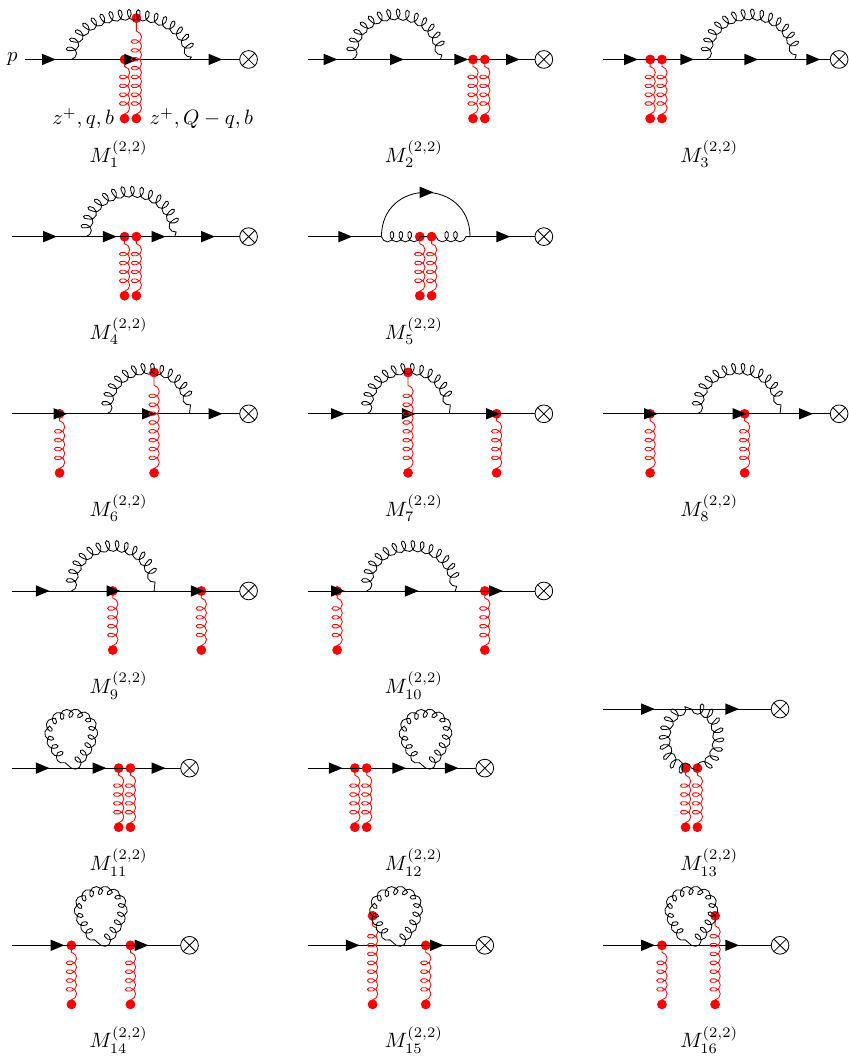}
\caption{ One-loop amplitudes with two contact Glauber interactions. Only the first five diagrams contribute. The 6-10th, 15th, and 16th diagrams are zero because the contour integration of the ``minus'' momentum component has two poles on the same side of the real axis. The 11-14th diagrams are zero due to the scaleless $\bfk$ integration.}
\label{fig:VV-all}
\end{figure}

\paragraph{Type IV: virtual correction with virtual Glauber collision.}
Lastly, we evaluate the loop diagrams with two Glauber gluon exchanges at the same space-time point.  For the first diagram, the $\sim \nbar^\mu\nbar^\nu$ part of the gluon propagator does not contribute because they are contracted with the Glauber vertex
\begin{align}
M_1^{(2,2)} &= \int\frac{dk^+d^2\bfk}{2(2\pi)^3} \int\frac{dk^-}{2\pi}\int\frac{d\ell^-}{2\pi}e^{i\ell^-z^+/2}\int\frac{dq^-}{2\pi} \frac{\sln}{2}\frac{i}{p^-+\ell^--\frac{\bfp^2}{p^+}+\frac{i\eta}{p^+}} \nonumber\\
&  \hspace*{0.4cm} \times igt^a\frac{\SP^*_\mu(\bfQ_3)}{k^+(p^+-k^+)/p^+}\frac{\slnbar}{2} \frac{\sln}{2}\frac{i}{p^-+q^--k^--\frac{(\bfp+\bfq-\bfk)^2}{p^+-k^+} + \frac{i\eta}{p^+-k^+}} \nonumber\\
& \hspace*{0.4cm} \times igt^b A^-(\bfq) \frac{\slnbar}{2}  \frac{\sln}{2}\frac{i}{p^--k^--\frac{(\bfp-\bfk)^2}{p^+-k^+} + \frac{i\eta}{p^+-k^+}}  igt^c\frac{\SP_\nu(\bfQ_1)}{k^+(p^+-k^+)/p^+}\frac{\slnbar}{2} \frac{\sln}{2}\chi_{n}(p) \nonumber\\
& \hspace*{0.4cm} \times  \frac{i\sum_\lambda \e_\lambda^\mu(k+Q-q)\e_\lambda^{\mu'}(k+Q-q)}{(k+Q-q)^2+i\eta} gf^{abc} A^{-*}(\bfq) k^+ g_{T, \mu'\nu'} \frac{i\sum_\theta \e_\theta^\nu(k)\e_\theta^{\nu'}(k)}{k^2+i\eta} \nonumber \\
&=  - i g_s^4 f^{abc} t^a t^b t^c |A^-(\bfq)|^2 \Theta(-z^+)\int_0^1\frac{dx}{1-x} \int\frac{d^2\bfk}{2(2\pi)^3}\left[1-e^{i\omega_6 z^+/2}\right] \nonumber \\ 
&\hspace*{0.4cm} \times  \frac{\sum_\lambda \SP^*_\lambda(\bfQ_3) \SP_\lambda(\bfQ_1)}{\bfQ_3^2 \bfQ_1^2} \frac{\sln}{2}\chi_{n}(p) .
\end{align}

The second diagram $M_2^{(2,2)}$ is another wave-function renormalization and involves scaleless integration when expanded in the power counting $b^2\Lambda_{\rm QCD}^2\ll 1$, so $M_2^{(2,2)}=0$. 

In the third diagram $M_3^{(2,2)}$, the $\nbar^\mu\nbar^\nu$ part of the gluon propagator will render the $\bfk$ integration scaleless after $k^-$ integration, so 
\begin{align}
M_3^{(2,2)} &= \int\frac{dk^+d^2\bfk}{2(2\pi)^3} \int\frac{dk^-}{2\pi}\int\frac{d\ell^-}{2\pi}e^{i\ell^-z^+/2}\int\frac{dq^-}{2\pi} \frac{\sln}{2}\frac{i}{p^-+\ell^--\frac{\bfp^2}{p^+}+\frac{i\eta}{p^+}} \nonumber\\
& \hspace*{0.4cm} \times  \sum_\lambda igt^a\frac{\SP^*_\lambda(\bfQ_1)}{k^+(p^+-k^+)/p^+}\frac{\slnbar}{2} \frac{\sln}{2}\frac{i}{p^-+\ell^--k^--\frac{(\bfp-\bfk)^2}{p^+-k^+} + \frac{i\eta}{p^+-k^+}} \frac{i}{k^2+i\eta} \nonumber\\
& \hspace*{0.4cm} \times  igt^a\frac{\SP_\lambda(\bfQ_1)}{k^+(p^+-k^+)/p^+}\frac{\slnbar}{2} \frac{\sln}{2}\frac{i}{p^-+\ell^--\frac{\bfp^2}{p^+}+\frac{i\eta}{p^+}} \nonumber\\
&\hspace*{0.4cm} \times   igt^b A^{-*}(\bfq) \frac{\slnbar}{2}  \frac{\sln}{2}\frac{i}{p^-+q^--\frac{(\bfp+\bfq)^2}{p^+}+\frac{i\eta}{p^+}} igt^b A^-(\bfq) \frac{\slnbar}{2}  \frac{\sln}{2}\chi_{n}(p) \nonumber  \\
&= \frac{1}{2} g_s^4 t^a t^a t^b t^b |A^-(\bfq)|^2\Theta(-z^+)\int_0^1\frac{dx}{1-x} \int\frac{d^2\bfk}{2(2\pi)^3}\nonumber\\
& \hspace*{0.4cm}\times \left[1-e^{i\omega_4 z^+/2}\right] \frac{\sum_\lambda \SP^*_\lambda(\bfQ_1) \SP_\lambda(\bfQ_1)}{\bfQ_1^2 \bfQ_1^2} \frac{\sln}{2}\chi_{n}(p) . 
\end{align}

For the fourth diagram $M_4^{(2,2)}$, the $ \nbar^\mu\nbar^\nu$ part of the gluon propagator will yield zero because after $q^-$ integration, the two poles of the $k^-$ integration are on the same side of the real axis:
\begin{align}
M_4^{(2,2)} &= \int\frac{dk^+d^2\bfk}{2(2\pi)^3} \int\frac{dk^-}{2\pi}\int\frac{d\ell^-}{2\pi}e^{i\ell^-z^+/2}\int\frac{dq^-}{2\pi} \frac{\sln}{2}\frac{i}{p^-+\ell^--\frac{\bfp^2}{p^+}+\frac{i\eta}{p^+}} \nonumber\\
& \hspace*{0.4cm} \times\sum_\lambda igt^a\frac{\SP^*_\lambda(\bfQ_1)}{k^+(p^+-k^+)/p^+}\frac{\slnbar}{2} \frac{\sln}{2}\frac{i}{p^-+\ell^--k^--\frac{(\bfp-\bfk)^2}{p^+-k^+} + \frac{i\eta}{p^+-k^+}} \nonumber\\
& \hspace*{0.4cm} \times igt^b A^{-*}(\bfq) \frac{\slnbar}{2} \frac{\sln}{2}\frac{i}{p^-+q^--k^--\frac{(\bfp-\bfk+\bfq)^2}{p^+-k^+} + \frac{i\eta}{p^+-k^+}} \nonumber\\
& \hspace*{0.4cm} \times igt^b A^-(\bfq) \frac{\slnbar}{2} \frac{\sln}{2}\frac{i}{p^--k^--\frac{(\bfp-\bfk)^2}{p^+-k^+} + \frac{i\eta}{p^+-k^+}}  igt^a\frac{\SP_\lambda(\bfQ_1)}{k^+(p^+-k^+)/p^+}\frac{\slnbar}{2}   \frac{\sln}{2}\chi_{n}(p) \frac{i}{k^2+i\eta}  \nonumber \\
&= -\frac{1}{2}g_s^4 t^a t^b t^b t^a |A^-(\bfq)|^2\Theta(-z^+)\int_0^1\frac{dx}{1-x} \int\frac{d^2\bfk}{2(2\pi)^3}\nonumber\\
& \hspace*{0.4cm}\times \left[1-e^{i\omega_4 z^+/2}\right] \frac{\sum_\lambda \SP^*_\lambda(\bfQ_1) \SP_\lambda(\bfQ_1)}{\bfQ_1^2 \bfQ_1^2} \frac{\sln}{2}\chi_{n}(p) .
\end{align}

Considering the fifth diagram $M_5^{(2,2)}$, the Glauber vertex eliminates the $\sim \nbar^\mu\nbar^\nu$ component of the gluon propagator, leaving 
\begin{align}
M_5^{(2,2)} &= \int\frac{dk^+d^2\bfk}{2(2\pi)^3} \int\frac{dk^-}{2\pi}\int\frac{d\ell^-}{2\pi}e^{i\ell^-z^+/2}\int\frac{dq^-}{2\pi} \frac{\sln}{2}\frac{i}{p^-+\ell^--\frac{\bfp^2}{p^+}+\frac{i\eta}{p^+}} \nonumber\\
& \hspace*{0.4cm} \times \sum_\lambda igt^a\frac{\SP^*_\mu(\bfQ_1)}{k^+(p^+-k^+)/p^+}\frac{\slnbar}{2} \frac{\sln}{2}\frac{i}{p^--k^--\frac{(\bfp-\bfk)^2}{p^+-k^+} + \frac{i\eta}{p^+-k^+}}  \nonumber\\
&\hspace*{0.4cm} \times igt^d\frac{\SP_{\rho'}(\bfQ_1)}{k^+(p^+-k^+)/p^+}\frac{\slnbar}{2} \frac{\sln}{2}\chi_{n}(p) \nonumber\\
& \hspace*{0.4cm} \times \frac{i\sum_\lambda \e_\lambda^{\mu}(k+Q)\e_\lambda^{\mu'}(k+Q)}{(k+Q)^2+i\eta}
gf^{abc}A^{-*}(\bfq)k^+ g_{T, \mu'\nu}
\frac{i\sum_\theta \e_\theta^{\nu}(k+q)\e_\theta^{\nu'}(k+q)}{(k+q)^2+i\eta}\nonumber\\
&\hspace*{0.4cm} \times  gf^{cbd}A^{-}(\bfq)k^+ g_{T, \nu'\rho}
\frac{i\sum_\phi \e_\phi^{\rho}(k)\e_\phi^{\rho'}(k)}{k^2+i\eta} \nonumber \\
&= \frac{1}{2}g_s^4 f^{abc}t^af^{cbd}t^d |A^-(\bfq)|^2\Theta(-z^+)\int_0^1\frac{dx}{1-x} \int\frac{d^2\bfk}{2(2\pi)^3}\nonumber\\
& \hspace*{0.4cm}\times\left[1-e^{i\omega_4 z^+/2}\right] 
 \frac{\sum_\lambda \SP^*_\lambda(\bfQ_1) \SP_\lambda(\bfQ_1)}{\bfQ_1^2 \bfQ_1^2} \frac{\sln}{2}\chi_{n}(p) .
\end{align}

\begin{figure}
    \centering
    \includegraphics[scale=1]{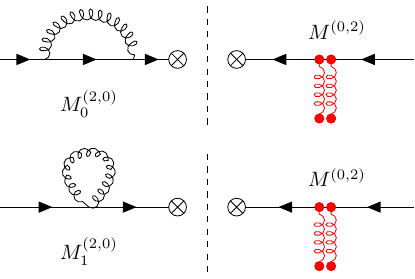}
    \caption{Intereference of wave function renormalization diagram and the double-Glauber exchange diagram.}
    \label{fig:wave-function-renormalizatino}
\end{figure}

Finally, apart from the diagrams shown in figure~\ref{fig:VV-all}, there is an additional contribution from the interference between the wave-function renormalization diagram and the double-Glauber exchange diagram, shown as figure~\ref{fig:wave-function-renormalizatino}. 
In the perturbative regime where the off-shellness of the initial-state quark is much less than the interested transverse momentum, to leading power of $p_0^2 b^2$ such diagrams are scaleless,
\begin{align}
&\left(M_1^{(2,0)*}+M_2^{(2,0)*}\right)M^{(0,2)}  = 0 . 
\end{align}

Summing up all interference terms and take the ensemble average of the medium, the final form of the Type-IV contribution is
\begin{align}
& \frac{1}{p^+}\int dz^+ \int\frac{d^2\bfq}{(2\pi)^2} \left\langle 2\mathfrak{Re}\left\{M^{(0,0)*}\sum_{i=1}^5M^{(2,2)}_i+M^{(0,2)*} M^{(2,0)}\right\} \right\rangle_{\rm med} \nonumber\\
&= \int_{-\infty}^0 dz^+ \rho(z^+) \int\frac{d^2\bfk}{(2\pi)^3} g_s^2 P_{qq}(x', \epsilon) \rho_G(z^+)\int\frac{d^2\bfq}{(2\pi)^2} \frac{g_s^2}{\bfq^4} e^{i\bfb\cdot\bfp} \nonumber\\
& \hspace*{0.4cm} \times \left\{C_A\frac{\bfQ_3}{\bfQ_3^2}\cdot\frac{\bfQ_1}{\bfQ_1^2}\phi_3
- C_A \frac{1}{\bfQ_1^2}\phi_1
\right\} .
\end{align}

\paragraph{Final results} 

Putting all type-1, type-2, type-3, and type-4 contributions together and integrating over the path length, we find 
\begin{align}
& \rho_0^-L^+ J_{q/q, R}^{(1)}\otimes \Sigma_{RT}^{(0)} \otimes \mathcal{N}_T^{(0)}  \nonumber \\
&= \int_{-\infty}^0 dz^+\rho^-(z^+) g_s^2\frac{C_F}{2\pi} P_{qq}(x)  \int\frac{d^2\bfk}{(2\pi)^3}   \int\frac{d^2\bfq}{(2\pi)^2} g_s^2\frac{C_T}{d_A}\frac{1}{(\bfq^2+\xi^2)^2} g_s^2\nonumber\\
&\hspace*{0.4cm} \times  \left\{
e^{-i\bfb\cdot(\bfp-\bfk+\bfq)} \left[
C_F\frac{1}{\bfQ_1^2} + 2C_F\frac{\bfQ_2}{\bfQ_2^2}\cdot\left(\frac{\bfQ_2}{\bfQ_2^2}-\frac{\bfQ_1}{\bfQ_1^2}\right)\phi_2 + 
\frac{C_A}{\bfQ_3^2}\right.\right.\nonumber\\
&\left.\left.\phantom{ssssssssssss} \ \ -C_A\frac{\bfQ_1}{\bfQ_1^2}\cdot\frac{\bfQ_3}{\bfQ_3^2} 
+ C_A\frac{\bfQ_2}{\bfQ_2^2}\cdot\left( \frac{\bfQ_1}{\bfQ_1^2}- \frac{\bfQ_3}{\bfQ_3^2}\right)\phi_2
\right]
\right.\nonumber\\
& \hspace*{0.2cm} \phantom{s}+\left. e^{-i\bfb\cdot(\bfp-\bfk)} \left[
-C_F\frac{1}{\bfQ_1^2} + C_A\frac{\bfQ_1}{\bfQ_1^2}\cdot\left( \frac{\bfQ_1}{\bfQ_1^2} -\frac{\bfQ_3}{\bfQ_3^2}\right)\left(\phi_1-1\right)
\right]\right\}   \nonumber\\
& \hspace*{0.4cm} + \delta(1-x) \int_{-\infty}^0 dz^+ \rho^-(z^+)\int_0^1 dx' g_s^2 \frac{C_F}{2\pi} P_{qq}(x') \int\frac{d^2\bfk}{(2\pi)^3} \int\frac{d^2\bfq}{(2\pi)^2} g_s^2\frac{C_T}{d_A}\frac{1}{(\bfq^2+\xi^2)^2}g_s^2  \nonumber\\
& \hspace*{0.4cm} \times \left\{e^{-i\bfb\cdot(\bfp+\bfq)}\left[- 2C_F\frac{1}{\bfQ_2^2}\phi_2 + (2C_F-C_A) \frac{\bfQ_2}{\bfQ_2^2}\cdot \frac{\bfQ_1}{\bfQ_1^2}\phi_2
+ C_A \frac{\bfQ_4}{\bfQ_4^2}\cdot \frac{\bfQ_1}{\bfQ_1^2} \phi_4
\right] \right.\nonumber\\
&\left.\phantom{ss}+ e^{-i\bfb\cdot\bfp} \left[C_A\frac{\bfQ_3}{\bfQ_3^2}\cdot\frac{\bfQ_1}{\bfQ_1^2}\phi_3 - C_A \frac{1}{\bfQ_1^2}\phi_1\right]\right\} .
\label{eq:NLO-collinear-bare-form}
\end{align}
To properly define this integral, we can further use dimensional regularization and insert the rapidity regulator. Upon completing the $z^+$ integral using a step function for the density $\rho^-(z^+)$, we recover Eq.~(\ref{eq:NLO-collinear-matching-coeff-full}).

\section{Separation of the collinear and rapidity divergences in $\mathcal{J}_{q/q,A}^{(1)}$}
\label{app:separation_Cnu_from_Cmu}

From Eq.~(\ref{eq:NLO-collinear-matching-coeff-full}) and table~\ref{tab:NLO-collinear}, we can write down the expression for $\mathcal{J}_{q/q,A}^{(1)}$. Taking its convolution with the leading order Glauber cross-section and anti-collinear sector and transforming to the impact parameter space yields
\begin{align}
& \mathcal{J}_{q/q,A}^{(1)}\otimes\Sigma_{AT}^{(0)} \otimes \mathcal{N}_T^{(0)} \nonumber\\
&= \frac{g_s^2 C_F}{2\pi}\left[\frac{(1-x)p^+}{\nu}\right]^{-\tau}P_{qq, \epsilon}(x)\int \frac{d^{2-2\epsilon}\bfk}{(2\pi)^{2-2\epsilon}}\int \frac{d^{2-2\epsilon}\bfq}{(2\pi)^{2-2\epsilon}} \frac{g_s^2 C_A g_s^2 C_T}{d_A q^4}\nonumber\\
&\hspace*{.4cm}\times\left\{ e^{i\bfb\cdot(\bfk-\bfq)}\left[\frac{\bfC}{\bfC^2}\cdot\left(\frac{\bfC}{\bfC^2}-\frac{\bfA}{\bfA^2}\right) + \frac{\bfB}{\bfB^2}\cdot\left(\frac{\bfA}{\bfA^2}-\frac{\bfC}{\bfC^2}\right)\Phi_B\right]\right.\nonumber\\
&\hspace*{1cm}\left.+ e^{i\bfb\cdot\bfk}\left[\frac{\bfA}{\bfA^2}\cdot\left(\frac{\bfA}{\bfA^2}-\frac{\bfC}{\bfC^2}\right)\left(\Phi_C-1\right)\right] \right\} \nonumber\\
&\hspace*{.4cm}+ \delta(1-x)\frac{g_s^2 C_F}{2\pi}\int_0^1 dx' \left[\frac{(1-x')p^+}{\nu}\right]^{-\tau}P_{qq, \epsilon}(x') \int \frac{d^{2-2\epsilon}\bfk}{(2\pi)^{2-2\epsilon}}\int \frac{d^{2-2\epsilon}\bfq}{(2\pi)^{2-2\epsilon}} \frac{g_s^2 C_A g_s^2 C_T}{d_A q^4}\nonumber\\
&\hspace*{.4cm}\times\left\{ e^{-i\bfb\cdot\bfq}\left[-\frac{\bfA'}{{\bfA'}^2}\cdot \frac{\bfB'}{{\bfB'}^2}\Phi_{B'} + \frac{\bfA'}{{\bfA'}^2}\cdot \frac{\bfD'}{{\bfD'}^2}\Phi_{D'} \right] + \left[-\frac{1}{{\bfA'}^2}\Phi_{A'} + \frac{\bfA'}{{\bfA'}^2}\cdot \frac{\bfC'}{{\bfC'}^2}\Phi_{C'}\right] \right\},
\end{align}
where the four square brackets corresponds to the contributions from the type-I to type-IV recoils demonstrated in figure~\ref{fig:four-types-of-diagrams}, respectively, which are also reflected by the argument of the phase factor.


The expression above contains both collinear and rapidity divergences.
The goal of this appendix is provide a detailed procedure to separate them for independent treatment
\begin{align}
\mathcal{J}_{q/q,A}^{(1)} = \mathcal{J}_{q/q,A}^{(1),\rm rap} + \mathcal{J}_{q/q,A}^{(1),\rm coll} + \cdots \, , 
\end{align}
up to some fixed-order terms contained in the ellipses.

To proceed, note that all the singularites are properly regulated using DR and the rapidity regulator. We can drop scaleless integrals and shift the transverse momentum integral variable for computational conveniences, arriving at a simplified result: 
\begin{align}
& \mathcal{J}_{q/q,A}^{(1)}\otimes\Sigma_{AT}^{(0)} \otimes \mathcal{N}_T^{(0)} \nonumber\\
&= \frac{g_s^2 C_F}{2\pi}\left[\frac{(1-x)p^+}{\nu}\right]^{-\tau}P_{qq, \epsilon}(x)\int \frac{d^{2-2\epsilon}\bfk}{(2\pi)^{2-2\epsilon}}\int \frac{d^{2-2\epsilon}\bfq}{(2\pi)^{2-2\epsilon}} \frac{g_s^2 C_A g_s^2 C_T}{d_A q^4}\nonumber\\
&\hspace*{.4cm}\times\left\{ e^{i\bfb\cdot(\bfk-\bfq)}\left[\frac{\bfB}{\bfB^2}\cdot\left(\frac{\bfA}{\bfA^2}-\frac{\bfC}{\bfC^2}\right)\Phi_B + \frac{\bfC}{\bfC^2}\cdot\left(\frac{\bfC}{\bfC^2}-\frac{\bfA}{\bfA^2}\right)\Phi_C\right] \right\} \nonumber\\
&+ \delta(1-x)\frac{g_s^2 C_F}{2\pi}\int_0^1 dx' \left[\frac{(1-x')p^+}{\nu}\right]^{-\tau}P_{qq, \epsilon}(x') \int \frac{d^{2-2\epsilon}\bfk}{(2\pi)^{2-2\epsilon}}\int \frac{d^{2-2\epsilon}\bfq}{(2\pi)^{2-2\epsilon}} \frac{g_s^2 C_A g_s^2 C_T}{d_A q^4}\nonumber\\
&\hspace*{.4cm}\times\left\{ e^{-i\bfb\cdot\bfq}\left[-\frac{\bfB'}{{\bfB'}^2}\cdot\left(\frac{\bfA'}{{\bfA'}^2}- \frac{\bfC'}{{\bfC'}^2}\right)\Phi_{B'}\right] + \left[-\frac{\bfC'}{{\bfC'}^2}\cdot\left(\frac{\bfC'}{{\bfC'}^2} - \frac{\bfA'}{{\bfA'}^2}\right)\Phi_{C'}\right] \right\}.
\end{align}
Specifically, we have performed the following manipulations: first shift $\bfk\rightarrow \bfk+\bfq$ and then reflect $\bfq\rightarrow -\bfq$ in the type-II contribution. This way, it can be merged with the type-I contribution.
For the type-III contribution, we shift the $\bfk$ integral for its second term such that $\bfA'\rightarrow \bfC'$ and $\bfD'\rightarrow \bfB'$. 
For the type-IV contribution, we shift the $\bfk$ integral in its first term such thaht $\bfA'\rightarrow \bfC'$. 
After these steps it is evident that the structures of the integrals for the real-gluon emissions (type-I and type-II contributions) and the virtual corrections (type-III and type-IV contributions) can be brought to a similar form.

Next, we decompose real-gluon emission terms using the plus prescription
\begin{align}
f(x) = \delta(1-x)\int_0^1 f(x') dx' + [f(x)]_+.
\end{align}
Now we can demonstrate that the delta function term combined with virtaul correction term does not contain collinear divergences but only rapidity divergences, while the plus function piece contains the remaining collinear divergence.

\paragraph{Extracting $\mathcal{J}_{q/q, A}^{(1), \rm coll}$.}
We define $\mathcal{J}_{q/q, A}^{(1), \rm coll}$ to be the piece with the plus prescription in $\mathcal{J}_{q/q, A}^{(1)}$. The ``plus'' procedure removes the rapidity divergence, so we drops the rapidity regulator from its definition
\begin{align}
&\mathcal{J}_{q/q,A}^{(1), \rm coll}\otimes\Sigma_{AT}^{(0)} \otimes \mathcal{N}_T^{(0)} \nonumber\\
& = \frac{g_s^2 C_F}{2\pi}\left[ P_{qq, \epsilon}(x)\int \frac{d^{2-2\epsilon}\bfk}{(2\pi)^{2-2\epsilon}}\int \frac{d^{2-2\epsilon}\bfq}{(2\pi)^{2-2\epsilon}} \frac{g_s^2 C_A g_s^2 C_T}{d_A q^4} \right.\nonumber\\
&\hspace*{0.4cm}\times \left.\left\{ e^{i\bfb\cdot(\bfk-\bfq)}\left[\frac{\bfB}{\bfB^2}\cdot\left(\frac{\bfA}{\bfA^2}-\frac{\bfC}{\bfC^2}\right)\Phi_B + \frac{\bfC}{\bfC^2}\cdot\left(\frac{\bfC}{\bfC^2}-\frac{\bfA}{\bfA^2}\right)\Phi_C\right] \right\}\right]_+ \, .
\end{align}
Note that the plus prescription applies on the whole expression, this is because the phase factors $\Phi_B$ and $\Phi_C$ and the kinematic variable $\bfB$ also depend on $x$. This is the expression quoted in Eq. (\ref{eq:coll-div-A}) in section~\ref{sec:TMD-NLO-matching:collinear-div}.

\paragraph{Extracting $\mathcal{J}_{q/q, A}^{(1), \rm rap}$.}
The summation of the delta-function piece of the real-emission term and the virtual correction is
\begin{align}
&\mathcal{J}_{q/q,A}^{(1)}\otimes\Sigma_{AT}^{(0)} \otimes \mathcal{N}_T^{(0)} \nonumber\\
&\supset \delta(1-x)  \frac{g_s^2 C_F}{2\pi}\int \frac{d^{2-2\epsilon}\bfk}{(2\pi)^{2-2\epsilon}}\int \frac{d^{2-2\epsilon}\bfq}{(2\pi)^{2-2\epsilon}} \frac{g_s^2 C_A g_s^2 C_T}{d_A q^4}\nonumber\\
&\hspace*{0.4cm}\times\left\{  \left(e^{i\bfb\cdot(\bfk-\bfq)}-e^{-i\bfb\cdot\bfq}\right) \left[\frac{p^+}{\nu}\right]^{-\tau}\int_0^1 dx' \frac{P_{qq,\epsilon}(x')}{(1-x')^{\tau}}\frac{\bfB'}{{\bfB'}^2}\cdot\left(\frac{\bfA'}{{\bfA'}^2}- \frac{\bfC'}{{\bfC'}^2}\right)\Phi_{B'}\right. \nonumber\\
&\hspace*{0.4cm}\left. + \left(e^{i\bfb\cdot(\bfk-\bfq)}-1\right)\left[\frac{p^+}{\nu}\right]^{-\tau}\int_0^1 dx' \frac{P_{qq, \epsilon}(x')}{(1-x')^{\tau}}\frac{\bfC'}{{\bfC'}^2}\cdot\left(\frac{\bfC'}{{\bfC'}^2} - \frac{\bfA'}{{\bfA'}^2}\right)\Phi_{C'} \right\}.
\label{eq:Jqq_A_delta_piece}
\end{align}
First, it does not contain collinear divergences as those appears in $\mathcal{J}_{q/q, A}^{(1), \rm coll}$. When we take $b\rightarrow 0$ limit, the expression vanishes to leading power in $b$. Still, it  contains soft divergences.

For conveniences, we write down the path-length averaged LPM phase factors
\begin{align}
\Phi_{B'} &= 1-\frac{\sin \frac{\bfB'^2 }{2x'(1-x')p^+/L^+}}{\frac{\bfB'^2 }{2x'(1-x')p^+/L^+}} \\
\Phi_{C'}  &= 1-\frac{\sin \frac{\bfC'^2}{2x'(1-x')p^+/L^+}}{\frac{\bfC'^2}{2x'(1-x')p^+/L^+}}.
\end{align}
The presence of the phase factor modifies the rapidity logarithm. When $p^+/L^+$ is smaller than any transverse momentum scales, to leading power the phase factor is unity $\Phi\approx 1$, and the rapidity logarithm is the same as that in the vacuum with a CS scale $\sqrt{\zeta_1}=p_1^+$. 
However, as is noted in Ref.~\cite{Vaidya:2021vxu}, the full phase factor imposes a maximum cut off to the CS scale. This happens when $p^+/L^+$ is much greater than the transverse momentum scale. Then, for a large range of $x'$, one can peform the expansion $\Phi_{\bfB'}\approx \frac{1}{6}\left[\frac{\bfB'^2 }{2x'(1-x')p^+/L^+}\right]^2$. This qualitatively modifies the soft behavior near $x'=1$ and destroy the rapidity logarithm for this region of $x'$.
We now demonstrate this argument mathematically using the first term in Eq. (\ref{eq:Jqq_A_delta_piece}) as an example. 
\begin{align}
&\left[\frac{p^+}{\nu}\right]^{-\tau}\int_0^1 dx' \frac{1+{x'}^2-\epsilon(1-x')^2}{(1-x')^{1+\tau}} \frac{\bfB'}{{\bfB'}^2}\cdot\left(\frac{\bfA'}{{\bfA'}^2}-\frac{\bfC'}{{\bfC'}^2}\right)\Phi\left(\frac{{\bfB'}^2 L^+}{2x(1-x)p^+}\right) \nonumber\\
& \qquad \qquad = \left(\frac{1}{\bfk^2}-\frac{\bfk\cdot(\bfk-\bfq)}{\bfk^2(\bfk-\bfq)^2}\right) \left[\frac{p^+}{\nu}\right]^{-\tau}\int_0^1 \frac{2\Phi\left(\frac{\bfk^2 L^+}{2(1-x')p^+}\right)}{(1-x')^{1+\tau}}  dx' +\cdots   \, .
\end{align}
In the second line, we have separate the function under the $x'$ integral using its asymptotic form in the $x'\rightarrow 1$ limit and the residual part is denoted by the ellipses. Because the residual part is the differences between the original function and its $x'\rightarrow 1$ asymptotic form, it does not contain rapidity divergences. Focusing only on the asymptotic term and performing a change of varaibles
\begin{align}
& \left[\frac{p^+}{\nu}\right]^{-\tau}\int_0^1 \frac{2\Phi\left(\frac{\bfk^2 L^+}{2(1-x')p^+}\right)}{(1-x')^{1+\tau}}  dx' = 2\left[\frac{\bfk^2 L^+}{2\nu}\right]^{-\tau}  \int_0^{\frac{2p^+/L^+}{\bfk^2}}\frac{1-u \sin(1/u)}{u^{1+\tau}} du  \nonumber \\
&\qquad \qquad =\begin{cases}
 \left(-\frac{2}{\tau}\right)\left[\frac{e^{\gamma_E-1}\bfk^2 L^+}{2\nu}\right]^{-\tau} + \mathcal{O}\left(\left(\frac{\bfk^2}{2p^+/L^+}\right)^2\right) + \mathcal{O}(\tau),~~2p^+/L^+\gg \bfk^2\\
  \left(-\frac{2}{\tau}\right)\left[\frac{p^+}{\nu}\right]^{-\tau} + \mathcal{O}\left(\frac{2p^+/L^+}{\bfk^2}\right) + \mathcal{O}(\tau) ,~~2p^+/L^+\ll \bfk^2
\end{cases}\, .
\end{align}
This expression gives the expected behavior from the qualitative argument, i.e., the CS scale is $\sqrt{\zeta_1}=p^+$ for large transverse momenta but is restricted by a scale of order $\bfk^2 L^+$ for small transverse momenta.
A simple prescription to capture both scenario and to provide an approximate interpolation is to use $\min\left\{p^+, \frac{e^{\gamma_E-1}\bfk^2 L^+}{2}\right\}$ as the CS scale for the first order in opacity calculation. Using this prescription, we can finally extract the rapidity divergent piece in $\mathcal{J}_{q/q,A}^{(1)}$, 
\begin{align}
&\mathcal{J}_{q/q,A}^{(1),\rm rap}\otimes\Sigma_{AT}^{(0)} \otimes \mathcal{N}_T^{(0)}  = \delta(1-x)  \frac{g_s^2 C_F}{2\pi}\int \frac{d^{2-2\epsilon}\bfk}{(2\pi)^{2-2\epsilon}}\int \frac{d^{2-2\epsilon}\bfq}{(2\pi)^{2-2\epsilon}} \frac{g_s^2 C_A g_s^2 C_T}{d_A q^4}\nonumber\\
&\qquad ~~ \times \left(-\frac{2}{\tau}\right)\left\{  \left(e^{i\bfb\cdot(\bfk-\bfq)}-e^{-i\bfb\cdot\bfq}\right) \left[\frac{\min\left\{p^+, e^{\gamma_E-1}\bfk^2 L^+/2\right\}}{\nu}\right]^{-\tau}\left(\frac{1}{{\bfk}^2}- \frac{\bfk\cdot(\bfk-\bfq)}{\bfk^2(\bfk-\bfq)^2}\right)\right. \nonumber\\
&\qquad ~~\left. + \left(e^{i\bfb\cdot(\bfk-\bfq)}-1\right)\left[\frac{\min\left\{p^+, e^{\gamma_E-1}(\bfk-\bfq)^2 L^+/2\right\}}{\nu}\right]^{-\tau}\left(\frac{1}{(\bfk-\bfq)^2} - \frac{\bfk\cdot(\bfk-\bfq)}{\bfk^2(\bfk-\bfq)^2}\right) \right\}.
\end{align}
This is the result quoted in Eq. (\ref{eq:Arap}) in section~\ref{sec:TMD-NLO-matching:rapidity-div}.

\section{NLO correction to the Glauber cross-section $\Sigma^{(1)}$}\label{app:soft}

\begin{figure}[ht!]
\centering
\includegraphics[scale=.85]{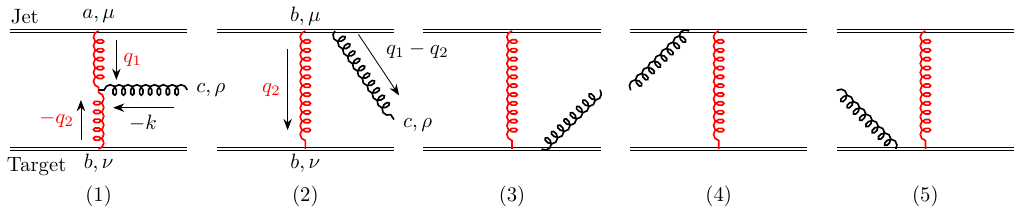}
\caption{Diagrams for real soft gluon emission due to Glauber interaction.}
\label{fig:Glauber-soft-real}
\end{figure}

\begin{figure}[ht!]
\centering
\includegraphics[scale=.9]{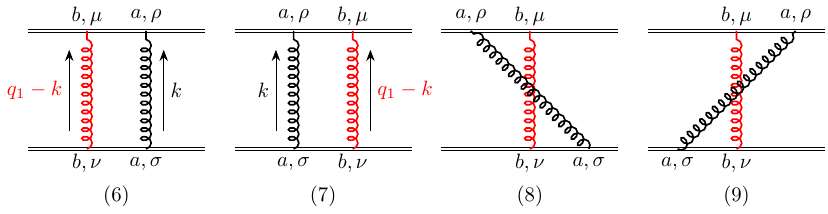}
\caption{virtual corrections that contain rapidity divergence.}
\label{fig:Glauber-soft-virtual-1}
\end{figure}

\begin{figure}[ht!]
\centering
\includegraphics[scale=.9]{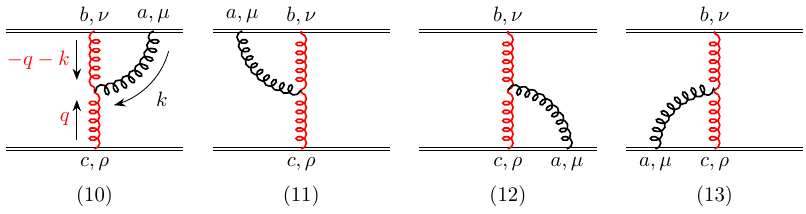}
\caption{
Virtual corrections that do not contain rapidity divergence: vertex correction.}
\label{fig:Glauber-soft-virtual-2}
\end{figure}

\begin{figure}[ht!]
\centering
\includegraphics[scale=.9]{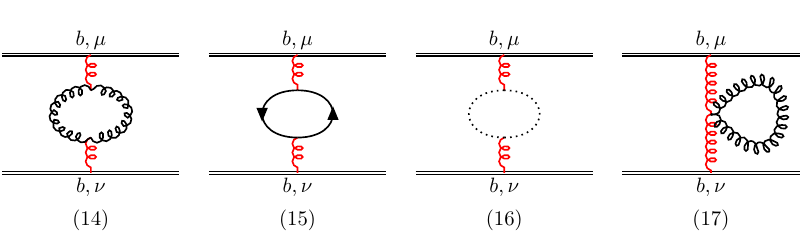}
\caption{
Virtual corrections that do not contain rapidity divergence: soft gluon contribution to the self-energy of the Glauber gluon.
}
\label{fig:Glauber-soft-virtual-3}
\end{figure}

We now turn to the question of how soft radiation can affect the Glauber gluon exchange cross section. The diagrams are shown in figures~\ref{fig:Glauber-soft-real}, \ref{fig:Glauber-soft-virtual-1}, \ref{fig:Glauber-soft-virtual-2}, and \ref{fig:Glauber-soft-virtual-3}.

\paragraph{Real emission diagrams.}
The real emission diagrams are shown in figure~\ref{fig:Glauber-soft-real}.
Make use of the following power counting. The scaling of the momentum of the Glauber gluons are determined as follows: the jet $j$ momentum retains the scaling $p_j^\mu\sim (1, \lambda^2, \lambda)$. The target $t$ momentum retains the scaling $p_t^\mu\sim (\lambda^2, 1, \lambda)$. The radiated gluon is soft, $k^\mu\sim (\lambda, \lambda, \lambda)$. Let $q_1$ be the total momentum taken away from the jet and $q_2$ to be the total momentum flowing into the target. From the power counting, we can deduce that $q_1^\mu \sim (\lambda, \lambda^2, \lambda)$ and $q_2^\mu \sim (\lambda^2, \lambda, \lambda)$.

For diagram $(1)$, we expand the propagator and the three-gluon vertex consistently with the outlined power counting.
The amplitude (before contracting with the gluon polarization vector $\epsilon_\lambda^\rho(k)$) is
\begin{align}
M_1^\rho &= J^a_n \frac{-gn^\mu}{q_1^2} J^b_{\bar{n}} \frac{-g\bar{n}^\nu}{q_2^2} g f^{abc} \left[g^{\mu\nu}(q_1+q_2)^\rho + g^{\nu\rho}(-q_2+k)^\mu + g^{\rho\mu}(-k-q_1)^\nu\right] \nonumber \\
&= g_s^2 J^a_n\frac{1}{\bfq_1^2}J^b_{\bar{n}} \frac{1}{\bfq_2^2} g f^{abc} \left[2(q_1+q_2)^\rho + \bar{n}^{\rho}(q_1-2q_2)^- + n^\rho(q_2-2q_1)^+\right] + \mathcal{O(\lambda)}  \nonumber \\
&= g_s^2 J^a_n\frac{1}{\bfq_1^2} J^b_{\bar{n}} \frac{1}{\bfq_2^2} 2g f^{abc} \left[q_1^\rho + q_2^\rho - \bar{n}^{\rho} q_2^-  - n^\rho q_1^+\right] + \mathcal{O(\lambda)} \nonumber \\
&= 2g_s^2 J^a_n J^b_{\bar{n}} \frac{1}{\bfq_1^2}  \frac{1}{\bfq_2^2} g f^{abc} \left[\bfq_1^\rho + \bfq_2^\rho - \frac{\bar{n}^{\rho} }{2}n\cdot q_2  - \frac{n^\rho }{2}\bar{n}\cdot q_1\right] + \mathcal{O(\lambda)},
\end{align}
where one also uses $k=q_1-q_2$ and only retains the leading contribution in $\lambda$. 
The large components of the jet and target currents are
\begin{align}
J^a_n = \bar{\xi}_nt^a\frac{\slashed{\bar{n}}}{2}\xi_n, \qquad
J^b_{\bar{n}} = \bar{\xi}_{\bar{n}}t^b\frac{\slashed{n}}{2}\xi_{\bar{n}},
\end{align}
consistent with Eq. (16) of Ref.~\cite{Fleming:2014rea}.
The other four diagrams have a very similar structure. It is convenient to present the sum $(2)+(4)$, where
\begin{align}
M_2^\rho &= \left[ig\bar{\xi}_{\bar{n}}t^b \frac{\slashed{n}} {2}\xi_{\bar{n}} \bar{n}_{\mu}\right]\left[ig\bar{\xi}_{n} t^c t^b \frac{\slashed{\bar{n}}} {2}\xi_{n} n_{\nu}\right] \frac{ig^{\mu\nu}}{\bfq_2^2} \frac{-g n^\rho}{n\cdot(q_1-q_2)} , \\
M_4^\rho &= \left[ig\bar{\xi}_{\bar{n}}t^b \frac{\slashed{n}} {2}\xi_{\bar{n}} \bar{n}_{\mu}\right]\left[ig\bar{\xi}_{n} t^b t^c \frac{\slashed{\bar{n}}} {2}\xi_{n} n_{\nu}\right] \frac{ig^{\mu\nu}}{\bfq_2^2} \frac{-g n^\rho}{-n\cdot(q_1-q_2)}
. \end{align}
Summing the two and using $[t^c, t^b] = if^{cbd} t^d$ gives
\begin{align}
M_2^\rho+M_4^\rho &= 2g_s^2\left[\bar{\xi}_{\bar{n}}t^b \frac{\slashed{n}} {2}\xi_{\bar{n}}\right]\left[\bar{\xi}_{n} t^d \frac{\slashed{\bar{n}}} {2}\xi_{n} \right] igf^{cbd} \frac{1}{\bfq_2^2} \frac{ n^\rho}{n\cdot(q_1-q_2)} \nonumber \\
&= 2g_s^2 J^d_{n} J^b_{\bar{n}}gf^{cbd}\frac{1}{\bfq_2^2} \frac{ n^\rho}{n\cdot q_2} = -2g_s^2 J^a_{n} J^b_{\bar{n}} gf^{abc}\frac{1}{\bfq_2^2} \frac{ n^\rho}{n\cdot q_2}, 
\end{align}
where we have relabeled the dummy indices such that the structure constants have indices $abc$ in the last equation.
Similarly, the sum of $(3)$ and $(5)$ is
\begin{align}
M_3^\rho+M_5^\rho &= -2g_s^2 J^a_{n} J^b_{\bar{n}} gf^{abc}\frac{1}{\bfq_1^2} \frac{\bar{n}^\rho}{\bar{n}\cdot q_1} .
\end{align}
Combining $(2), (3), (4), (5)$ gives
\begin{align}
\sum_{i=2}^5M_i^\rho = 2g_s^2 J^a_{n} J^b_{\bar{n}} gf^{abc} \frac{1}{\bfq_1^2}\frac{1}{\bfq_2^2} \left[ \frac{n^\rho \bfq_1^2}{n\cdot q_2} + \frac{\bar{n}^\rho \bfq_2^2}{\bar{n}\cdot q_1}\right],
\end{align}
also in agreement with Eq. (12) of Ref.~\cite{Fleming:2014rea}.
Finally, adding diagram $(1)$ one arrives at the effective vertex
\begin{align}
\sum_{i=1}^5M_i^\rho = 2g_s^2\frac{1}{\bfq_1^2}\frac{1}{\bfq_2^2} J^a_{n}J^b_{\bar{n}} gf^{abc}\left[\bfq_1^\rho + \bfq_2^\rho - \frac{\bar{n}^{\rho} }{2}n\cdot q_2  - \frac{n^\rho }{2}\bar{n}\cdot q_1-\frac{n^\rho \bfq_1^2}{n\cdot q_2} - \frac{\bar{n}^\rho \bfq_2^2}{\bar{n}\cdot q_1}\right],
\end{align}
and this is the Lipatov vertex for soft gluon emission.
Real emission contribution to the jet parton TMD distribution in the impact parameter space ($e^{i\bfb\cdot \bfq_1}$).
The squared amplitude (summing over polarization and color, and averaging over initial-state color and polarizations) is
\begin{align}
\sum_{i,j=1}^5(M_i^\rho)^* M_{j, \rho} &= -4g_s^4 \frac{1}{\bfq_1^4 \bfq_2^4} \frac{1}{d_j}J_n^{a*}J_n^{r} \frac{1}{d_t} J_{\bar{n}}^{b*}J_{\bar{n}}^{s} g_s^2 f^{abc} f^{rsc} \nonumber\\
& \hspace*{0.4cm}\times \left[-(\bfq_1 + \bfq_2)^2 +  q_2^- q_1^+ + \frac{4\bfq_1^2\bfq_2^2}{q_2^- q_1^+} + 2\bfq_1^2 + 2\bfq_2^2\right].
\end{align}
Taking into account  that $0 = k^2 = (q_1-q_2)^2 = (q_1^+-q_2^+)(q_1^--q_2^-) - (\bfq_1-\bfq_2)^2 = -q_1^+q_2^- - (\bfq_1-\bfq_2)^2$, such that $q_1^+q_2^- = - (\bfq_1-\bfq_2)^2$, 
we find
\begin{align}
\sum_{i,j=1}^5(M_i^\rho)^* M_{j, \rho} &= 4g_s^4 \frac{1}{\bfq_1^4 \bfq_2^4} \frac{1}{d_j}J_n^{a*}J_n^{r} \frac{1}{d_t} J_{\bar{n}}^{b*}J_{\bar{n}}^{s} g_s^2 f^{abc} f^{rsc}\frac{4\bfq_1^2\bfq_2^2}{(\bfq_1-\bfq_2)^2}  \nonumber \\
&=  4g_s^4  \frac{1}{d_j}J_n^{a*}J_n^{r} \frac{1}{d_t} J_{\bar{n}}^{b*}J_{\bar{n}}^{s} 4g_s^2 f^{abc} f^{rsc} \frac{1}{\bfq_1^2 \bfq_2^2} \frac{1}{(\bfq_1-\bfq_2)^2}.
\end{align}

Going to impact parameter space with $\bfk=\bfq_1-\bfq_2$
\begin{align}
S_{R}(b) &= \frac{J_n^{a*}J_n^{r} }{d_j}\frac{J_{\bar{n}}^{b*}J_{\bar{n}}^{s}}{d_t}  16g_s^4 f^{abc} f^{rsc}  \nonumber\\
& \hspace*{0.4cm}\times \int\frac{d^2\bfq_1}{(2\pi)^2}  \frac{e^{i\bfq_1\cdot \bfb}}{\bfq_1^2} g_s^2\int\frac{d^2\bfk} {(2\pi)^2}\frac{1}{\bfk^2(\bfq_1-\bfk)^2}\int_0^{\infty}\frac{1}{2(2\pi)}\frac{dk^+}{ k^+} \left|\frac{k_z}{\nu}\right|^{-\tau}.
\label{eq:softImpact}
\end{align}
The integral with the rapidity regulator gives
\begin{align}
\int_0^\infty \frac{dk^+}{k^+} \left|\frac{k^+-\frac{\bfk^2}{k^+}}{2\nu}\right|^{-\tau} = \left(\frac{|\bfk|}{2\nu}\right)^{-\tau} \mathrm{B}\left(1-\tau, \frac{\tau}{2}\right) = \left(\frac{|\bfk|}{\nu}\right)^{-\tau} \frac{\Gamma\left(\frac{1}{2}-\frac{\tau}{2}\right)\Gamma\left(\frac{\tau}{2}\right)}{\sqrt{\pi}} ,
\end{align}
and substituting in Eq.~(\ref{eq:softImpact}) yields
\begin{align}
S_{R}(b) &= \frac{J_n^{a*}J_n^{r} }{d_j}\frac{J_{\bar{n}}^{b*}J_{\bar{n}}^{s}}{d_t}  16g_s^4 f^{abc} f^{rsc}  \nonumber\\
& \hspace*{0.4cm}\times\int\frac{d^2\bfq_1}{(2\pi)^2} \frac{e^{i\bfq_1\cdot \bfb}}{\bfq_1^2} \frac{\Gamma\left(\frac{1}{2}-\frac{\tau}{2}\right)\Gamma\left(\frac{\tau}{2}\right)}{\sqrt{\pi}}\nu^\tau g_s^2\int\frac{d^2\bfk} {2(2\pi)^3}\frac{1}{\bfk^{2+\tau}(\bfq_1-\bfk)^2} \, .
\end{align}

\paragraph{Virtual correction with rapidity divergence.}

Diagrams in figure~\ref{fig:Glauber-soft-virtual-1} contains the rapidity divergence and are of particular relevance to the TMD observable under consideration.
Diagram $(6)$ reads
\begin{align}
M_6 &= \left[ig\bar{\xi}_{\bar{n}}t^a t^b \frac{\slashed{n}} {2}\xi_{\bar{n}} \bar{n}_{\mu}\right]_T\left[ig\bar{\xi}_{n} t^a t^b \frac{\slashed{\bar{n}}} {2}\xi_{n} n_{\nu}\right]_J\nonumber\\
&\hspace*{0.4cm}\times \int \frac{dk^+dk^-d^2\bfk}{2(2\pi)^4}\frac{ig^{\mu\nu}}{(\bfq-\bfk)^2} \frac{-g\nbar_\sigma}{\nbar\cdot k+i\eta}\frac{-gn_\rho}{-n\cdot k+i\eta} \frac{-ig^{\rho\sigma}}{k^2+i\eta} \left|\frac{k_z}{\nu}\right|^{-\tau} \nonumber \\
&= 4g_s^2\left[\bar{\xi}_{\bar{n}}t^a t^b \frac{\slashed{n}} {2}\xi_{\bar{n}}\right]_T \left[\bar{\xi}_{n} t^a t^b \frac{\slashed{\bar{n}}} {2}\xi_{n}\right]_J g_s^2\int \frac{d^2\bfk}{(2\pi)^3}\frac{1}{(\bfq-\bfk)^2} \int_{-\infty}^\infty dk_z \left|\frac{k_z}{\nu}\right|^{-\tau}\nonumber\\
& \hspace*{0.4cm}\times \frac{1}{2\pi} \int_{-\infty}^\infty d\omega \frac{1}{\omega+k_z+i\eta}\frac{1}{\omega-k_z-i\eta} \frac{1}{\omega-|k|+i\eta} \frac{1}{\omega+|k|-i\eta} 
\end{align}
with $|k|=\sqrt{\bfk^2+k_z^2}$. The gluon energy ($\omega$) and longitudinal momentum ($k_z$) integrals can be completed as
\begin{align}
&\int_{-\infty}^\infty dk_z \left|\frac{k_z}{\nu}\right|^{-\tau}\frac{1}{2\pi} \int_{-\infty}^\infty d\omega \frac{1}{\omega+k_z+i\eta}\frac{1}{\omega-k_z-i\eta} \frac{1}{\omega-|k|+i\eta} \frac{1}{\omega+|k|-i\eta} \nonumber\\
&=\int_{-\infty}^\infty dk_z \left|\frac{k_z}{\nu}\right|^{-\tau} (2i)\frac{1}{2|k|}\frac{1}{2k_z}\frac{1}{k_z-|k|} \nonumber \\
&=-2i\int_{0}^\infty dk_z \left|\frac{k_z}{\nu}\right|^{-\tau} \frac{1}{2|k|}\frac{1}{\bfk^2} = -i \frac{\nu^\tau}{\bfk^{2+\tau}} \frac{\Gamma\left(\frac{1}{2}-\frac{\tau}{2}\right)\Gamma\left(\frac{\tau}{2}\right)}{2\sqrt{\pi}}.
\end{align}
The other three diagrams only differ in the order of color indices and the sign of the Wilson-line propagator. Summing over diagrams (6),(7),(8),(9) we find 
\begin{align}
\sum_{i=6}^9 M_{i} &= 4i g_s^2 f^{abc} f^{abd} J_{\nbar}^c J_n^d  \frac{\Gamma\left(\frac{1}{2}-\frac{\tau}{2}\right)\Gamma\left(\frac{\tau}{2}\right)}{\sqrt{\pi}} g_s^2 \int \frac{d^2\bfk}{2(2\pi)^3}\frac{\nu^\tau}{\bfk^{2+\tau}(\bfq-\bfk)^2} .
\end{align}
Their interference with the single-Glauber exchange diagram ($M_s$) gives the virtual correction to the cross section
\begin{align}
\frac{1}{2}\times 2\mathfrak{Re}\left\{M_s^* \sum_{i=6}^9 M_{i} \right\} &=  -8g_s^4 f^{abc} f^{abd} \frac{J_{\nbar}^{r*} J_{\nbar}^c}{d_t} \frac{J_n^{r*} J_n^d}{d_j} \frac{\Gamma\left(\frac{1}{2}-\frac{\tau}{2}\right)\Gamma\left(\frac{\tau}{2}\right)}{\sqrt{\pi}} \nonumber\\
&\hspace*{0.4cm}\times  g_s^2 \frac{1}{\bfq^2}\int \frac{d^2\bfk}{2(2\pi)^3}\frac{\nu^\tau}{\bfk^{2+\tau}(\bfq-\bfk)^2}.
\end{align}
According to Ref.~\cite{Fleming:2014rea}, a symmetry factor of $1/2$ is multiplied in the end. This is because diagrams E and G are symmetric to diagrams F and H in the original theory. Its Fourier transformation to the impact parameter space defines $S_V(b)$.

\paragraph{Virtual correction without rapidity divergence}

Diagrams in figure~\ref{fig:Glauber-soft-virtual-2} are soft gluon corrections to the Glauber vertex. They do not contain rapidity divergence. We now demonstrate this point by calculating diagram (10) and the calculations for the other three diagrams are similar.
\begin{align}
M_{10} &= \int \frac{d^4 k}{(2\pi)^4}g_s f^{abc} \left[g^{\mu'\nu'}(2k+q)^{\rho'}+g^{\nu'\rho'}(-2q-k)^\mu+g^{\rho'\mu'}(q-k)^{\nu'}\right] \nonumber\\
& \times \left[\bar{\xi}_n t^a\frac{\slnbar}{2}ig_s t^b \xi_n n^\mu\right] \left[\bar{\xi}_{\nbar} ig_s t^c\frac{\sln}{2} \xi_{\nbar} \nbar^\nu\right] \frac{-g_s n^\rho}{n\cdot k + i\eta} \frac{-ig_{\nu\nu'}}{-(\bfk+\bfq)^2} \frac{-ig_{\rho\rho'}}{-\bfq^2} \frac{-i g_{\rho\rho'}}{k^2+i\eta} \\
&=  (-4i)g_s^4 f^{abc}
\left[\bar{\xi}_n t^a t^b\frac{\slnbar}{2}  \xi_n\right]\left[\bar{\xi}_{\nbar}t^c\frac{\sln}{2} \xi_{\nbar}\right] \frac{1}{\bfq^2}\int \frac{d^2 \bfk}{(2\pi)^3}\frac{1}{(\bfk+\bfq)^2}\nonumber\\
& \times \int_{-\infty}^\infty dk_z \int_{-\infty}^\infty d\omega\frac{1}{k^2+i\eta} \left|\frac{k_z}{\nu}\right|^{-\tau}
\end{align}
where we have introduced the rapidity regulator in the second equation and put the transverse integral into $2-2\epsilon$ dimensions. If there is no rapidity divergence and the coefficient of $1/\tau$ term of $M_{10}$ is zero. The final result is
\begin{align}
M_{10} &= -4g_s^4 f^{abc}
\left[\bar{\xi}_n t^a t^b\frac{\slnbar}{2}  \xi_n\right]\left[\bar{\xi}_{\nbar}t^c\frac{\sln}{2} \xi_{\nbar}\right] \frac{1}{\bfq^2} \nonumber\\
&\times \frac{\Gamma\left(\frac{1}{2}-\frac{\tau}{2}\right)\Gamma\left(\frac{\tau}{2}\right)}{\sqrt{\pi}} \int \frac{d^{2-2\epsilon} \bfk}{2(2\pi)^3}\frac{1}{(\bfk+\bfq)^2} \left(\frac{\nu}{|\bfk|}\right)^\tau
\end{align}
The $\Gamma\left(\frac{\tau}{2}\right)$ has a simple pole in $1/\tau$, but the $\bfk$ integral in the $\tau=0$ case is scaleless. Therefore, such diagrams do not display rapidity divergence.

Finally, the virtual diagrams in figure~\ref{fig:Glauber-soft-virtual-3}. Because the soft gluon Lagrangian is just a scaled-down version of the full gluon sector of the QCD Lagrangian, these diagrams just correspond to one-loop soft gluon correction to the self-energy of the Glauber gluon, which does not contain rapidity divergence. Scale divergences in the vertex correction and self-energy correction are canceled by the soft Lagrangian.

\paragraph{The soft contribution to the cross section.}
We sum over the final-state and average over the initial-state color and spin, then perform the Fourier transform to compute the virtual correction soft factor $S_V(b)$. The summation of real and virtual correction, multiplied by the area density of color sources, produces
\begin{align}
S(b) &= S_R(b)+S_V(b)  \nonumber\\
&= 4g_s^4 \frac{C_F^2 C_A}{d_A}\frac{\Gamma\left(\frac{1}{2}-\frac{\tau}{2}\right)\Gamma\left(\frac{\tau}{2}\right)}{\sqrt{\pi}} \left(\frac{\nu}{\mu_b}\right)^\tau\ g_s^2 \int\frac{d^2 \bfq}{(2\pi)^2}\frac{e^{i\bfq\cdot\bfb}}{\bfq^2}   \nonumber\\
& \hspace*{.4cm} \times \int \frac{d^2\bfk}{2(2\pi)^3}\left[\frac{\mu_b^\tau}{(\bfq-\bfk)^{2+\tau}}\frac{1}{\bfk^2} - \frac{1}{2}\frac{\mu_b^\tau\bfq^2}{\bfk^{2+\tau}(\bfq-\bfk)^2}\frac{1}{\bfq^2}  \right] \nonumber \\
&=  \frac{(g_s^2 C_F)^2 }{d_A}\left(\frac{2}{\tau}+\ln \frac{\nu^2}{\mu_b^2} + 2\ln 2 + \mathcal{O}(\tau)\right)\ \int\frac{d^2 \bfq}{(2\pi)^2}\frac{e^{i\bfq\cdot\bfb}}{\bfq^2} \left(\mathcal{C}\left[\frac{1}{\bfq^2}\right]+\mathcal{O}(\tau)\right)
 \label{app:eq:rapidity-div-2}
\end{align}

\section{NLO correction to the anti-collinear sector at  first order in opacity}
\label{app:anticollinear}

For the NLO correction to the (anti-collinear) medium color source, we only need to calculate the real-Glauber exchange. This is because we do not have the additional LPM scale. After the power expansion, both the initial and final-sate medium color source parton are taken to be on shell, and the only external scale after power expansion is the impact parameter $b$ of the collinear parton. Only the real Glauber exchange introduces the phase factor $e^{i\bfb\cdot \bfq}$, and the double-Glauber exchange diagrams are scaleless\footnote{See for example~\cite{Vitev:2007ve} for on-shell partons and focus on the first order in opacity.}.

\begin{figure}
    \centering
    \includegraphics[scale=1]{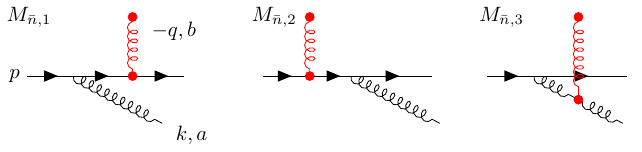}
    \caption{Real emission diagrams with a single Glauber exchange to the anti-collinear sector.}
    \label{fig:anti-collinear-real-emission}
\end{figure}
\paragraph{Real emission diagrams with single Glauber exchange.} 
The diagrams for anti-collinear quarks are shown in figure~\ref{fig:anti-collinear-real-emission}.
Their amplitudes are
\begin{align}
M_{\nbar, 1} = ig_s^2 t^bt^a \chibar_{\nbar,p}\frac{\SP_{\nbar,\lambda}(\bfQ_1)}{\bfQ_1^2}\frac{\slnbar}{2}\chi_{\nbar,p} 
, , 
\end{align}
where $\bfQ_1 = x\bfk-(1-x)(\bfp-\bfk)$.
The quark scattering in the final state gives
\begin{align}
M_{\nbar, 2} = -ig_s^2 t^at^b\chibar_{\nbar,p}\frac{\SP_{\nbar,\lambda}(\bfQ_6)}{\bfQ_6^2}\frac{\slnbar}{2}\chi_{\nbar,p} \, , 
\end{align}
where $\bfQ_6 = x\bfk-(1-x)(\bfp-\bfk-\bfq)$.
Finally, the gluon scattering
\begin{align}
M_{\nbar, 3} &= -g_s^2f^{abc}t^c\chibar_{\nbar,p}\frac{\SP_{\nbar,\lambda}(\bfQ_8)}{\bfQ_8^2}\frac{\slnbar}{2}\chi_{\nbar,p} = ig_s^2(t^at^b-t^bt^a)\chibar_{\nbar,p}\frac{\SP_{\nbar,\lambda}(\bfQ_8)}{\bfQ_8^2}\frac{\slnbar}{2}\chi_{\nbar,p} \, , 
\end{align}
where $\bfQ_8 = x(\bfk+\bfq)-(1-x)(\bfp-\bfk-\bfq)$.

\paragraph{Virtual correction diagrams with single Glauber exchange.} 
\begin{figure}
    \centering
    \includegraphics[scale=1]{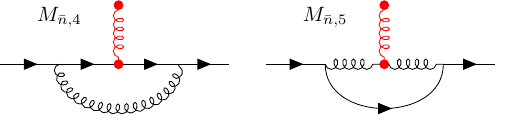}
    \caption{Virtual diagrams with a single Glauber exchange to the anti-collinear sector.}
    \label{fig:anti-collinear-virtual-emission}
\end{figure}
There are two non-vanishing diagrams as shown in figure~\ref{fig:anti-collinear-virtual-emission}. For quark scattering we have
\begin{align}
M_{\bar{n},4} &= \int \frac{dk^-dk^+d^2\bfk}{2(2\pi)^4} \frac{i}{k^2+i\eta}\left[\sum_\lambda\epsilon_\lambda^\mu\epsilon_\lambda^\nu-\frac{4k^2}{(k^-)^2}n^\mu n^\nu\right] \nonumber\\
& \hspace*{0.4cm}\times  \chibar_{\nbar,p} \frac{igt^a\SP_{\nbar,\mu}^*(\bfQ_6)}{k^-(p^--k^-)/p^-}\frac{\sln}{2}\frac{\slnbar}{2}\frac{i}{(p-q)^+-k^+-\frac{(\bfp-\bfk-\bfq)^2}{p^--k^-}+\frac{i\eta}{p^--k^-}} \nonumber\\
& \hspace*{0.4cm}\times  igt^b\frac{\sln}{2}\frac{\slnbar}{2}\frac{i}{p^+-k^+-\frac{(\bfp-\bfk)^2}{p^--k^-}+\frac{i\eta}{p^--k^-}} \frac{igt^a\SP_{\nbar,\nu}(\bfQ_1)}{k^-(p^--k^-)/p^-}\frac{\sln}{2}\chi_{\nbar,p} \nonumber \\
&= ig_s^3 t^at^bt^a \int_0^1\frac{dx}{(1-x)}\int \frac{d^2\bfk}{2(2\pi)^3}\chibar_{\nbar,p}\frac{\sum_\lambda\SP^*_{\nbar,\lambda}(\bfQ_6)\SP_{\nbar,\lambda}(\bfQ_1)}{\bfQ_6^2\bfQ_1^2}\frac{\sln}{2}\chi_{\nbar,p} \, ,
\end{align}
where $\bfQ_6 = x\bfk-(1-x)(\bfp-\bfk-\bfq)$.

For gluon scattering we find
\begin{align}
M_{\bar{n},5} &= \int \frac{dk^-dk^+d^2\bfk}{2(2\pi)^4} \frac{i}{(k-q)^2+i\eta}\sum_\lambda\epsilon_\lambda^\mu\epsilon_\lambda^{\rho}
gf^{abc}k^+g_{T,\rho\rho'}
\frac{i}{k^2+i\eta}\sum_\theta\epsilon_\theta^{\rho'}\epsilon_\theta^{\nu} \nonumber \\
& \hspace*{0.4cm}\times  \chibar_{\nbar,p} \frac{igt^a\SP_{\nbar,\mu}^*(\bfQ_7)}{k^-(p^--k^-)/p^-}\frac{\sln}{2}\frac{\slnbar}{2}\frac{i}{p^+-k^+-\frac{(\bfp-\bfk)^2}{p^--k^-}+\frac{i\eta}{p^--k^-}} \frac{igt^c\SP_{\nbar,\nu}(\bfQ_1)}{k^-(p^--k^-)/p^-}\frac{\sln}{2}\chi_{\nbar,p} \nonumber\\
&= -g_s^3 f^{abc}t^at^c\int_0^1\frac{dx}{(1-x)}\int \frac{d^2\bfk}{2(2\pi)^3}\chibar_{\nbar,p}\frac{\sum_\lambda\SP^*_{\nbar,\lambda}(\bfQ_7)\SP_{\nbar,\lambda}(\bfQ_1)}{\bfQ_7^2\bfQ_1^2}\chi_{\nbar,p} \, , 
\end{align}
where $\bfQ_7=x(\bfk-\bfq)-(1-x)(\bfp-\bfk)$.

\paragraph{Contribution to the TMD cross section.}
The above calculation is similar for the case of a gluonic medium color source.
Taking the squared amplitude with medium quarks as an example, we get the NLO correction to the collinear source density
\begin{align}
C_{\nbar}(x,b) &=\delta(1-x)
\frac{(4\pi \alpha_s C_F)^2 }{d_A}
\mathcal{N}_q L
\int \frac{d^2\bfq}{(2\pi)^2}\frac{e^{i\bfq\cdot \bfb}}{\bfq^4}  \int_0^1 dx' g_s^2 P_{qq}(x') \int\frac{d^2\bfk}{(2\pi)^3} \nonumber\\
& \hspace*{0.4cm}\times  \left\{C_F\left|\frac{\bfQ_8}{\bfQ_8^2}-\frac{\bfQ_6}{\bfQ_6^2}\right|^2 + C_F\left|\frac{\bfQ_8}{\bfQ_8^2}-\frac{\bfQ_1}{\bfQ_1^2}\right|^2\right. \nonumber\\
&\hspace*{0.4cm} -(2C_F-C_A) \left(\frac{\bfQ_8}{\bfQ_8^2}-\frac{\bfQ_6}{\bfQ_6^2}\right)\cdot\left(\frac{\bfQ_8}{\bfQ_8^2}-\frac{\bfQ_1}{\bfQ_1^2}\right)\nonumber\\
& \hspace*{0.4cm} \left. +(2C_F-C_A)\frac{\bfQ_1\bfQ_6}{\bfQ_1^2\bfQ_6^2} + C_A \frac{\bfQ_1\bfQ_7}{\bfQ_1^2\bfQ_7^2}\right\}\, .
\end{align}

 After shifting integration variables and dropping scaleless integrals, for a quark source we find
\begin{align}
C_{\nbar}(x,b) &\supset \delta(1-x)\frac{(4\pi \alpha_s C_F)^2 }{d_A}\mathcal{N}_q L 
 \left[\frac{p^-}{\nu}\right]^{-\tau} \left(-\frac{1}{\tau}+\frac{1}{2}\frac{\tau-3}{\tau^2-3\tau+2}\right) \nonumber
 \\
& \hspace*{0.4cm} \times \int\frac{d^2\bfq}{(2\pi)^2} \frac{e^{i\bfq\cdot\bfb}}{\bfq^2}   \frac{\alpha_s^{(0)} C_A}{\pi} \int \frac{d^2\bfk}{\pi} \left[ \frac{1}{(\bfk+\bfq)^2}v(\bfk^2)-\frac{1}{2} \frac{\bfq^2}{\bfk^2(\bfk+\bfq)^2}v(\bfq^2)\right] \nonumber \\
&\supset \delta(1-x)\alpha_s C_F \frac{4\pi\alpha_s C_F }{d_A} \mathcal{N}_q L
 \left[-\frac{1}{\tau}+\ln \frac{p^- e^{-3/4}}{\nu}+\mathcal{O}(\tau)\right]  \int \frac{d^2\bfq}{(2\pi)^2} \frac{e^{i\bfq\cdot\bfb}}{\bfq^2}  \hat{\mathcal{C}}\left[v(\bfq^2)\right].
 \label{app:eq:rapidity-div-3}
\end{align}
Adding up the rapidity divergent term from the collinear sector in Eq. (\ref{eq:A-rap-final-form}), the soft sector Eq. (\ref{app:eq:rapidity-div-2}), and Eq. (\ref{app:eq:rapidity-div-3}) for the anti-collinear sector, the rapidity divergences in the form of $1/\tau$ poles are canceled.

\end{document}